\documentclass[aps,prd,amsfonts,amssymb,amsmath,nofootinbib,showpacs,superscriptaddress,twocolumn,floatfix]{revtex4-1}
\usepackage{graphicx}
\usepackage{color}
\usepackage{mathrsfs}
\usepackage[caption=false]{subfig}
\usepackage{dcolumn}
\usepackage[breaklinks=true]{hyperref}
\usepackage{textcomp}

\newcommand{\eqnref}[1]{Eq.~(\ref{eq:#1})}
\newcommand{\figref}[1]{Fig.~\ref{fig:#1}}
\newcommand{\Figref}[1]{Figure~\ref{fig:#1}}
\newcommand{\secref}[1]{Sec.~\ref{sec:#1}}
\newcommand{\apref}[1]{Appendix~\ref{sec:#1}}

\newcommand{\units}[1]{\ensuremath{~\mathrm{#1}}}

\newcommand{\sub}[1]{\ensuremath{_\mathrm{#1}}}

\newcommand{\dd}{\ensuremath{\mathrm{d}}}
\newcommand{\diff}[2]{\ensuremath{\dfrac{\dd {#1}}{\dd {#2}}}}

\newcommand{\diffop}[1]{\ensuremath{\dfrac{\dd}{\dd {#1}}}}

\newcommand{\partialdiff}[2]{\ensuremath{\dfrac{\partial {#1}}{\partial {#2}}}}

\newcommand{\partialdiffop}[1]{\ensuremath{\dfrac{\partial}{\partial {#1}}}}
\newcommand{\recip}[1]{\ensuremath{\frac{1}{#1}}}

\newcommand{\intd}[4]{\ensuremath{\int_{#1}^{#2}{#3}\,\dd{#4}}}
\newcommand{\order}[1]{\ensuremath{\mathcal{O}({#1})}}

\newcommand{\overlap}[2]{\ensuremath{\left(#1\middle|#2\right)}}

\DeclareMathOperator{\sgn}{sgn}

\begin{document}


\title{Importance of transient resonances in extreme-mass-ratio inspirals}

\author{Christopher P.L. Berry}
\email[]{cplb@star.sr.bham.ac.uk}
\affiliation{School of Physics and Astronomy, University of Birmingham, Edgbaston, Birmingham B15 2TT, United Kingdom}
\affiliation{Institute of Astronomy, Madingley Road, Cambridge, CB3 0HA, United Kingdom}
\author{Robert H. Cole}
\affiliation{Institute of Astronomy, Madingley Road, Cambridge, CB3 0HA, United Kingdom}
\author{Priscilla Ca\~{n}izares}
\affiliation{Institute of Mathematics, Astrophysics and Particle Physics, Radboud University, Heyendaalseweg 135, 6525 AJ Nijmegen, The Netherlands}
\affiliation{Institute of Astronomy, Madingley Road, Cambridge, CB3 0HA, United Kingdom}
\author{Jonathan R. Gair}
\affiliation{School of Mathematics, University of Edinburgh, Peter Guthrie Tait Road, Edinburgh EH9 3FD, United Kingdom}
\affiliation{Institute of Astronomy, Madingley Road, Cambridge, CB3 0HA, United Kingdom}

\date{27 December 2016}

\begin{abstract}
The inspiral of stellar-mass compact objects, like neutron stars or stellar-mass black holes, into supermassive black holes provides a wealth of information about the strong gravitational-field regime via the emission of gravitational waves.
In order to detect and analyse these signals, accurate waveform templates which include the effects of the compact object's gravitational self-force are required. For computational efficiency, adiabatic templates are often used. These accurately reproduce orbit-averaged trajectories arising from the first-order self-force, but neglect other effects, such as transient resonances, where the radial and poloidal fundamental frequencies become commensurate.
During such resonances the flux of gravitational waves can be diminished or enhanced, leading to a shift in the compact object's trajectory and the phase of the waveform.
We present an evolution scheme for studying the effects of transient resonances and apply this to an astrophysically motivated population.
We find that a large proportion of systems encounter a low-order resonance in the later stages of inspiral; however, the resulting effect on signal-to-noise recovery is small as a consequence of the low eccentricity of the inspirals. Neglecting the effects of transient resonances leads to a loss of $4\%$ of detectable signals.
\end{abstract}


\maketitle

\section{Introduction}
 
In the prologue to his classic monograph, Chandrasekhar~\cite{Chandrasekhar1992} celebrates the simplicity of black holes (BHs). The Kerr solution is defined by just two parameters: mass and spin. Despite the baldness of the BH metrics, great intricacies manifest in their properties. This is made evident when a second body is introduced. The two-body problem in general relativity (GR) is well studied. It is of paramount importance for gravitational-wave (GW) astronomy, where binary systems are the dominant source of radiation. Correctly modelling the dynamics of these systems is necessary to interpret and extract information from gravitational waveforms.

We have made progress in understanding the general relativistic two-body problem in recent years. Bodies of comparable mass can be studied using numerical relativity. Rapid advances in this field have been made following breakthroughs in 2005~\cite{Pretorius2005,Campanelli2006,Baker2006}; it is now possible to simulate hundreds of orbits~\cite{Szilagyi2015}. However, the computational cost of numerical-relativity simulations means that other approaches must be used to generate the large number of waveforms required for GW detection and analysis. Analytic relativity approaches such as post-Newtonian (PN) theory~\cite{Blanchet2014,Buonanno2009}, which can be used to model the early inspiral where the gravitational field is still relatively weak, and the effective-one-body formalism~\cite{Buonanno1999,Buonanno2000,Damour2009,Barausse2010}, which can incorporate merger and ringdown, allow us to generate less expensive waveform approximants. These approximants can be calibrated to match numerical relativity results for improved accuracy~\cite{Taracchini2014,Pan2014,Husa2015,Khan2015,Schmidt2015}, and the resulting waveforms allow us to understand comparable-mass binary BHs.

Stellar-mass BH mergers are targets for ground-based GW detectors, such as Advanced LIGO~\cite{Aasi2015} and Advanced Virgo~\cite{Acernese2015}, the in-construction KAGRA~\cite{Aso2013}, and the proposed Einstein Telescope~\cite{Punturo2010a}. The first direct observations of GWs came from the coalescences of two stellar-mass BHs~\cite{Abbott2016,Abbott2016e,Abbott2016d}, and analysis of their properties~\cite{Abbott2016f,Abbott2016h,Abbott2016d} (plus subsequent inferences about their astrophysical origin~\cite{Abbott2016i,Abbott2016d} and tests of GR~\cite{Abbott2016j,Yunes2016,Abbott2016d}) relied upon our knowledge of binary BH waveforms. 

Systems of unequal masses are more challenging to evolve numerically as they complete a larger number of orbits, and it is necessary to resolve two different scales. Calculations can instead be performed perturbatively. The paradigm unequal-mass system has a stellar-mass BH orbiting a supermassive BH (SMBH), such as those expected to be found at the centres of galaxies~\cite{Kormendy1995,Ferrarese2005,Boehle2016}. These extreme-mass-ratio inspirals (EMRIs) produce GWs that are a promising signal for space-borne detectors like the evolving Laser Interferometer Space Antenna (eLISA)~\cite{Amaro-Seoane2007,Amaro-Seoane2012a}. EMRIs provide a chance to measure the properties of SMBHs~\cite{Barack2004,Arun2009}, their evolution~\cite{Gair2010b,Gair2010a} and environment~\cite{Yunes2011a,Barausse2014b}, and also test for deviations from the predictions of GR~\cite{Barack2007,Gair2012a}. To detect and analyse EMRI signals we must have waveforms for generic orbits which are accurate for the $\sim10^4$--$10^5$ cycles of the inspiral.

To improve our understanding of extreme-mass-ratio systems, efforts are concentrated on modelling the gravitational self-force~\cite{Barack2009,Poisson2004,Pound2015}. In the test-particle limit, the smaller body follows an exact geodesic of the SMBH's spacetime. Including the effects of the smaller body's finite mass, the background spacetime is perturbed. The backreaction from this deformation alters the small body's orbital trajectory, and can be modelled as a self-force that moves the body from its geodesic. The self-force can be divided into two pieces, dissipative and conservative~\cite{Sago2008,Barack2009}. The former encapsulates the slow decay of the orbital energy and angular momentum (constants of the motion in the test particle limit) through radiation of GWs. The latter shifts the orbital phases inducing precession. The dissipative piece is time asymmetric and has the larger effect on the evolution of the orbital phase; the conservative piece is time symmetric and has a smaller influence on the phase, although this can accumulate over many orbits. Being able to accurately model the influence of the self-force allows us to create reliable waveform models.

Flanagan and Hinderer~\cite{Flanagan2012} highlighted a previously overlooked phenomenon that occurs in the general relativistic two-body problem, that of transient resonances. Geodesic orbits in GR have three associated frequencies: the radial frequency $\Omega_r$, the polar frequency $\Omega_\theta$ and the azimuthal frequency $\Omega_\phi$.\footnote{In the strong-field regime, it is possible to have isofrequency pairings, where two different orbits share the same orbital frequencies~\cite{Warburton2013}. The evolution of the frequencies still differ, such that orbital trajectories can be reconstructed from the frequencies.} The first two describe libration and the third rotation (except in the case of polar orbits where $\Omega_\theta$ also describes rotation)~\cite{Goldstein2002}. 
In the weak-field limit, these all tend towards the Keplerian frequency; in the strong-field regime they may differ significantly. For EMRIs, the evolution timescale is much longer than the orbital period such that the motion of the smaller body is approximately geodesic over orbital timescales. The inspiral of the orbit can be approximated as a series of geodesics using the osculating element formalism~\cite{Pound2008,Gair2011a}. During this evolution, the frequencies may become commensurate: resonances occur when the radial and polar frequencies are rational multiples of each other:
\begin{equation}
\nu \equiv \frac{\Omega_r}{\Omega_\theta} = \frac{n_\theta}{n_r},
\end{equation}
where $n_r$ and $n_\theta$ are integers (with no common factors). During resonance, terms in the self-force that usually average to zero can combine coherently, significantly impacting the orbital motion~\cite{Flanagan2012a}.

Resonances involving the azimuthal motion do not produce a comparable effect because of the axisymmetry of the background spacetime. However, both $\theta$--$\phi$ resonances~\cite{Hirata2011} and $r$--$\phi$ resonances~\cite{VanDeMeent2013} can lead to extrinsic effects; the GWs from such systems are not emitted isotropically and the imbalance produces a kick velocity that is, in some cases with moderate mass ratios, sufficient to eject the central BH from its host~\cite{VanDeMeent2014}.

Geodesic motion in Kerr spacetime can be described by use of the action--angle formalism~\cite{Goldstein2002}. 
We consider a body of mass $\mu$ orbiting a BH of mass $M$, with $\eta = \mu/M \ll 1$,\footnote{To first order, the mass ratio $\eta$ is the same as the symmetric mass ratio $\mu M/(\mu+M)^2$.} and describe the motion in the directions of the standard Boyer--Lindquist coordinates $\{t,r,\theta,\phi\}$~\cite{Boyer1967} using generalised angle variables $q_\alpha = \{q_t,q_r,q_\theta,q_\phi\}$~\cite{Hinderer2008}. We denote the first integrals of the geodesic motion, the generalised action variables, by $J_\alpha$. These are some combination of the energy per unit mass $E$ and the axial angular momentum per unit mass $L_z$ of the orbit, which arise from isometries of the metric in $t$ and $\phi$, and the Carter constant per unit mass squared $Q$~\cite{Carter1968}, which is related to the separability of the equations of motion in $r$ and $\theta$. The system evolves following~\cite{Flanagan2012}
\begin{subequations}
\label{eq:Mino-E-o-M}
\begin{align}
\diff{q_\alpha}{\lambda} = {} & \omega_\alpha(\boldsymbol{J}) + \eta g_\alpha^{(1)}(q_r,q_\theta,\boldsymbol{J}) + \order{\eta^2}, \\
\diff{J_\alpha}{\lambda} = {} & \eta G_\alpha^{(1)}(q_r,q_\theta,\boldsymbol{J}) + \order{\eta^2},
\end{align}
\end{subequations}
where $\lambda$ is Mino time~\cite{Mino2003}, and the forcing functions $g_\alpha^{(1)}$ and $G_A^{(1)}$ originate from the first-order self-force.\footnote{For a discussion of the second-order self-force, see \cite{Rosenthal2006,Pound2012,Gralla2012}.} By working with $\lambda$ instead of proper time $\tau$, the radial and polar motions decouple. At zeroth order in the mass ratio we recover the limit of purely geodesic motion: the integrals of the motion are actually constants and the angle variables evolve according to their associated frequencies $\omega_\alpha$.

The leading-order dissipative correction to geodesic motion is calculated following the adiabatic prescription~\cite{Hinderer2008}: by dropping the forcing term $g_\alpha^{(1)}$ (and all higher-order terms) and replacing the forcing term $G_\alpha^{(1)}$ with $\langle G_\alpha^{(1)}\rangle_{q_r,\,q_\theta}$, its average over the $2$-torus parametrized by $q_r$ and $q_\theta$~\cite{Drasco2005}. For most orbits this is sufficient, $G_\alpha^{(1)}$ is given by its average value plus a rapidly oscillating component~\cite{Arnold1988}. 
However, this averaging fails when the ratio of frequencies is the ratio of integers. In this case the trajectory does not ergodically fill the $2$-torus, but instead traces out a $1$-dimensional subspace.\footnote{For illustrations, see Grossman, Levin and Perez-Giz~\cite{Grossman2012}.} There are then contributions to the self-force that no longer average out beyond $\langle G_A^{(1)}\rangle_{q_r,\,q_\theta}$. Intuitively, we expect that this effect is more important for ratios of small integers since when the integers are large the orbit comes close to all points on the $2$-torus.

In this work we seek to characterise the importance of these resonances for the purposes of modelling EMRIs. The amplitude of expected signals is below the level of noise in a space-based GW detector. However, systems remain in band for many hundreds of thousands of cycles and so may be detected using a matched filter, provided we have sufficiently accurate waveform templates. Ensuring the accuracy of EMRI templates requires calculating the impact that passing through a resonance has on the orbital evolution and discovering for which resonances this is significant.

We show how the properties of resonances can be understood from the properties of the orbit. The effects of passing through resonance depend sensitively on the phase at resonance, making them difficult to predict without detailed calculation. The low-order resonances, such as the $1$:$2$ and $2$:$3$ resonances, can leave a noticeable imprint on the waveform. However, since most EMRIs have a low eccentricity when passing through these resonances, we find that for an astrophysical population of EMRIs there should not be a significant reduction in detectable signals when using adiabatic waveforms. The effect of resonances on parameter estimation is yet to be investigated.

In \secref{problem}, we formulate the specific problem: that of geodesic motion in Kerr spacetime, perturbed  by the gravitational self-force. We then study generic properties of transient resonances in \secref{properties}, detailing their location in parameter space, the timescales over which they affect the motion and the resulting GW flux enhancements. Specific examples are considered to illustrate the effects of resonances in \secref{waveforms}, before finally turning to an astrophysical population in \secref{astrophysics}. Our conclusions can be found in \secref{conclusion}.

We use geometric units with $G = c = 1$ throughout. We always use $M$ for the mass of the central SMBH and $a$ as its Kerr spin parameter. We also use the dimensionless spin $a_\ast \equiv a/M$; we take the convention that $0 \leq a_\ast < 1$. We assume a standard cosmology with $\Omega_\Lambda = 0.7$, $\Omega\sub{m} = 0.3$ and $H_0 = 70\units{km\, s^{-1}\, Mpc^{-1}}$ and do not expect the exact details of the cosmology to significantly alter our results~\cite{Mapelli2012}.

\section{The problem of EMRI transient resonances}
\label{sec:problem}

The evolution of an extreme-mass-ratio ($\eta \ll 1$) system is slow. Instantaneously, the motion of the orbiting mass can be described as geodesic, with the integrals of the motion changing on timescales of many orbital periods. It is therefore necessary to develop an understanding of the Kerr geodesics (\secref{geodesic}; those familiar with calculating orbits in Kerr may skip this section.). Transient resonances occur when the radial and polar frequencies become commensurate (\secref{frequencies}); we analyse the behaviour of resonances within the osculating element framework, where the trajectory is described by a sequence of geodesics that each match onto the motion at a particular instance (\secref{forced-motion}). The osculating elements formalism allows for the orbital evolution to be driven by a force, here, a particular model for the self-force (\secref{self-force}) and its adiabatic average (\secref{adiabatic}). In following sections, we study the differences between the adiabatic and full orbital evolutions.

\subsection{Kerr geodesics}
\label{sec:geodesic}

Central to understanding transient resonances is a knowledge of orbits in Kerr spacetime, and hence we begin with details of evolving Kerr geodesics. The geodesic equations may be written as~\cite{Carter1968, Chandrasekhar1992} 
\begin{subequations}
\begin{align}
\diff{t}{\lambda} = {} & a\left(L_z - aE\sin^2 \theta\right) + \frac{r^2 + a^2}{\Delta}\mathcal{T},\\
\diff{r}{\lambda} = {} & \pm \sqrt{V_r},\\
\diff{\theta}{\lambda} = {} & \pm \sqrt{V_\theta},\\
\diff{\phi}{\lambda} = {} & \frac{L_z}{\sin^2 \theta} - aE + \frac{a}{\Delta}\mathcal{T},
\end{align}
\end{subequations}
where $\Delta = r^2 - 2M r + a^2$; the signs of the $r$ and $\theta$ equations can be chosen independently, and we have introduced potentials
\begin{subequations}
\begin{align}
\mathcal{T} = {} & E\left(r^2 +a^2\right) - aL_z,\\
V_r = {} & \mathcal{T}^2 - \Delta\left[r^2 + \left(L_z -aE\right)^2 + Q\right],\\
V_\theta = {} & Q - \cos^2 \theta\left[a^2\left(1 - E^2\right) + {\displaystyle \frac{L_z^2}{\sin^2\theta}}\right].
\end{align}
\end{subequations}
As an affine parameter, we have used Mino time which is related to the proper time $\tau$ by~\cite{Mino2003}
\begin{equation}
\tau = \intd{}{}{r^2 + a^2 \cos^2\theta}{\lambda}.
\end{equation}
Using Mino time allows us to decouple the $r$ and $\theta$ motions.

We only consider bound motion~\cite{Wilkins1972}: the radial motion covers a range $r\sub{p} \leq r \leq r\sub{a}$, where the turning points are the periapsis $r\sub{p}$ and apoapsis $r\sub{a}$. Drawing upon Keplerian orbits we parametrize the motion using
\begin{equation}
r = \frac{p M}{1+e\cos\psi},
\end{equation}
introducing eccentricity $e$, (dimensionless) semilatus rectum $p$ and relativistic anomaly $\psi$~\cite{Darwin1961,Drasco2004}. While $r$ oscillates between its maximum and minimum values, $\psi$ increases secularly, increasing by $2\pi$ across an orbit. The polar motion covers a range $\theta_- \leq \theta \leq \pi - \theta_-$. We also parametrize this motion in terms of an angular phase $\chi$, according to~\cite{Hughes2000}
\begin{equation}
\cos\theta = \cos\theta_-\cos\chi.
\end{equation}
While $\psi$ and $\chi$ are $2\pi$ periodic they are not the canonical action--angle variables~\cite{Schmidt2002}; they are, however, easy to work with.

The geodesic motion can equally be described by $\{E,L_z,Q\}$ or $\{p,e,\theta_-\}$~\cite{Schmidt2002}. Converting between them requires finding the solutions of $V_r = 0$ and $V_\theta = 0$. We employ a slightly different parameter set of $\{p,e,\iota\}$ where we have introduced the inclination~\cite{Ryan1996,Glampedakis2002}
\begin{equation}
\tan \iota = \frac{\sqrt{Q}}{L_z}.
\end{equation}
This is $0 \leq \iota < \pi/2$ for prograde orbits and $\pi/2 < \iota \leq \pi$ for retrograde orbits. Equatorial orbits ($\theta_- = \pi/2$) have $\iota = 0$ or $\pi$ and polar orbits ($\theta_- = 0$) have $\iota = \pi/2$. While formulae exist for conversion between the different parameters, these are complicated and uninsightful, so we do not reproduce them here.\footnote{In practice we find turning points numerically.}

\subsection{Orbital resonances}
\label{sec:frequencies}

The radial and polar orbital periods in Mino time are given by
\begin{subequations}
\begin{align}
\Lambda_r = {} & 2\intd{r\sub{p}}{r\sub{a}}{\recip{\sqrt{V_r}}}{r} = \intd{-\pi}{\pi}{\diff{\lambda}{\psi}}{\psi}, \\
\Lambda_\theta = {} & 4\intd{\theta_-}{\pi/2}{\recip{\sqrt{V_\theta}}}{\theta} = \intd{-\pi}{\pi}{\diff{\lambda}{\chi}}{\chi}.
\end{align}
\end{subequations}
The orbital frequencies are thus~\cite{Fujita2009}
\begin{equation}
\Upsilon_r = \frac{2\pi}{\Lambda_r}, \quad \Upsilon_\theta = \frac{2\pi}{\Lambda_\theta}.
\end{equation}
The geodesic equations for coordinate time $t$ and azimuthal angle $\phi$ are just functions of $r$ and $\theta$, hence their evolutions can be expressed as Fourier series~\cite{Drasco2004}
\begin{subequations}\label{eq:Mino-Fourier}
\begin{align}
\diff{t}{\lambda} = {} & \sum_{k_r,\,k_\theta}T_{k_r,\, k_\theta}\exp\left[-i\left(k_r\Upsilon_r + k_\theta\Upsilon_\theta\right)\lambda\right], \\
\diff{\phi}{\lambda} = {} & \sum_{k_r,\,k_\theta}\Phi_{k_r,\, k_\theta}\exp\left[-i\left(k_r\Upsilon_r + k_\theta\Upsilon_\theta\right)\lambda\right].
\end{align}
\end{subequations}
The $(0,\,0)$ coefficients in these series give the average secular rate of increase of these quantities. We define
\begin{equation}
\Gamma = T_{0,\,0}, \quad \Upsilon_\phi = \Phi_{0,\,0}
\end{equation}
to act as Mino-time frequencies. We can now convert to coordinate-time frequencies with~\cite{Drasco2004}
\begin{equation}
\Omega_r = \frac{\Upsilon_r}{\Gamma}, \quad \Omega_\theta = \frac{\Upsilon_\theta}{\Gamma}, \quad \Omega_\phi = \frac{\Upsilon_\phi}{\Gamma}.
\end{equation}

Transient resonances occur when the radial and poloidal motions are commensurate, when
\begin{equation}
\nu = \frac{\Upsilon_r}{\Upsilon_\theta} = \frac{\Omega_r}{\Omega_\theta} = \frac{n_\theta}{n_r}
\end{equation}
is the ratio of small integers. At this point, any Fourier series like those in \eqnref{Mino-Fourier} goes from being an expansion in two frequencies to being an expansion in a single frequency~\cite{Bosley1992}.

For a general nonresonant orbit there is no fixed correlation between the radial and polar coordinates. After a sufficiently long time, the trajectory comes arbitrarily close to every point in the range of motion (with $r\sub{p} \leq r \leq r\sub{a}$ and $\theta_- \leq \theta \leq \pi - \theta_-$); on account of the orbital precession, the whole space is densely covered. This does not happen on resonance, as the radial and polar motions are locked together such that we can express one as a function of the other, and so the trajectory keeps cycling over the same path. The points visited are controlled by the relative phases of the $r$ and $\theta$ motions. To represent this, we use the $r$ phase at the $\theta$ turning point $\psi_{\theta_-} = \psi(\chi = 0)$. Varying $\psi_{\theta_-}$ across its full range allows every point in the range of motion to be reached. Hence averaging over all values of $\psi_{\theta_-}$ for resonant orbits is equivalent to averaging over the $\psi$--$\chi$ $2$-torus for nonresonant orbits.

One might be concerned about the nature of resonances following the inclusion of the self-force: true geodesic motion only exists at zeroth order in $\eta$ and, while it is a good approximation over short timescales, for small $\eta$ there is a small disparity. The conservative piece of the self-force induces extra precession which leads to a slight shift in the orbital frequencies~\cite{Warburton2012}.\footnote{The Kolmogorov--Arnold--Moser (KAM) theorem states that when an integrable Hamiltonian (i.e.\ the case for motion in Kerr) is subject to a small perturbation the form of the orbits is preserved albeit slightly deformed~\cite{Arnold1963,Moser1973}. 
This should ensure that, in general, there are only small shifts in the orbital frequencies. However, the KAM theory is only valid for sufficiently incommensurate orbits: close to resonance it does not apply~\cite{Moser1973}. 
This is a further reason why resonances merit an in-depth investigation.}
The dissipative piece causes the frequencies to evolve and, hence, the resonance cannot persist for multiple orbits (without some feedback coupling). In effect, we are really considering a period of time about the resonant crossing. The instantaneous orbital frequencies oscillate back and forth around their averaged values. However, there is a time span when the frequencies are consistently close to being commensurate. During this time, the trajectory appears similar to a resonant trajectory, filling only a smaller region of the parameter space. It is this time period that is of interest for transient resonances~\cite{Bosley1992}.

\subsection{Osculating elements and forced motion}
\label{sec:forced-motion}

For generic EMRIs, there are two characteristic timescales: the fast orbital motion, related to the fundamental frequencies $\sim1/\Omega$, and the slow inspiral, related to the change in fundamental frequencies $\sim\Omega/\dot{\Omega}$, where an overdot denotes a derivative with respect to coordinate time $t$. These, along with the resonance timescale, are discussed more in \secref{res-time}. The two-timescale nature of the problem makes it ideally suited to the method of osculating elements~\cite{Pound2008,Gair2011a}: on short timescales, we analyse the unperturbed system resulting in geodesic motion, and then the long-term evolution is described by a sequence of instantaneous geodesics.

We require, at each instant in time, that the chosen geodesic matches the true position and velocity of the particle. This amounts to a specific choice of the orbital shape parameters (for example, the set $\{E,L_z,Q\}$ or the generalised action variables $J_\alpha$) and some initial phases at $t=t_0$ (for example, the set $\{\psi_0,\chi_0,\phi_0\}$). Collectively, these are referred to as \emph{osculating elements} and we denote them by $I^A(t)$, making explicit the variation with time. For a sequence of geodesics of a background spacetime, where the evolution is forced by some external acceleration (in our case from the self-force), we can calculate the evolution of the osculating elements $\dot{I}^A$. The specific equations for motion in Kerr are derived by Gair {\it{et al}}.~\cite{Gair2011a}. 


\subsection{Gravitational self-force model}
\label{sec:self-force}

To follow the evolution of the inspiral we must have a means of prescribing the forcing acceleration which causes the orbit to deviate from a single geodesic. We work directly with the gravitational self-force, using the same PN approximation as Flanagan and Hinderer~\cite{Flanagan2012}. For comparison, Flanagan, Hughes and Ruangsri~\cite{Flanagan2012a} use a Teukolsky-equation calculation of GW fluxes to account for the inspiral due to radiation reaction.

The self-force model uses the first-order PN terms of the dissipative self-force formulated in Flanagan and Hinderer~\cite{Flanagan2007} and the conservative force formulated in Iyer and Will~\cite{Iyer1993}, and Kidder~\cite{Kidder1995}. Since only the first PN terms are used, this prescription is of limited validity in strong fields. Both pieces of the self-force are computed assuming that the SMBH's spin is small: the dissipative piece contains terms to $\order{a_\ast^2}$ and the conservative piece to $\order{a_\ast}$. This is suboptimal for high spins. We also find that this particular implementation of the self-force model marginally overestimates the adiabatic inspiral rate with respect to direct PN evolutions by a factor of $\mathcal{O}(1)$, even for systems in the weak field and with low values of the spin. While this approximate self-force is not perfect, it should serve as a guide for the behaviour of the full self-force, allowing us to assess the qualitative impact of resonances on EMRI detection.


\subsection{Adiabatic evolution}
\label{sec:adiabatic}

Beyond geodesic motion in the Kerr spacetime, a test particle follows an accelerated trajectory determined by \eqnref{Mino-E-o-M}. This may be approximated by the adiabatic prescription~\cite{Hinderer2008} by dropping the forcing term $g_\alpha^{(1)}$ (and all higher-order terms) and replacing $G_\alpha^{(1)}$ with its average over the $2$-torus parametrized by $q_r$ and $q_\theta$, $\langle G_\alpha^{(1)}\rangle_{q_r,\,q_\theta}$~\cite{Drasco2005,Grossman2011}. The averaged force can be computed from the radiative field~\cite{Galtsov1982,Mino2003,Sago2006,Ganz2007}. This piece is purely dissipative~\cite{Pound2005} and determines how the inspiral evolves due to the radiation of GWs.

To construct an adiabatic trajectory we need the $2$-torus-averaged fluxes of our osculating elements. To guarantee consistency, we average our instantaneous self-force. Computing an average of a quantity over the $\{q_r, q_\theta\}$ is trivial if it is parametrized in terms of these variables,
\begin{equation}
\left\langle \diff{X}{\lambda}\right\rangle_{q_r,\,q_\theta} = \frac{1}{(2\pi)^2}\intd{0}{2\pi}{\intd{0}{2\pi}{\diff{X}{\lambda}}{q_r}}{q_\theta}.
\end{equation}
However, we are using $\psi$ and $\chi$, as these are simpler to evolve; furthermore, we compute instantaneous coordinate-time fluxes $\dot{X}$, not Mino-time fluxes. Changing variables gives an average of \cite{Drasco2005}
\begin{widetext}\begin{align}
\left\langle \diff{X}{\lambda}\right\rangle_{q_r,\,q_\theta} = {} & \frac{1}{\Lambda_r \Lambda_\theta}\intd{0}{2\pi}{\intd{0}{2\pi}{\left(\diff{\psi}{t}\right)^{-1} \left(\diff{\chi}{t}\right)^{-1} \left(\diff{t}{\lambda}\right)^{-2} \diff{X}{\lambda}}{\psi}}{\chi} \\
 = {} & \frac{1}{\Lambda_r \Lambda_\theta}\intd{0}{2\pi}{\intd{0}{2\pi}{\left(\diff{\psi}{t}\right)^{-1} \left(\diff{\chi}{t}\right)^{-1} \left(\diff{t}{\lambda}\right)^{-1} \dot{X}}{\psi}}{\chi}.
\label{eq:2torus-average}
\end{align}\end{widetext}
This average describes the Mino-time rate of change of the quantity $X$ over an orbit. To convert to a coordinate flux of the averaged quantity, we simply divide by the period $\Gamma$~\cite{Flanagan2012a}, defining
\begin{equation}
\dot{\left\langle X\right\rangle}_{q_r,\,q_\theta} = \frac{1}{\Gamma}\left\langle \diff{X}{\lambda}\right\rangle_{q_r,\,q_\theta}.
\end{equation}
It is convenient to calculate $\Gamma$ as
\begin{equation}
\Gamma = \left\langle \diff{t}{\lambda}\right\rangle_{q_r,\,q_\theta},
\end{equation}
using \eqnref{2torus-average}, as this allows us to eliminate $\Lambda_r$ and $\Lambda_\theta$ from the calculation.\footnote{We compute these integrals using a $300\times 300$ grid of $\{\psi,\chi\}$ values and employing a Newton--Cotes approximation in each dimension. The procedure requires $\mathcal{O}(10^5)$ separate evaluations of the derivatives at each time step of an evolution, and so is computationally expensive to perform. However, the adiabatic derivatives vary on much longer timescales than the orbital motion (see \secref{res-time}), and so in practice, we can interpolate.}
The averaged fluxes successfully describe the leading-order secular evolution of the trajectory (as illustrated in \figref{good-traj}).

The combination of a full instantaneous evolution and an adiabatic evolution allows us to systematically study the effect of transient resonances on EMRIs over the course of an inspiral. Before approaching this problem, we first investigate the properties of the resonances themselves.

\section{Properties of transient resonances}
\label{sec:properties}

The first step in studying the effect of transient resonances is to locate orbital parameters for which the frequencies are commensurate. We can calculate the frequencies and so we are left with the problem of solving $\Omega = n_r \Omega_r - n_\theta \Omega_\theta = 0$ numerically. When considering the full parameter set of $\{p,e,\iota,a_\ast,\nu\}$, it is apparent that the search for resonances becomes expensive as a consequence of the dimensionality. It is therefore useful to have a guide of where to look. In \apref{location} we build a simple approximate model as a starting point for the numerical search. The resonances occur at relatively small periapses, corresponding to regions of strong-field gravity. Having located where in an inspiral we can expect to encounter a transient resonance, we must now consider its impact. In \secref{res-time} we determine the characteristic timescales describing resonance, and in \secref{flux-enhance} we calculate the impact of passing through a resonance on the evolution of the orbit.

\subsection{Timescales}\label{sec:res-time}

When analysing resonances it is useful to refer to a number of characteristic timescales.  We always use coordinate time $t$ for these, as this corresponds to what is measured by an observer at infinity. Translation to Mino time can be done with an appropriate factor of $\Gamma$. We use the orbital period $T$, the evolution timescale $\tau\sub{ev}$, the precession timescale $\tau\sub{pres}$ and the resonance timescale $\tau\sub{res}$.

The simplest timescales are the orbital periods $T_r = 2\pi/\Omega_r$, $T_\theta = 2\pi/\Omega_\theta$ and $T_\phi = 2\pi/\Omega_\phi$. These are the shortest in our set. We use $T$ to denote a timescale of the same order as the orbital periods.

We define the evolution timescale as
\begin{equation}
\tau\sub{ev} = \frac{\nu}{\dot{\nu}},
\end{equation}
where an overdot denotes a derivative with respect to $t$. In general, away from resonance, we take $\nu \equiv \Omega_r/\Omega_\theta < 1$. This timescale sets the period over which there is a significant change in the frequencies. It acts as an inspiral timescale. It is long in all cases we study, $\tau\sub{ev} \sim \order{T/\eta}$. It is this property which makes EMRIs interesting, as we can follow the waveform for many cycles, accruing high signal-to-noise ratios (SNRs). This is also what allows us to use the adiabatic prescription, as it means the trajectory moves slowly through different orbital parameters.

We use the precession timescale
\begin{equation}
\tau\sub{pres}(t) = \frac{2\pi}{|\Omega(t)|},
\label{eq:t-pres}
\end{equation}
with $\Omega(t) = n_r \Omega_r(t) - n_\theta \Omega_\theta(t)$, where the frequencies are calculated instantaneously and the integers are for the resonance of interest. This timescale becomes infinite exactly on resonance, but decreases as we get further from resonance, eventually becoming $\order{T}$. It measures the relative precession rate of the radial and polar motions and hence gives an indication of how long it takes to fill the entire $\psi$--$\chi$ $2$-torus.

We also use the resonance timescale (cf.\ \cite{Ruangsri2014})
\begin{equation}
\tau\sub{res} = \left[\frac{2\pi}{\left|\left\langle\dot{\Omega}(0)\right\rangle_{q'}\right|}\right]^{1/2}.
\label{eq:t-res}
\end{equation}
Here $\dot{\Omega}(0)$ is the rate of change of $\Omega$ at resonance, which we take to be at $t = 0$. The instantaneous $\dot{\Omega}$ depends upon the orbital phase and oscillates about its mean trend over an orbit. We are interested in the averaged behaviour, not the periodic modulations about this, which is why we use the time average $\langle\dot{\Omega}\rangle_{q'}$; here we use $q'$ to represent a phase that varies over an orbit with period of order $T$.\footnote{On resonance, we are interested in the relative $r$--$\theta$ orbital phase ($\psi_{\theta_-}$ or equivalent), which sets the resonant trajectory in the $r$--$\theta$ plane, but not the exact phase of the orbit around this loop.} Close to resonance, $\Omega(t)$ is well approximated by a first-order Taylor expansion, decreasing linearly with time; hence we make the approximation
\begin{equation}
\left|{\Omega(t)}\right| \simeq \left|\left\langle\dot{\Omega}(0)\right\rangle_{q'} t\right|.
\label{eq:Taylor-Omega}
\end{equation}
The resonant timescale should give an indication of the time over which we expect the effects of the resonance to be felt~\cite{Bosley1992}. Consider the phase of the Mino-time Fourier expansion on resonance; neglecting the constant, the resonant Fourier component has form
\begin{equation}
\varphi_{n_r,\,-n_\theta} \simeq \left(n_r\Upsilon_r - n_\theta\Upsilon_\theta\right)\lambda + \left(n_r\dot{\Upsilon}_r - n_\theta\dot{\Upsilon}_\theta\right)\lambda^2 + \ldots
\end{equation}
Typically, the first term is nonzero and this gives the familiar oscillation. On resonance, it is zero, leaving the next-order term to govern the behaviour~\cite{Flanagan2012,Ruangsri2014}. Only once we have moved far enough away from resonance for the first term to dominate the second do we recapture the nonresonant behaviour. The first term (translating from Mino time to coordinate time) sets $\tau\sub{pres}$, the second sets $\tau\sub{res}$.

Since we have argued that the effect of resonance can be thought of as a consequence of not densely covering the $\psi$--$\chi$ $2$-torus, we might expect that $\tau\sub{pres}$, as well as $\tau\sub{res}$, could be used for setting the resonance duration: the resonance ends once sufficient time has elapsed that the $2$-torus could be filled. This is indeed the case. Let $t\sub{pres}$ be the time taken to fill the torus, then
\begin{align}
t\sub{pres} = {} & \tau\sub{pres}(t\sub{pres}) \\
 \simeq {} & \dfrac{2\pi}{\left|\left\langle\dot{\Omega}(0)\right\rangle_{q'} t\sub{pres}\right|}, \nonumber 
\end{align}
using \eqnref{t-pres} and \eqnref{Taylor-Omega}. Rearranging and using \eqnref{t-res} gives
\begin{equation}
t\sub{pres} \simeq \tau\sub{res}.
\end{equation}
The two timescales are equivalent: we preferentially use $\tau\sub{res}$ to denote the resonance width. It is shorter than the inspiral timescale, but longer than an orbital period, $\tau\sub{res} \sim \order{\eta^{1/2}\tau\sub{ev}} \sim \order{\eta^{-1/2}T}$~\cite{Flanagan2012,Gair2011a}; it therefore acts as a bridge between the two timescales~\cite{Hinderer2008}.

Since we shall be considering Fourier decompositions, in anticipation of future results, we also define a timescale for the $s$-th resonant frequency harmonic
\begin{align}
\tau_{\mathrm{res},\,s} = {} & \left[\frac{2\pi}{\left|s\left\langle \dot{\Omega}(0)\right\rangle_{q'}\right|}\right]^{1/2}.
\label{eq:T-res-s}
\end{align}
This assumes that $s$ is a nonzero integer.

\subsection{Resonant flux enhancement}
\label{sec:flux-enhance}

Evolving through a resonance can lead to an enhancement (or decrement) of fluxes relative to the adiabatic prescription. After crossing the resonance region, the orbital parameters are different from those calculated from an adiabatic evolution. Flanagan and Hinderer~\cite{Flanagan2012} gave an expression for this deviation. If we denote the orbital parameters by $\mathcal{I}^a = \{E,L_z,Q\}$, then the change across resonance is
\begin{align}
\Delta \mathcal{I}^a = {} & \eta\sum_{s\,\neq\,0}F_{a,\,s}^{(1)}\left[\dfrac{2\pi}{\left|s \left\langle\dot{\Omega}\right\rangle_{q'}\right|}\right]^{1/2} \nonumber \\*
 {} & \times {} \exp\left[i\left(s \widehat{\kappa}_0 + \dfrac{\pi}{4} \sgn s\dot{\Omega}\right)\right]. 
\end{align}
Here $\widehat{\kappa}_0$ is the phase on resonance, which sets the resonant trajectory in the $r$--$\theta$ plane similarly to $\psi_{\theta_-}$, and $F_{a,\,s}^{(1)}$ is the $s$-th harmonic of the first-order self-force on resonance, defined such that\footnote{Since the geodesic equations decouple in Mino time rather than coordinate time, this is true only in an average sense.}
\begin{equation}
\diff{\mathcal{I}^a}{t} = \eta\sum_s F_{a,\,s}^{(1)}(\boldsymbol{\mathcal{I}})\exp(is q) + \order{\eta^2}.
\end{equation}
A derivation is presented in \apref{res-asymptotic}, which contains a more comprehensive explanation of the various terms. This employs matched asymptotic expansions to track the evolution through resonance, following the approach of Kevorkian~\cite{Kevorkian1987}.

To explain the form of this expression we substitute in our expression for the resonance width from \eqnref{T-res-s},
\begin{equation}
\label{eq:delta-I-a}
\Delta \mathcal{I}^a = \eta\sum_{s\,\neq\,0}F_{a,\,s}^{(1)}\tau_{\mathrm{res},\,s}\exp\left[i\left(s \widehat{\kappa}_0 + \dfrac{\pi}{4} \sgn s\dot{\Omega}\right)\right]. 
\end{equation}
Schematically, this can then be understood as the magnitude of the forcing function on resonance $\sim \eta F_{a,\,s}$ multiplied by the time on resonance $\sim \tau_{\mathrm{res},\,s}$ and a function that varies with the phase $\widehat{\kappa}_0$. Averaging over all values of $\widehat{\kappa}_0$ is equivalent to averaging over all values of $\psi_{\theta_-}$, and has the same effect as averaging over the $\psi$--$\chi$ $2$-torus~\cite{Grossman2011}; this gives an average discrepancy relative to the adiabatic evolution of
\begin{equation}
\left\langle \Delta \mathcal{I}^a \right\rangle_{\hat{\kappa}_0} = 0,
\end{equation}
exactly as expected.

Knowing where resonances are found in parameter space, how long they last, and how great an effect they are likely to have, enables us to study and interpret the observable effects of resonances on EMRI waveforms (\secref{waveforms}) and the population of observable inspirals (\secref{astrophysics}).

\section{The impact of resonances on EMRIs}
\label{sec:waveforms}

Having built an understanding of the properties of transient resonances, we now consider their impact on GW signals. In \secref{kludge} we discuss our chosen waveform generation scheme, giving a demonstration of its accuracy for evolutions which avoid low-order resonances in \secref{nonres}. In \secref{effres-phase} we detail the impact of resonances on the match between waveforms computed from adiabatic and fully instantaneous evolutions, and in \secref{effres-jump} we look at the changes in orbital parameters across resonance.

\subsection{Waveform model and analysis}
\label{sec:kludge}

One of the targets of pre-eLISA research is to generate a bank of waveform templates $\{h(t;\boldsymbol{\Theta}_i)\}$ across a range of parameter space $\{\boldsymbol{\Theta}_i\}$. These can be compared to data to search for the presence of GWs. Templates must accurately reproduce what we expect to observe in nature, without being too computationally expensive. Ideally, we would like to use waveform templates from EMRI systems that are evolved under the instantaneous self-force model, but these are computationally challenging. The alternative is to use cheaper adiabatic waveforms, but these do not include the effect of resonances. To assess the impact of this choice, we compare data $s(t;\bar{\boldsymbol{\Theta}})$ generated using the full self-force model (\secref{forced-motion}) to templates $h(t;\boldsymbol{\Theta})$ generated using our $2$-torus averaged self-force model (\secref{adiabatic}).

To generate gravitational waveforms, we employ the numerical kludge (NK) method of Babak {\it{et al}}.~\cite{Babak2007}, augmented to include evolution of the positional elements. We first compute inspiral trajectories and then separately (and not necessarily consistently) calculate the GW emission sourced by a compact object moving along that trajectory. This is quicker and easier to calculate than waveforms using Teukolsky-based methods (currently the most accurate prescription available) and yet gives similar results; agreement between Teukolsky-based and the best NK waveforms is typically $95\%$ or higher for a variety of orbits~\cite{Babak2007, Berry2013}.

Following the NK method, we first compute the inspiral trajectory of the compact object around a central Kerr BH with evolution driven by the dissipative part of self-force either calculated instantaneously or following the adiabatic prescription. We then map the Boyer--Lindquist coordinates to spherical polar coordinates in Minkowski, facilitating the use of a flat-spacetime waveform generation technique, the standard quadrupole formula~\cite{Misner1973}.

We expect these NK waveforms to be sufficiently accurate for our purposes, given the approximate (PN, low-spin) self-force model (\secref{self-force}). Results can be straightforwardly refined as developments are made in computing more comprehensive self-force models.

The similarity of two waveforms, $s(t)$ and $h(t)$, can be evaluated using the noise-weighted inner product~\cite{Finn1992}
\begin{equation}
\label{eq:innerprod}
\overlap{s}{h} = 2 \intd{0}{\infty}{\frac{\tilde{s}(f)\tilde{h}^*(f)+\tilde{s}^*(f)\tilde{h}(f)}{S_n(f)}}{f},
\end{equation}
where $\tilde{s}(f)$ represents the Fourier transform of $s(t)$ and similarly for $\tilde{h}(f)$, and $S_n(f)$ is the one-sided noise power spectral density (PSD)~\cite{Moore2014a}.\footnote{We use cubic interpolation to construct NK waveforms with identical time sampling. 
A Planck-taper window function~\cite{McKechan2010} is applied to reduce unwanted spectral leakage in the Fourier transforms. 
}

We use the analytic approximation for the eLISA PSD of Amaro-Seoane {\it{et al}}.~\cite{Amaro-Seoane2012a}. Following the success for the LISA Pathfinder mission~\cite{Armano2016}, this sensitivity should be achievable in a future mission.

We wish to test whether there exists some set of parameters $\boldsymbol{\Theta}$ such that the resulting adiabatic waveform is sufficiently similar to the full waveform. We do this by evaluating the SNR for each (normalised) waveform template,
\begin{equation}
\rho\left[h\right] = \max_t \frac{\overlap{s}{h}}{\sqrt{\overlap{h}{h}}}.
\end{equation}
We maximise over the time offset for the template to find the best fit to the data. If a template exactly matches the data, it would produce an SNR of $\sqrt{\overlap{s}{s}}$, hence the overlap
\begin{equation}
\label{eq:overlap}
\mathbb{O}\left[h\right] = \max_t \frac{\overlap{s}{h}}{\sqrt{\overlap{s}{s}}\sqrt{\overlap{h}{h}}},
\end{equation}
which ranges from $0$ to $1$ provides a convenient indication of how well matched the template is to the data.

\subsection{The nonresonant case}
\label{sec:nonres}

\begin{figure*}
\centering
\includegraphics[width=\textwidth]{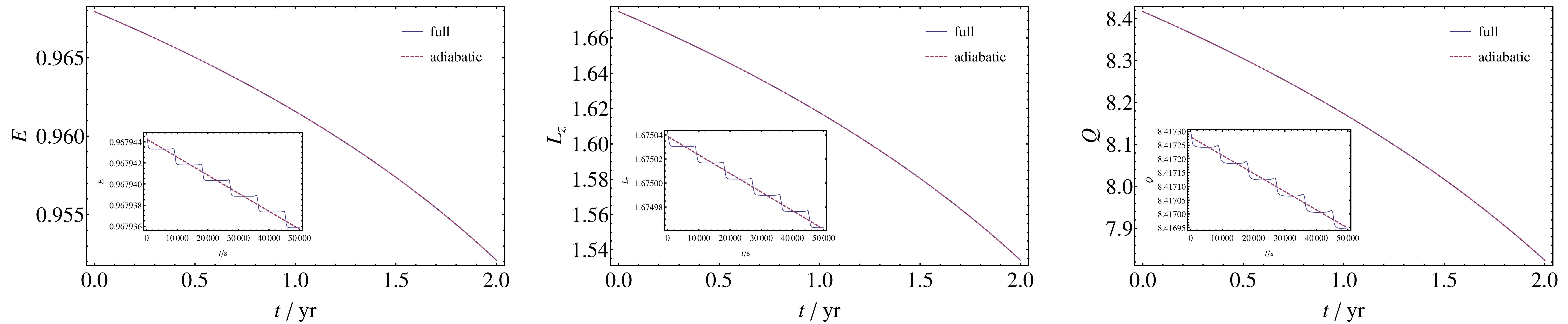}
\caption{\label{fig:good-traj}The evolution of the orbital parameters $E$ (left), $L_z$ (center) and $Q$ (right) under the full (solid line) and adiabatic (dashed) models for an illustrative EMRI system that does not encounter any significant resonances. The inset plots show the behaviour on short timescales, where the fast orbital oscillations can be seen. This system has $\mu = 10 M_{\odot}$, $\eta = 3\times 10^{-6}$, $a_\ast=0.95$, initial semilatus rectum $p_0 = 7.5$, initial eccentricity $e_0 = 0.7$, initial inclination $\cos \iota_0 = 0.5$ and redshift $z=0.204$.}
\end{figure*}

Before investigating the impact of resonances on an EMRI signal, we first compare results from the full instantaneous evolution and the $2$-torus averaged adiabatic evolution over $2~\mathrm{yr}$ for an inspiral which avoids any significant resonances. Example evolutions of the orbital parameters $E$, $L_z$ and $Q$ are shown in \figref{good-traj}, which shows that the adiabatic evolution closely matches the full evolution on longer inspiral timescales. The inset plots show the start of the evolution, on a timescale associated with the orbital motion of the compact object; the $2$-torus averaging explicitly smooths out the visible structure on this scale. The two approaches are in good agreement.

Using the two trajectories, we can calculate the corresponding NK waveforms. The two waveforms exhibit good agreement in both amplitude and phase across the entire duration. We find an overlap $\mathbb{O} = 0.993$, illustrating that adiabatic models can safely be used when resonances are not encountered.

\subsection{The effect of resonances: Dephasing and overlap}
\label{sec:effres-phase}

We now study a system that does pass through a resonance during its $2~\mathrm{yr}$ evolution. Specifically, we choose the initial conditions to be the same as in \secref{nonres}, but with an initial semilatus rectum $p_0 = 7.85$. This system passes through the $2$:$3$ resonance; the effect is to cause a shift in the orbital parameters (and hence the fundamental frequencies) that is not replicated by the adiabatic evolution, thus resulting in a rapid dephasing of the waveforms.

\begin{figure}
\centering
\includegraphics[width=0.4\textwidth]{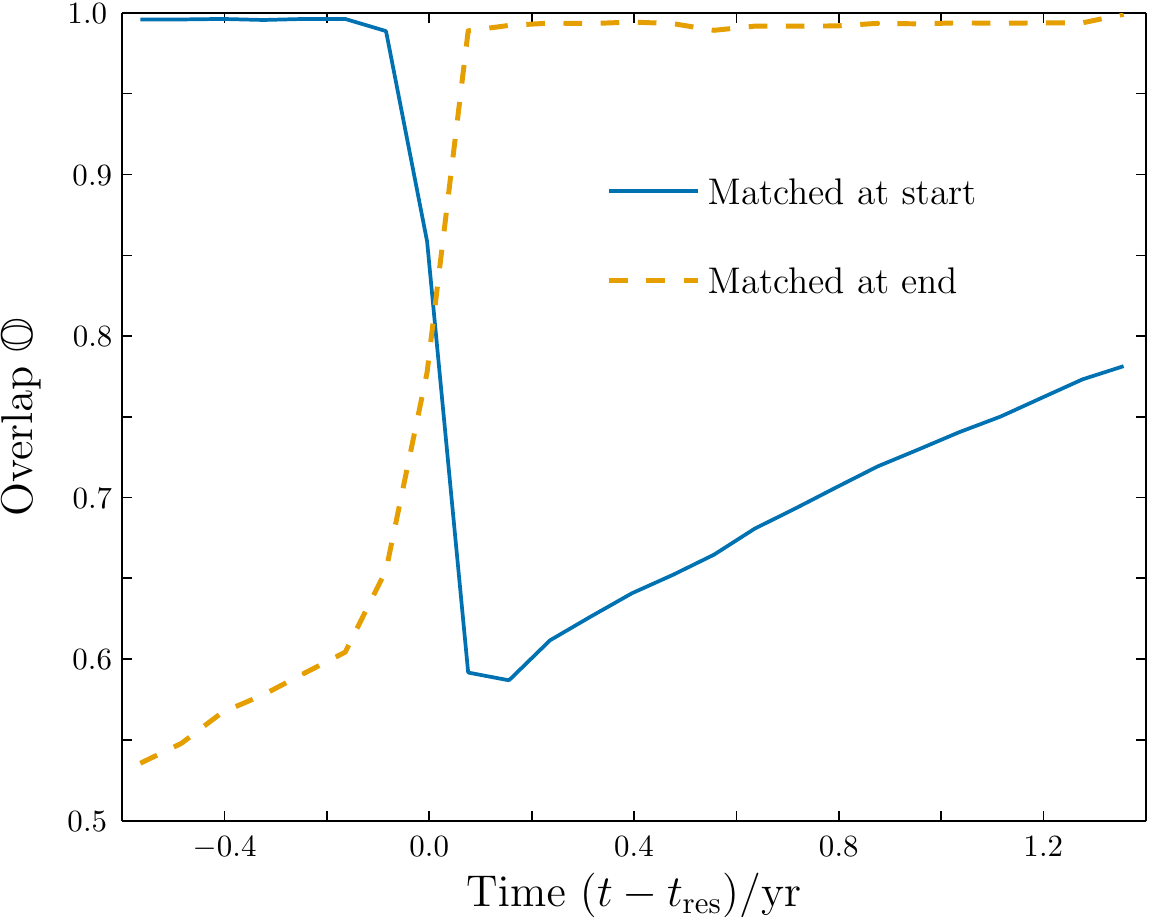}
\caption{\label{fig:overlap-dephasing}The overlap computed between the full evolution and an adiabatic evolution for an illustrative EMRI system with $p=7.85$, as a function of time, including only the parts of the waveforms within some $\Delta t$ of $t$, arbitrarily chosen to give $25$ independent (nonoverlapping) calculations. The time $t = t\sub{res}$ is when the full evolution crosses the $2$:$3$ resonance ($p \simeq 7.67$, $e \simeq 0.674$, $\cos(\iota) \simeq 0.497$).}
\end{figure}

To illustrate the dephasing, we calculate as a function of time $t$ the shortened overlap between the two models, defined as the overlap obtained by including only the part of the waveform within $\Delta t$ of $t$. We choose $\Delta t$ such that we can calculate $25$ nonoverlapping shortened overlaps. Before the resonance occurs, the adiabatic model provides a good match to the full evolution, but the overlap is reduced near to the resonance and never fully recovers afterwards. This is shown in \figref{overlap-dephasing}, which is centred on the time at which the full evolution crosses the $2$:$3$ resonance. Also shown is the shortened overlap computed between the full evolution and a different adiabatic evolution that is chosen to match the full evolution at the end of the integration. In this case, we see similar behaviour: the adiabatic waveform has a high overlap where it is constructed to match the full evolution, but this is disrupted by the resonance. Passing through resonance can adversely affect the overlap of adiabatic templates.

To be able to detect signals, we must have templates which match the signals. We have seen that overlaps between adiabatic and full instantaneous evolutions dephase following a resonance. However, this does not necessarily mean that no adiabatic evolution has a high overlap with the observed signal. It is possible that a difference between the instantaneous and adiabatic waveforms could be ameliorated by changing the parameters of the template $\boldsymbol{\Theta}$. In this case, the waveform mismatch would not limit detectability of the signals, but would lead to errors in parameter estimation, a stealth bias caused by incorrect waveforms~\cite{Cutler2007}. We leave an investigation into the impact of transient resonances on parameter estimation to future work.\footnote{If resonances were successfully included in the waveforms used for parameter estimation, the sharp nature of jumps may help construct precise inferences of the source parameters~\cite{Mandel2014}.} However, we consider the possibility of obtaining better waveform matches by varying the parameters of the EMRI.

The large parameter space of adiabatic waveforms, coupled with the expensive nature of our $2$-torus averaging routine, renders a brute-force approach prohibitively expensive. For this preliminary investigation, we focus on a small subset of parameters that we suspect will produce a large overlap, and make the assumption that a good adiabatic model \emph{exactly} matches the full model at some time $t\sub{match}$. This reduces the search to a $1$-parameter family of waveforms that can easily be computed concurrently with the full evolution.


The problem of searching over adiabatic templates now reduces to the task of choosing appropriate values of $t\sub{match}$. To demonstrate how changing the matching time affects the overlap, we use $5\tau\sub{res}$ after each resonance of interest, namely the low-order $1$:$2$ and $2$:$3$ resonances.\footnote{We adjust the values of $t\sub{match}$ so that they correspond to times of apoapsis. This ensures that the adiabatic model intersects with the full instantaneous model close to the centre of its oscillatory envelope, as demonstrated by the inset plots in \figref{good-traj}. This generally obtains better matches than if we match close to the extrema of the envelope, which are less representative of the average behaviour.} These matching times lie in a portion of the evolution that is not affected by a resonance, and so should allow for a large overlap with the adiabatic model for that region of the inspiral. For comparison, we also consider templates that match at the start and end of the evolution, and that match exactly on each of the resonances.

\begin{figure*}
\centering
\includegraphics[width=\textwidth]{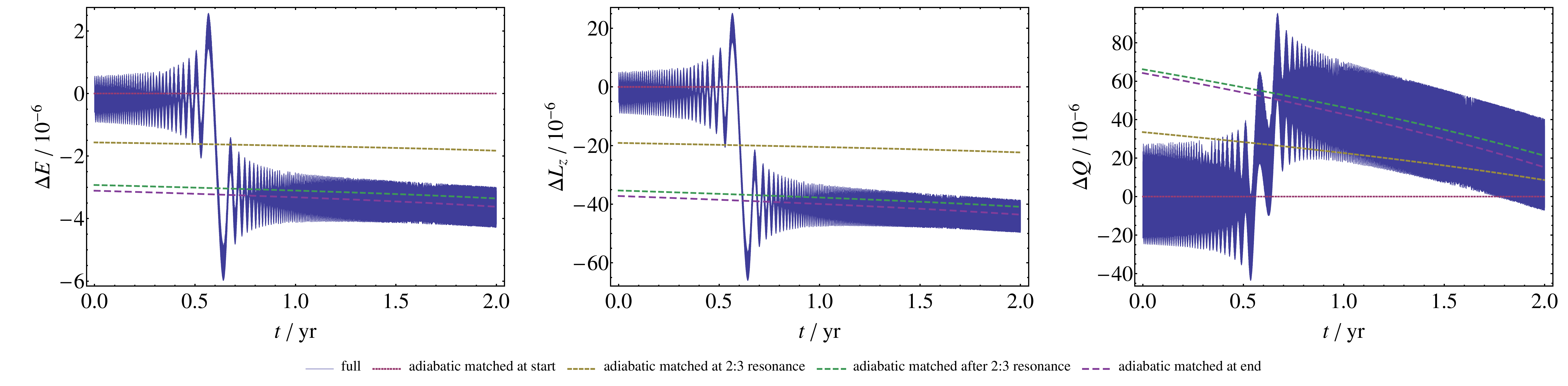}
\caption{\label{fig:res-diff-traj}The differences in orbital parameters $E$ (left), $L_z$ (centre) and $Q$ (right) between each evolution scheme and the adiabatic model that matches at the start. The solid line shows the full evolution and the dashed lines show the different adiabatic evolutions, which match the full evolution at different times throughout the inspiral (numbers in parentheses give the overlap with the full evolution): at the start ($0.207$), at the $2$:$3$ resonance ($0.432$), after the $2$:$3$ resonance ($0.258$) and at the end ($0.677$).}
\end{figure*}

We have computed this family of adiabatic evolutions for our illustrative resonance. In \figref{res-diff-traj}, we plot the difference in the orbital parameters ($E$, $L_z$ and $Q$) between the various trajectories and the adiabatic evolution that matches at the start. The jumps in the orbital parameters due to the $2$:$3$ resonance can be clearly seen, as can the fast orbital oscillations present in the full instantaneous evolution but absent in the adiabatic evolutions.

None of the adiabatic models presented here give a particularly high overlap with the entire signal, because of the effects of the resonance. In this case, the best-performing adiabatic model was that which matched at the end, giving $\mathbb{O} = 0.677$, while the model that matches at the start gives only $\mathbb{O} = 0.207$. These are similar overlaps from to the adiabatic evolutions matched before and after resonance respectively. These values can be explained qualitatively by the relative lengths of the adiabatic-like regions on either side of the resonance in the full evolution ($69.8\%$ of the inspiral is post-resonance): there is not an exact equivalence because of the frequency dependence of the PSD.

If we could construct an adiabatic model that includes the jump across resonances, it may give a good overall fit to the signal.

\subsection{The effect of resonances: Jump sizes}
\label{sec:effres-jump}

As explained in \secref{flux-enhance} and illustrated in \figref{res-diff-traj}, the full instantaneous evolution undergoes a rapid change in the orbital parameters (with respect to the adiabatic evolution) when passing through resonance. The size of the jump influences the subsequent orbital evolution.

\begin{figure}
\centering
\includegraphics[width=0.46\textwidth]{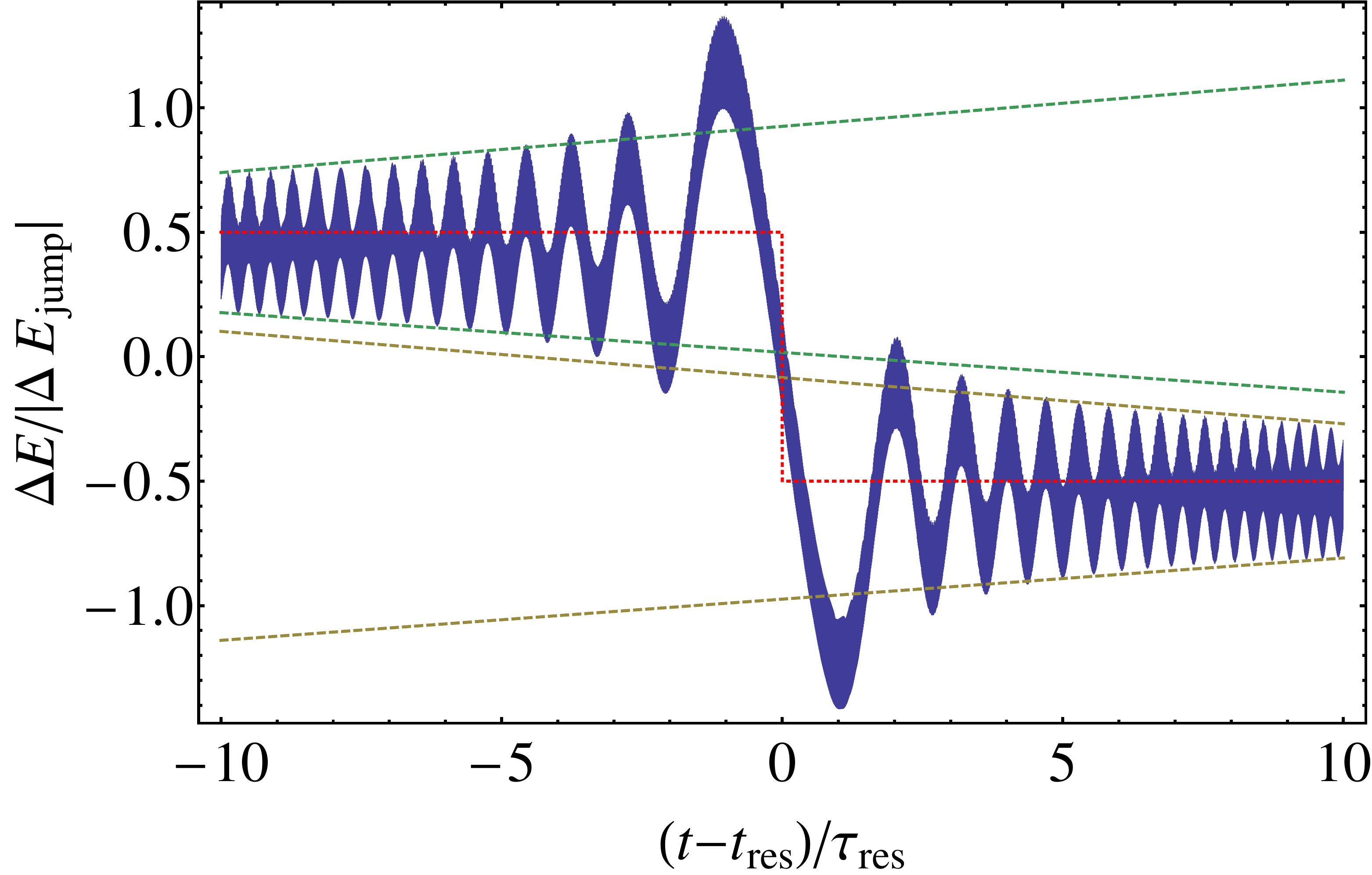}
\caption{\label{fig:res-jump-calc}The difference in energy between the instantaneous model and adiabatic model that matches at the $2$:$3$ resonance, scaled by the magnitude of the $2$:$3$ resonance jump. The apparent thickness of the line is because of oscillations on the short orbital timescale, which are too fast to be resolved on on the plotted scale. The dashed green (yellow) lines show the bounding fits to the data before (after) the resonance, used to numerically estimate the size of the jump. The dotted red line indicates the computed size of the jump. The time axis is centred on the $2$:$3$ resonance and is scaled by the resonance timescale $\tau\sub{res}$.}
\end{figure}

To extract the magnitude of this jump from the trajectory data, we must account for the fast orbital oscillations as well as the general average (adiabatic) evolution~\cite{ColeThesis2015}. We first computed the difference $\Delta \mathcal{I}^a \equiv \mathcal{I}^a\sub{full} - \mathcal{I}^a\sub{ad}[t\sub{match} = t\sub{res}]$ between the orbital parameters calculated using from the full instantaneous evolution and from an adiabatic evolution matched to the parameters at resonance. We then fit linear bounds to the oscillating envelope, both before and after the resonance, using data $5\tau\sub{res}$ to $10\tau\sub{res}$ away. These are averaged to give general pre- and post-resonance trends, which are extrapolated to the time of resonance. The difference at the time of resonance gives an estimate for the jump $\Delta \mathcal{I}^a\sub{jump}$ in orbital parameter $\mathcal{I}^a$.\footnote{As a cross-check, comparable values for the jump are obtained by averaging (over an integer number of radial and poloidal orbits) the flux on a resonant geodesic~\cite{ColeThesis2015}.} \Figref{res-jump-calc} illustrates how the size of the jump is calculated.

The absolute size of the resonance jump is often not particularly useful, especially when comparing different systems. Instead, we use the fractional enhancement relative to the adiabatic evolution,
\begin{equation}
\label{eq:res-jump-ratio}
\delta \mathcal{I}^a = \frac{\Delta \mathcal{I}^a\sub{jump}}{\dot{\mathcal{I}}^a\sub{ad}(t\sub{res})\tau\sub{res}}.
\end{equation}
Relative enhancements may be of the order of a few percent.

\begin{figure}
\centering
\includegraphics[width=0.4\textwidth]{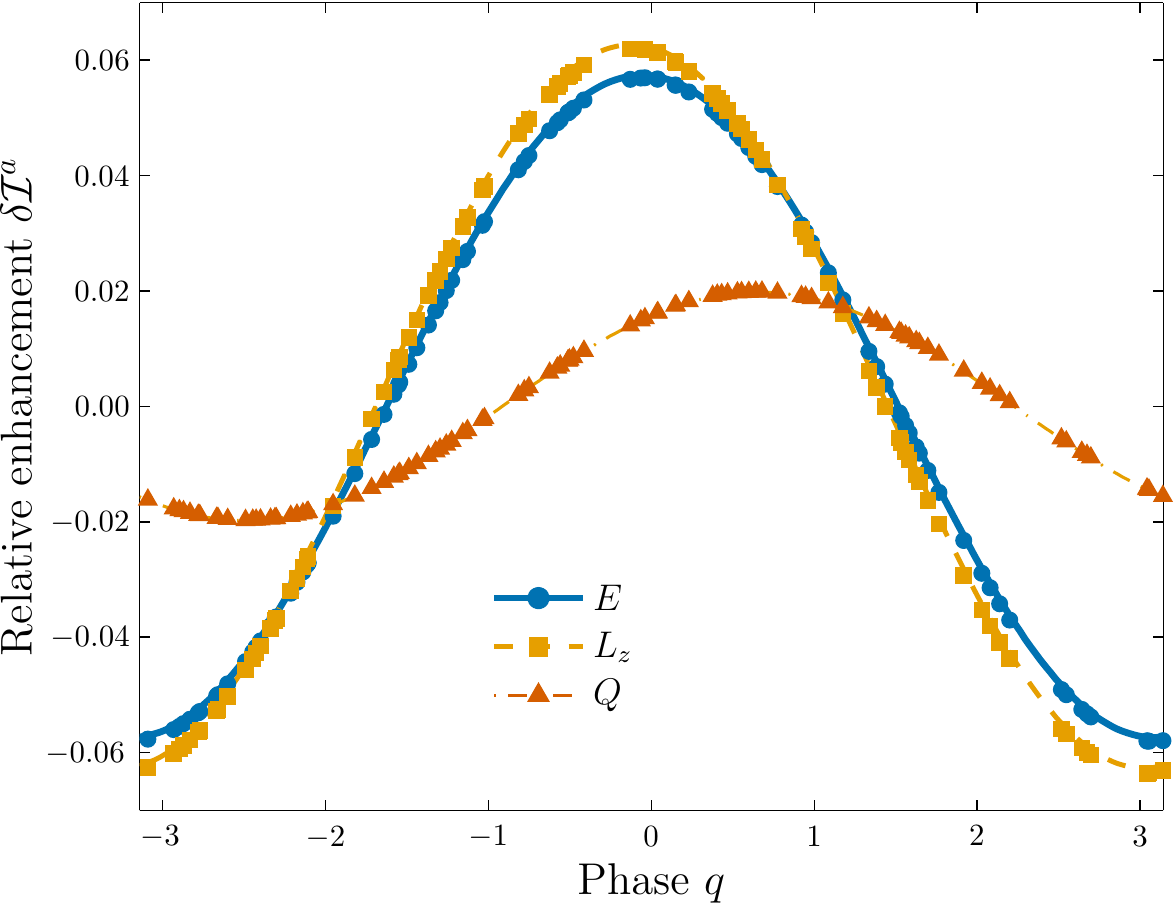}
\caption{\label{fig:resjump-vs-q}The magnitude of the resonance jump for our illustrative system as a function of the extracted phase parameter $q$ (cf.\ Fig.~3 of \cite{Flanagan2012a}). The jump is relative to the adiabatic change in each parameter across resonance. The individual jumps as well as a sinusoidal fit are plotted for the energy $E$, angular momentum $L_z$ and Carter constant $Q$.}
\end{figure}

The jump in orbital parameters depends sensitively on the relative phase of radial and poloidal motions. This can be illustrated using an ensemble of orbits with different orbital phases $\psi_{\theta_-}$. According to \eqnref{delta-I-a}, the jumps should oscillate as a function of the phase $q$.\footnote{The overlaps for all of these different phases are similar, clustering around $0.7$, as expected from matching the post-resonance region of the waveform.} Therefore, assuming that the lowest harmonic dominates, plotting a jump in one parameter against another should trace out an ellipse. We find that this is the case; in \figref{resjump-vs-q} we plot resonant jumps as a function of the phase $q$ constructed from the $\Delta E\sub{jump}$--$\Delta Q\sub{jump}$ ellipse. The jumps in $E$ and $L_z$ are approximately in phase, but those in $Q$ are found to be offset~\cite{Flanagan2012a}; this means that for every value of $q$, there is always a resonance jump in at least one parameter.

\begin{figure}
\centering
\includegraphics[width=0.4\textwidth]{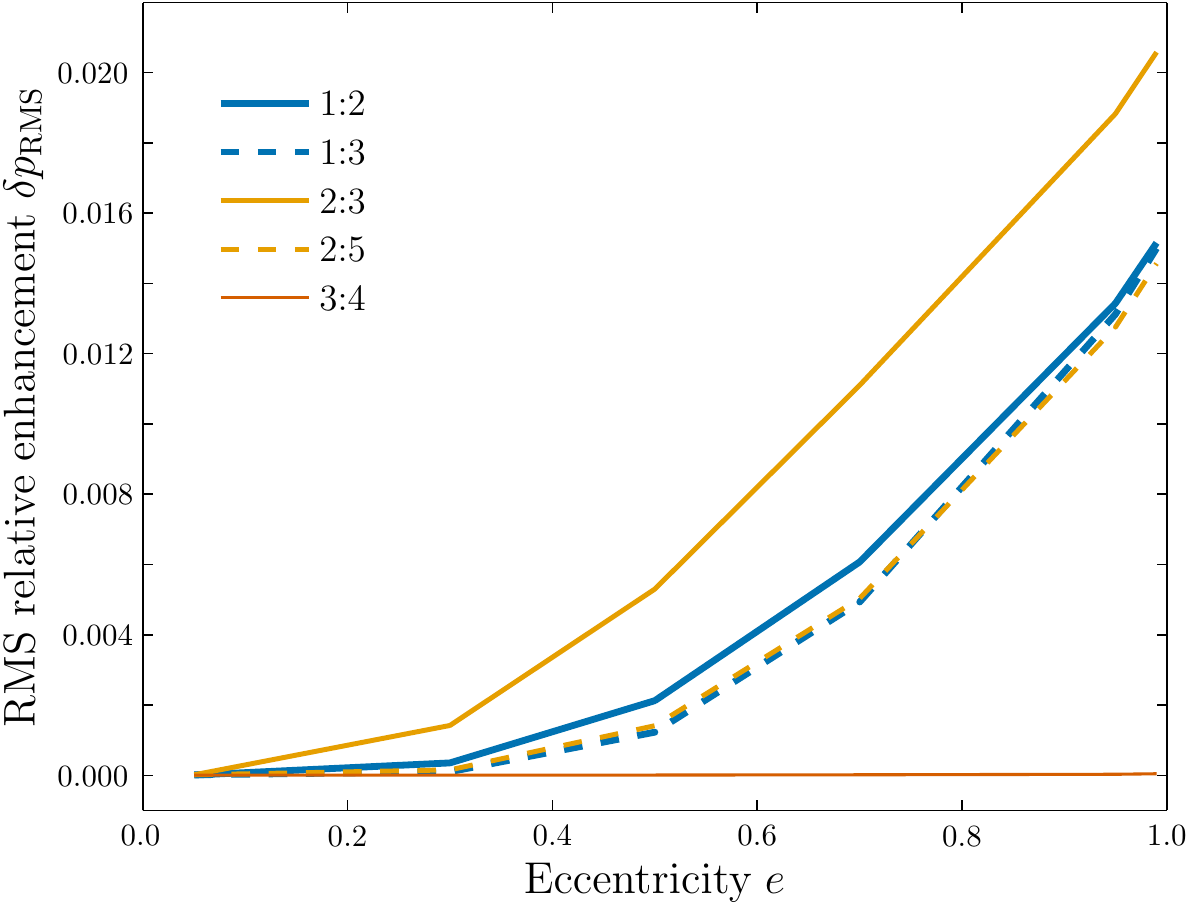}
\caption{\label{fig:res-flux-rms-p}Relative flux enhancements for the semilatus rectum $p$, as a function of $e$, marginalised over $a_\ast$ and $\cos\iota$ by taking the root-mean-square of a grid of values. Resonances with $n_\theta = 1$ ($2$; $3$) are coloured blue (gold; red). The $2$:$3$ resonance has the largest relative flux enhancement.}
\end{figure}

Let us now consider jumps for systems other than our illustrative $2$:$3$ resonance. We expect nearly circular orbits to encounter smaller jumps than more eccentric orbits~\cite{Flanagan2012} because they have a smaller $r$--$\theta$ $2$-torus, meaning that resonant orbits come closer to every allowed point, nearer approximating a nonresonant orbit. In \figref{res-flux-rms-p}, we show the root-mean-square (over a grid of $a_\ast$ and $\cos\iota$ values) relative resonant jump for the semilatus rectum $p$ as a function of eccentricity $e$ for a selection of low-order resonances. The other orbital shape parameters show a similar trend with increasing eccentricity~\cite{ColeThesis2015}. Across the grid of $a_\ast$ and $\cos\iota$ values there may be an order of magnitude variation, but the same general trends are observed. Larger eccentricities do give rise to larger jumps, matching our expectations and Teukolsky-based calculations by Flanagan, Hughes and Ruangsri~\cite{Flanagan2012a}.

\begin{figure}
\centering
\includegraphics[width=0.4\textwidth]{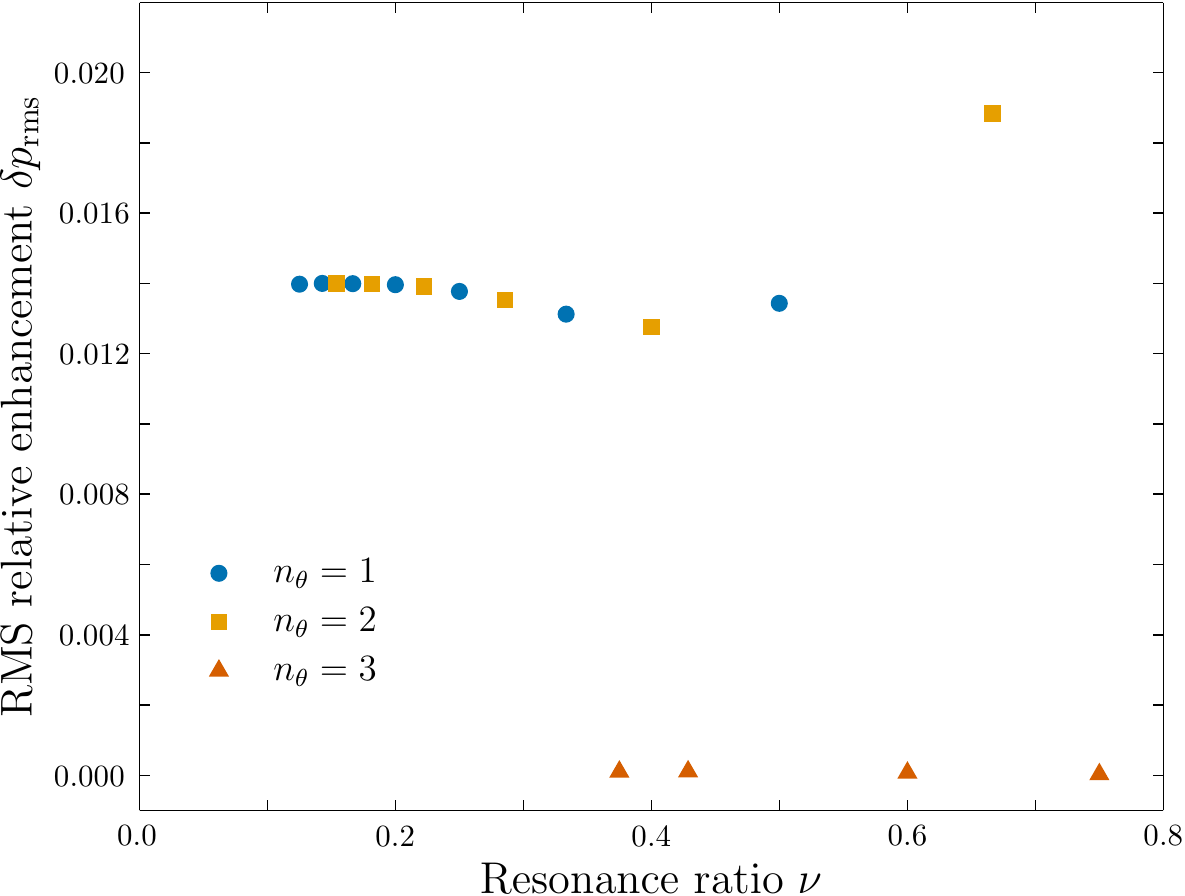}
\caption{\label{fig:res-flux-rms-p-vsnu}Relative flux enhancements for the semilatus rectum $p$ marginalised over $a_\ast$ and $\cos\iota$ by taking the root-mean-square. Resonances with $n_\theta = 1$ ($2$; $3$) are coloured blue (gold; red) and use circular (square; triangular) points.}
\end{figure}

We also expect that the effects of passing through resonance depend upon the particular resonance. Intuitively, we would expect that when $n_\theta$ and $n_r$ are large, the effects of resonance will be small, since the orbit comes close to all points on the $2$-torus. We have already seen that this is the case, as the adiabatic evolution is a good match to the instantaneous evolution up until it hits a resonance which is the ratio of small integers (like $2$:$3$). In \figref{res-flux-rms-p-vsnu} we plot the root-mean-square (over $a_\ast$ and $\cos\iota$) relative resonant jump for the semilatus rectum as a function of the resonance ratio $\nu$ for various $n_\theta$, the eccentricity is set to $e=0.95$ to emphasise the variation.

While we might naively expect jumps with $n_\theta = 1$ to be greatest, we see that this is not the case. Instead the $n_\theta = 1$ and $n_\theta = 2$ jumps form a single continuum: we can treat $1$:$x$ resonances as \emph{de facto} $2$:$2x$ resonances. This suggests that more insight into resonance jumps could be gained from considering the motion across two radial periods, even for resonances with $n_\theta = 1$. Moving beyond $n_\theta = 2$, we see that the magnitude of jumps decreases rapidly for $n_\theta = 3$~\cite{Flanagan2012a}. For larger $n_\theta$, jumps are so small that cannot accurately calculate them.

\section{Astrophysical implications}
\label{sec:astrophysics}

Strong resonances can limit our ability to recover SNR from waveforms using adiabatic templates; they partition the inspiral, splitting up the total SNR into different adiabatic regions, which may be individually undetectable. In order to assess the impact of this on future GW missions, we need to analyse the waveforms from a population of detectable EMRIs. In \secref{EMRI-population}, we detail our procedure for generating an astrophysically motivated distribution of EMRIs, before turning to the effect of resonances on the detectability of EMRIs in \secref{population-SNR}.

\subsection{Sample EMRI population}
\label{sec:EMRI-population}

The EMRI event rate depends on the exact composition of the population of compact objects around each SMBH, the stellar density profile for each species of compact object, and the mass and spin of each SMBH~\cite{Alexander2005}, all of these properties are highly uncertain, even for our own Galaxy. Here, we sketch out a model which includes most of the important effects, which should suffice for illustrating the potential impact of transient resonances on the rate of detection.

\subsubsection{Population model}

To generate a representative EMRI population, we need to establish plausible distributions for the parameters defining the EMRI: the properties of the orbiting compact objects, the properties of the central SMBHs, and the properties of the orbits. 

EMRIs require a compact object such as a white dwarf, a neutron star or a stellar-mass BH to orbit around a SMBH. Main-sequence stars are tidally disrupted before they can complete the inspiral~\cite{Rees1988,Sigurdsson1997}. We expect the EMRI event rate to be dominated by BHs as the most massive species of compact object~\cite{Gair2004}. First, as BHs are more massive than white dwarfs or neutron stars, their GW signal is louder~\cite{Peters1963,Barack2004} and hence detectable EMRIs can come from a larger volume. Second, dynamical friction in the dense nuclear star clusters~\cite{Chandrasekhar1960,Antonini2011} also leads to mass segregation, causing the most massive objects concentrate closer to the SMBH~\cite{Bahcall1977,Freitag2006,Alexander2009}. We therefore expect BHs to dominate the inner regions of nuclear star clusters, making them the most probable candidate to inspiral. We adopt a fiducial mass for the compact object of $\mu = 10 M_\odot$, corresponding to a typical mass for stellar BHs~\cite{Casares2014,Corral-Santana2016,Tetarenko2016,Abbott2016d}.\footnote{If the $\sim30 M_\odot$ BHs of GW150914~\cite{Abbott2016f,Abbott2016d} are common, then this would enhance the EMRI rate.}

We take the central object to be a typical SMBH at the centre of a galaxy~\cite{Kormendy1995,Ferrarese2005}; we are interested in galaxies that possess a SMBH with a mass in the range $(10^{4}$--$10^7) M_\odot$, as these give rise to EMRIs in the frequency band of space-based GW detectors~\cite{Amaro-Seoane2007,Amaro-Seoane2012a}. We consider the EMRI event rate as the combination of two pieces: the comoving number density of SMBHs in the Universe, which is the same as that of galaxies if we assume all galaxies host a single SMBH~\cite{Lynden-Bell1971,Soltan1982}, and the intrinsic event rate per SMBH $\mathcal{R}$, the number of inspiral events per unit time for a given galaxy.

The comoving number density of galaxies is challenging to estimate because of the effects of local structure in the Universe, the evolution of that structure, and properties of the SMBHs themselves. We simplify the problem by assuming a homogeneous distribution that does not evolve with redshift, which is reasonable for the typical scales considered by GW detectors. We also neglect correlations between the SMBH mass and spin~\cite{Volonteri2010,Dotti2013,Sesana2014}, and impose a power-law scaling relation for the comoving number density
\begin{equation}
\diff{n}{\,\ln M} = n_0 \left(\frac{M}{M_0}\right)^\beta.
\end{equation}
There is significant uncertainty in the SMBH mass function, but this simple functional form is found to be in good agreement with observations from the Sloan Digital Sky Survey for the mass range of interest; we use $\beta = 0$ and $n_0 = 0.002~\mathrm{Mpc}^{-3}$ for SMBHs with $M < \order{10^7 M_\odot}$~\cite{Greene2007, Gair2010b}.

Simple estimates of the intrinsic rate $\mathcal{R}$ have been carried out using Monte-Carlo methods to count the number of compact objects from isothermal distributions that spiral in to a SMBH without plunging~\cite{Merritt2013}. The result is a scaling law for each species of compact object of the form
\begin{equation}
\label{eq:EMRI-intrinsic-rate}
\mathcal{R}(M) = \mathcal{R}_0 \left(\frac{M}{M_0}\right)^\alpha,
\end{equation}
where $M$ is the mass of the SMBH and $M_0 = 3\times 10^6 M_\odot$ is a fiducial mass. Hopman~\cite{Hopman2009a} finds that $\alpha = \{-0.15,\,-0.25,\,-0.25\}$ for BHs, neutron stars and respectively, with respective event rates $\mathcal{R}_0 = \{400,\, 7,\, 20\}~\mathrm{Gyr}^{-1}$ for each component, showing how BHs dominate the event rate. Amaro-Seoane and Preto~\cite{Amaro-Seoane2011d} studied the effects of mass segregation on the intrinsic EMRI event rate, using direct-summation $N$-body simulations to calibrate a Fokker--Planck description for the bulk properties of the stellar distribution. They found a better fit for the power-law spectral index for BHs of $\alpha = -0.19$, which we use here. The simple power-law description does not incorporate the effects of either resonant relaxation~\cite{Rauch1996,Rauch1998,Merritt2011} or anomalous relaxation~\cite{Hamers2014,Merritt2015c}, ignores the spin of the SMBH~\cite{Amaro-Seoane2012b}, and assumes that the $M$--$\sigma$~\cite{Ferrarese2000,Graham2016} relation holds for all SMBH masses (cf.\ \cite{Hlavacek-Larrondo2012,Ferre-Mateu2015,VanLoon2015,Trakhtenbrot2015,King2016}). Each of these is likely to  impact the event rate, but \eqnref{EMRI-intrinsic-rate} can still be used as a rough guide to the expected number of events.

Combining the intrinsic event rate with the comoving number density, the mass of the SMBH for the EMRI population is then sampled from a power law with a probability distribution function $f(M) \propto M^{\alpha+\beta-1}$.

We distribute the SMBH's dimensionless spin $a_\ast$ uniformly between its limiting values of $0$ and $1$. X-ray measurements show that SMBH spins can take a range of values~\cite{Reynolds2013a,Patrick2012,Walton2013,BerryThesis2013}; there is an observed preference for larger ($> 0.9$) spin values, but this may be a selection effect~\cite{Brenneman2011}. Therefore, the uniform prior is a safe choice given our ignorance of the true distribution. The direction of the spin is uniformly distributed across the surface of the unit sphere.

We distribute SMBHs uniformly throughout the Universe. We sample redshift of the source uniformly from comoving volume out to maximum redshift of $z\sub{max} = 1.5$, beyond which we cannot detect EMRIs. Sources are uniformly distributed across the sky. 

To describe the orbit we need to specify the inclination distribution, the eccentricity distribution and the phase at periapsis. The inclination is uniformly distributed across all orientations (uniform in $\cos \iota$); the poloidal and azimuthal phases at periapsis are uniform between $0$ and $2\pi$. The eccentricity distribution is more complicated.

Eccentricities for EMRIs are uncertain, and depend strongly on the formation scenario being considered. Here, we adopt a fit to a distribution computed using Monte-Carlo simulations by Hopman and Alexander~\cite{Hopman2005}, who model the scattering process of compact objects onto inspiral orbits around a $3 \times 10^6 M_\odot$ Schwarzschild BH. We assume this can be extended to provide a rough estimate of the distribution around SMBHs of other masses and spins. At the point in the inspiral when the orbital period takes a fiducial value $T_0 = 10^4~\mathrm{s}$, we find that the Monte-Carlo eccentricity probability distribution function is well described by a power law with an exponential cutoff
\begin{equation}
\label{eq:EMRI-e-distribution}
f(e) \propto 
	\begin{cases}
		\left(e\sub{m}-e\right)^{b(e\sub{m}-e\sub{p})} \exp\left[b(e-e\sub{m})\right] & 0 \leq e \leq e\sub{m}\\
		0 & \mathrm{Otherwise}
	\end{cases},
\end{equation}
where $e\sub{m} = 0.81$ is the maximum observed eccentricity, $e\sub{p} = 0.69$ is the peak of the distribution, and $b=11$ is the exponential index~\cite{ColeThesis2015}. The mean eccentricity at this period is $0.60$, slightly below that expected for a thermal distribution. Significant eccentricity is retained as EMRIs enter the eLISA frequency band~\cite{Merritt2015c}. 

To evolve the orbits, we start with orbital periods of $T_0$ and then use the analytic kludge (AK) prescription of Barack and Cutler~\cite{Barack2004}. This is similar to the NK approach, but uses a series of Keplerian ellipses rather than Kerr geodesics. Evolution of the orbit includes the effects of periapse precession, Lense--Thirring
precession, and radiation reaction calculated using PN expressions.\footnote{While the AK approach does include these relativistic effects, it does not capture the full nature of the evolution: for example, it assumes that the angle between the orbital angular momentum and the SMBH spin is constant~\cite{Barack2004}, whereas radiation reaction should push the orbital plane towards being antialigned~\cite{Flanagan2007}. The approximate nature of the AK evolution should not effect our results more than the uncertainty in the initial conditions for EMRI orbits (for example, how loss-cone dynamics are modified by the SMBH's spin~\cite{Amaro-Seoane2012b}).} AK waveforms are less computationally expensive than NK waveforms, allowing us to simulate a large population of EMRIs. We follow inspirals until the last stable orbit (LSO).\footnote{The LSO is determined numerically by calculating the roots of $V_r(r) = 0$, which we denote in ascending order by $r_4 \leq r_3 \leq r\sub{p} \leq r\sub{a}$, and stopping the evolution when $r_3 = r\sub{p}$, which designates the orbit as marginally stable. This ignores the (small) influence of the self-force~\cite{Isoyama2014}.} Each of these systems is then evolved backwards for some time $t\sub{insp}$, chosen uniformly from the range $[0,t\sub{life}]$ for a mission lifetime $t\sub{life}$, and the expected GWs are calculated using the AK formalism.

\subsubsection{Population results}

The estimated size of the EMRI population can be by evaluating from~\cite{Gair2009}
\begin{equation}
\label{eq:EMRI-number}
N\sub{EMRI} = \intd{z\,=\,0}{z\sub{max}}{ \intd{M\,=\,M\sub{min}}{M\sub{max}}{\mathcal{R}t\sub{life}\diff{n}{\,\ln M} \diff{V\sub{c}}{z}}{\,\ln M}}{z},
\end{equation}
where $V\sub{c}(z)$ is the comoving volume at redshift $z$, and limits $z\sub{max} = 1.5$, $M\sub{min} = 10^4 M_\odot$ and $M\sub{max} = 10^7 M_\odot$ are chosen to encompass the range of detectable signals. For a mission lifetime of $t\sub{life} = 2\units{yr}$, the integral gives a total of $6330$ EMRI systems. This is a lower bound for $N\sub{EMRI}$ as we are neglecting EMRIs that merge outside the observation window but nevertheless accumulate sufficient SNR during this time to be observable.

\begin{figure*}
\centering
\includegraphics[width=0.37\textwidth]{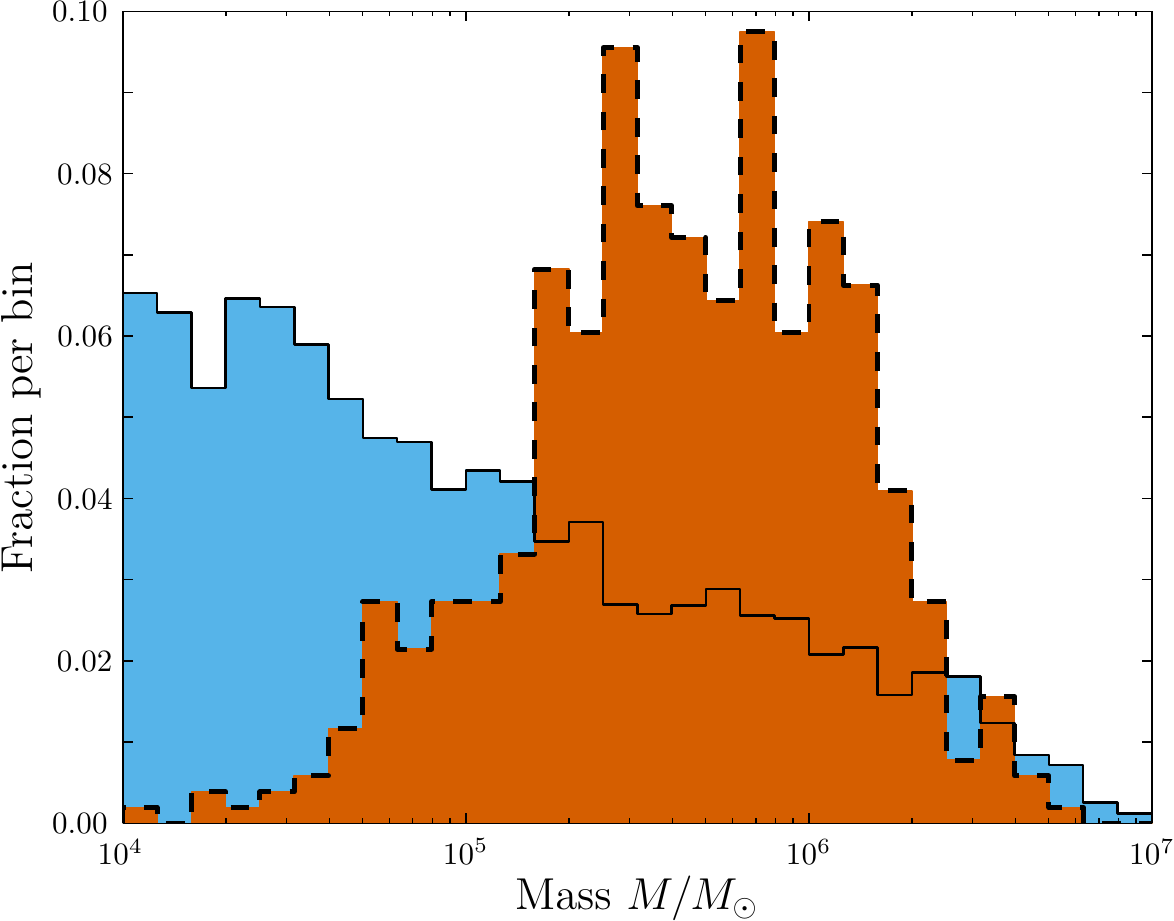} \quad
\includegraphics[width=0.37\textwidth]{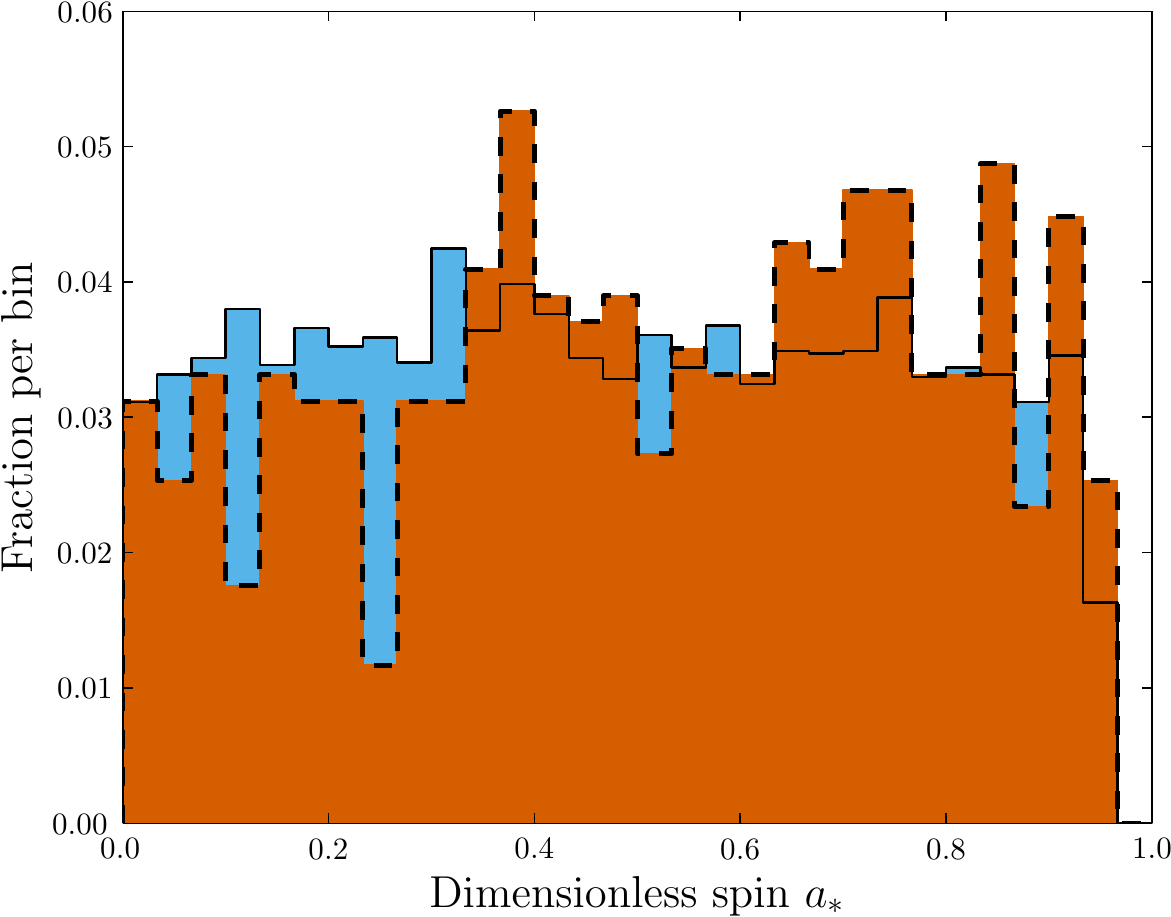} \\ \vspace{0.1cm}
\includegraphics[width=0.37\textwidth]{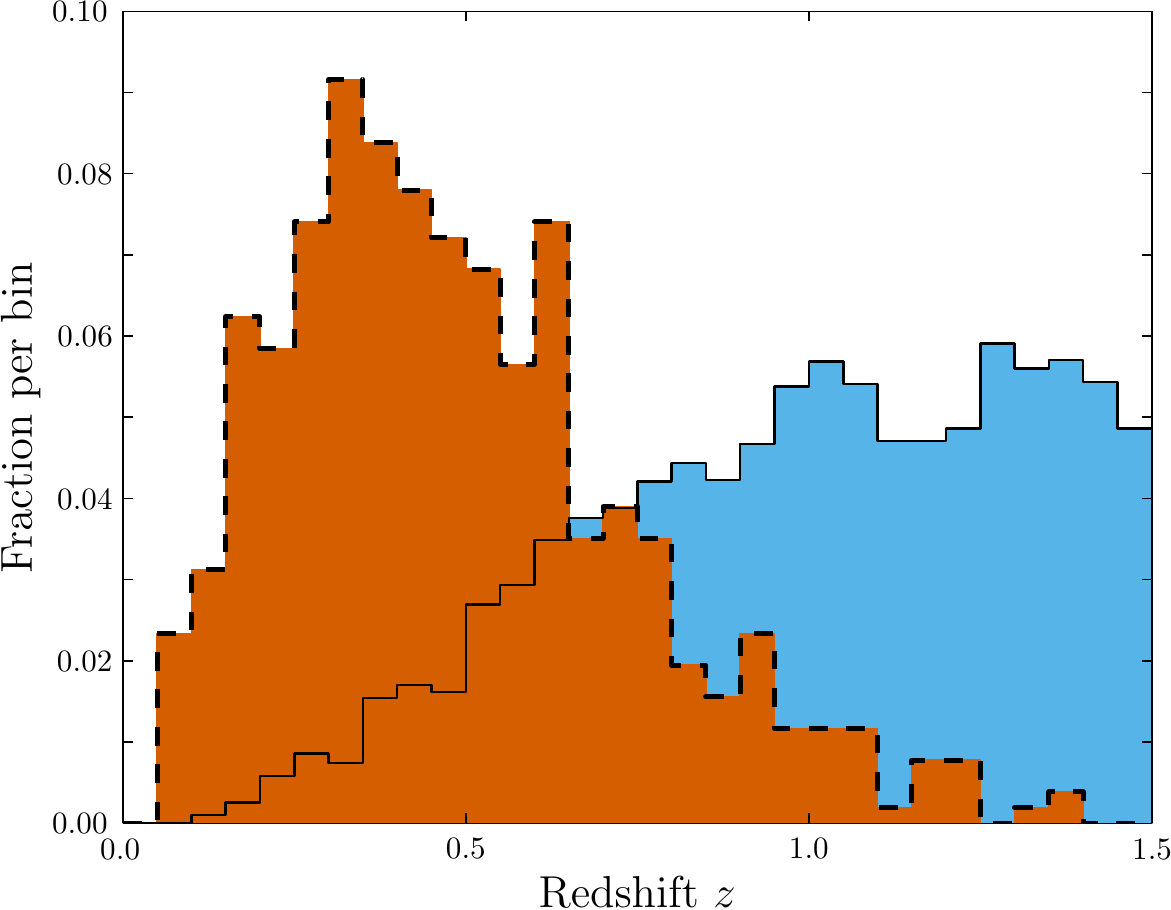} \quad
\includegraphics[width=0.37\textwidth]{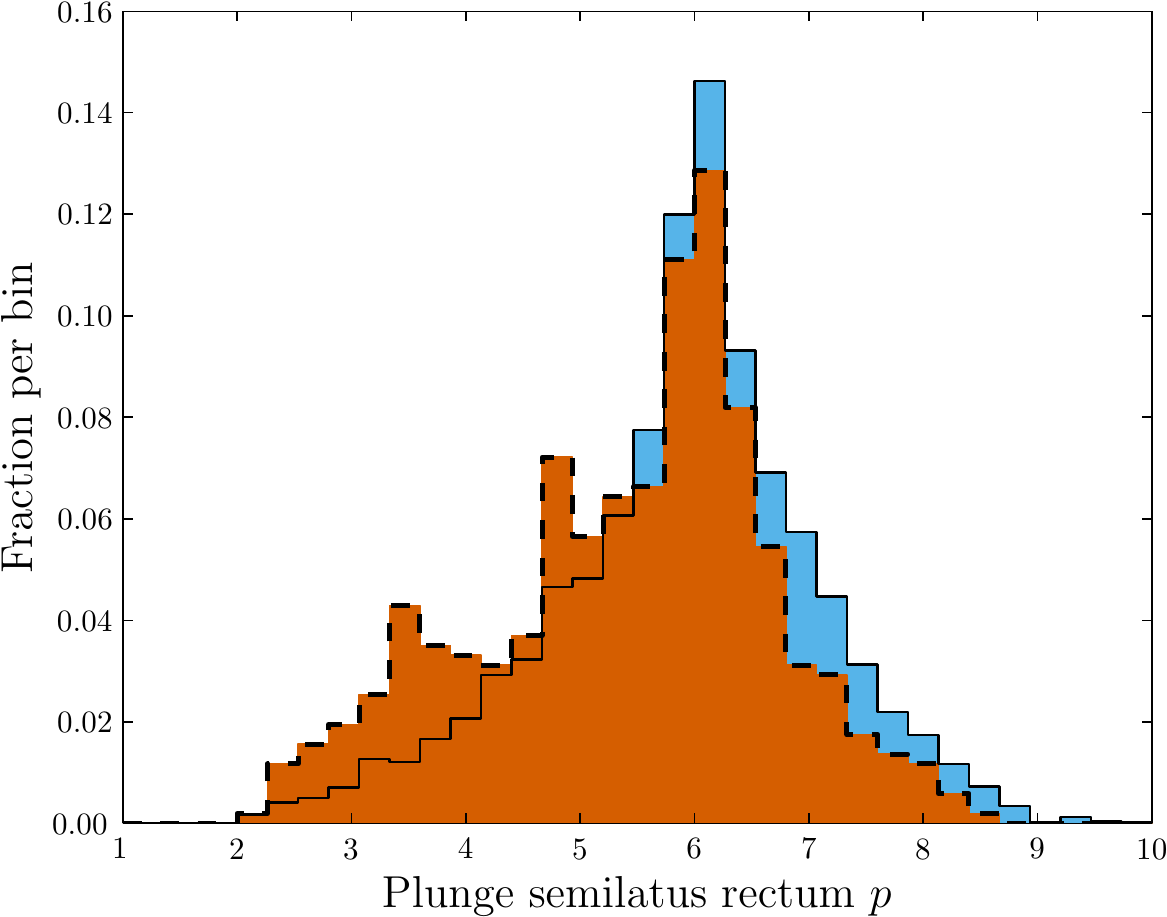} \\ \vspace{0.1cm}
\includegraphics[width=0.37\textwidth]{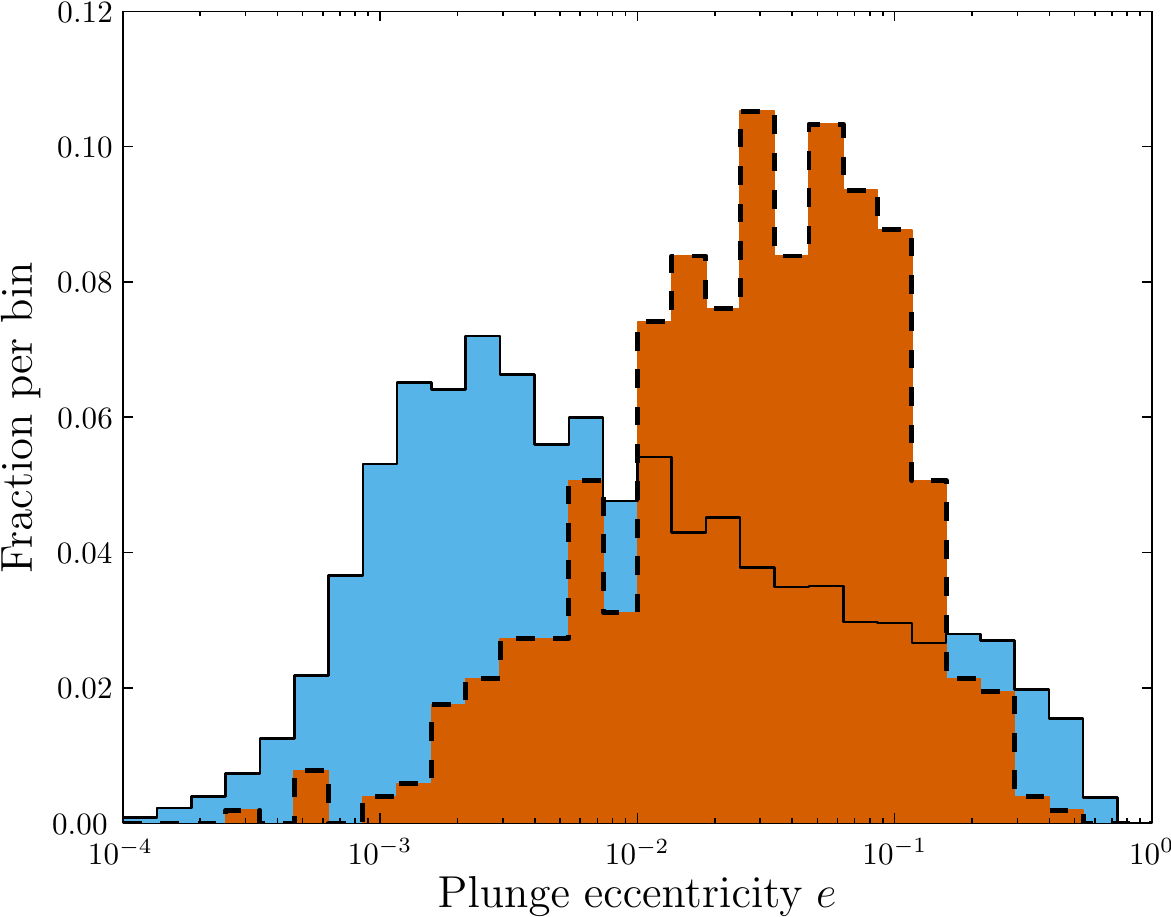} \quad
\includegraphics[width=0.37\textwidth]{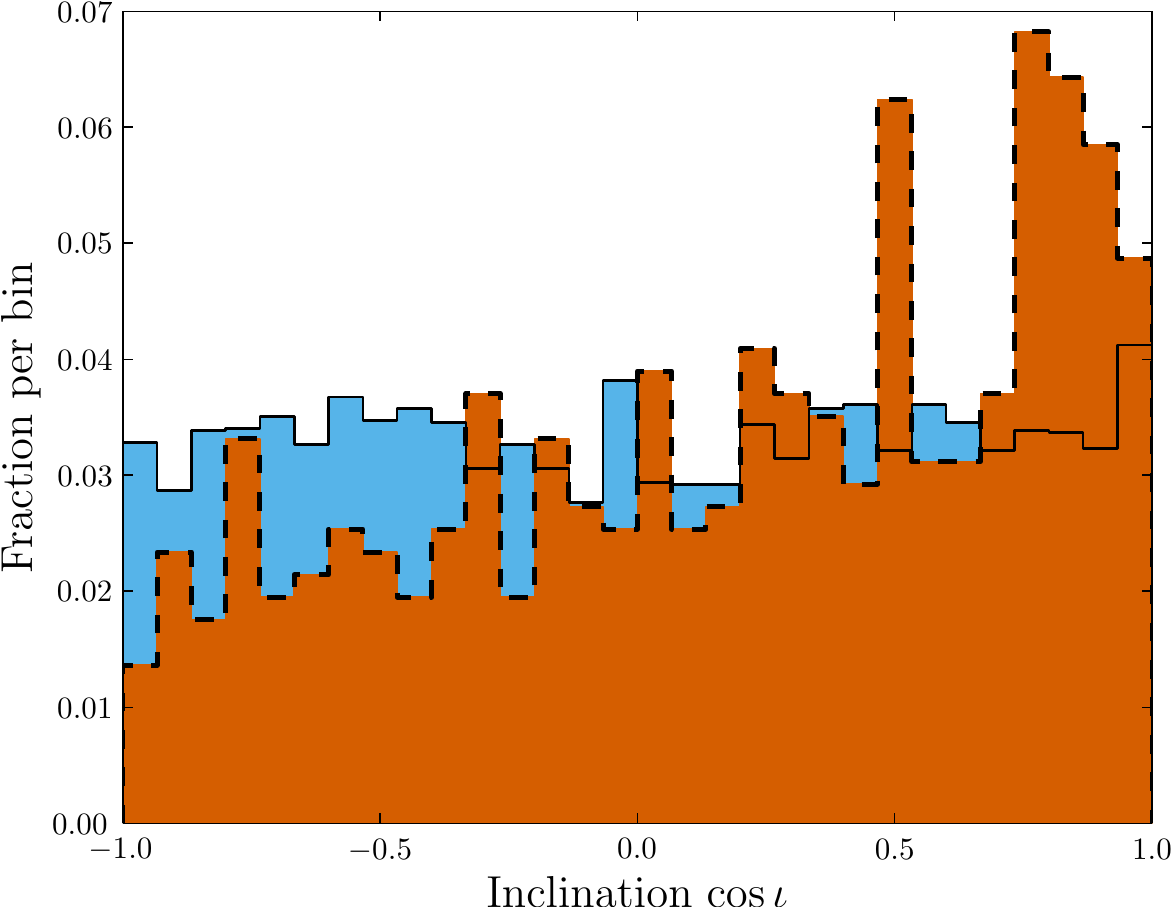} \\ \vspace{0.1cm}
\includegraphics[width=0.37\textwidth]{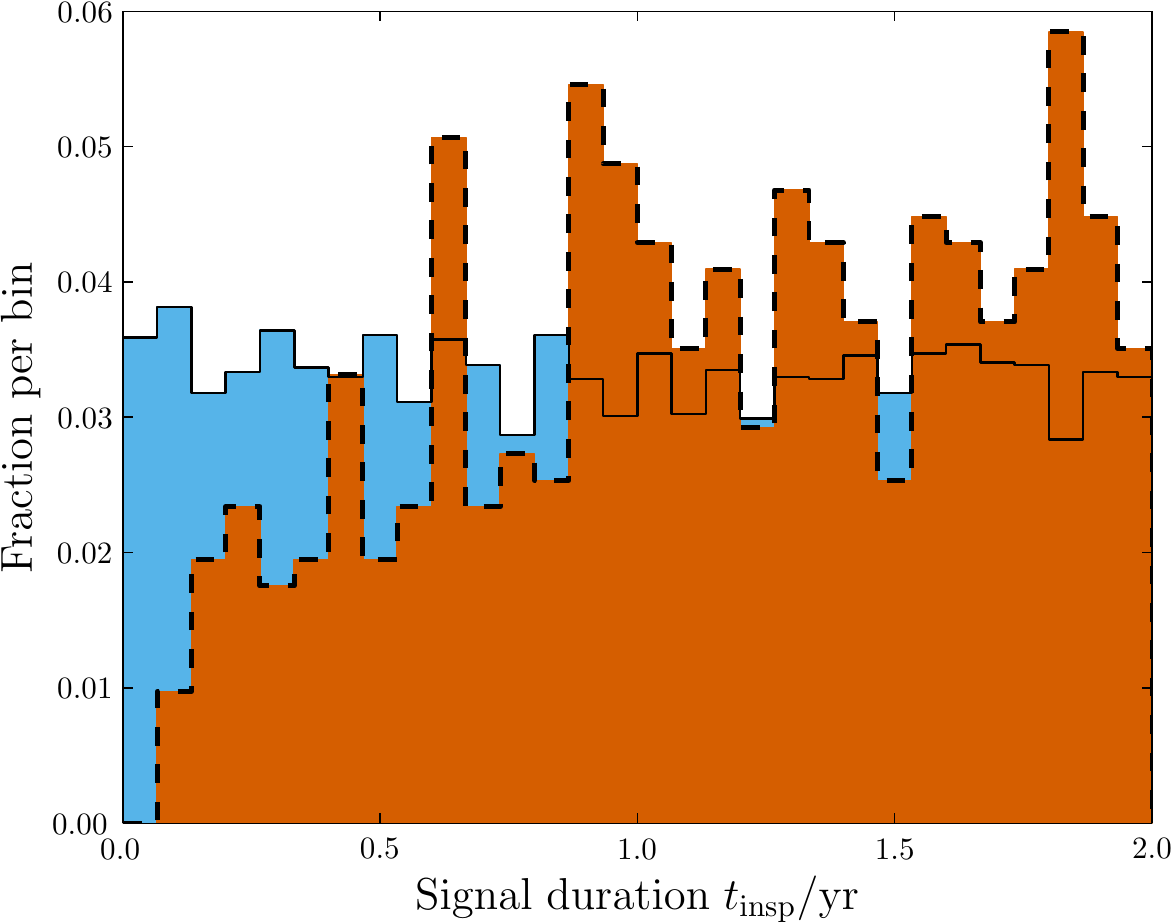} \quad
\includegraphics[width=0.37\textwidth]{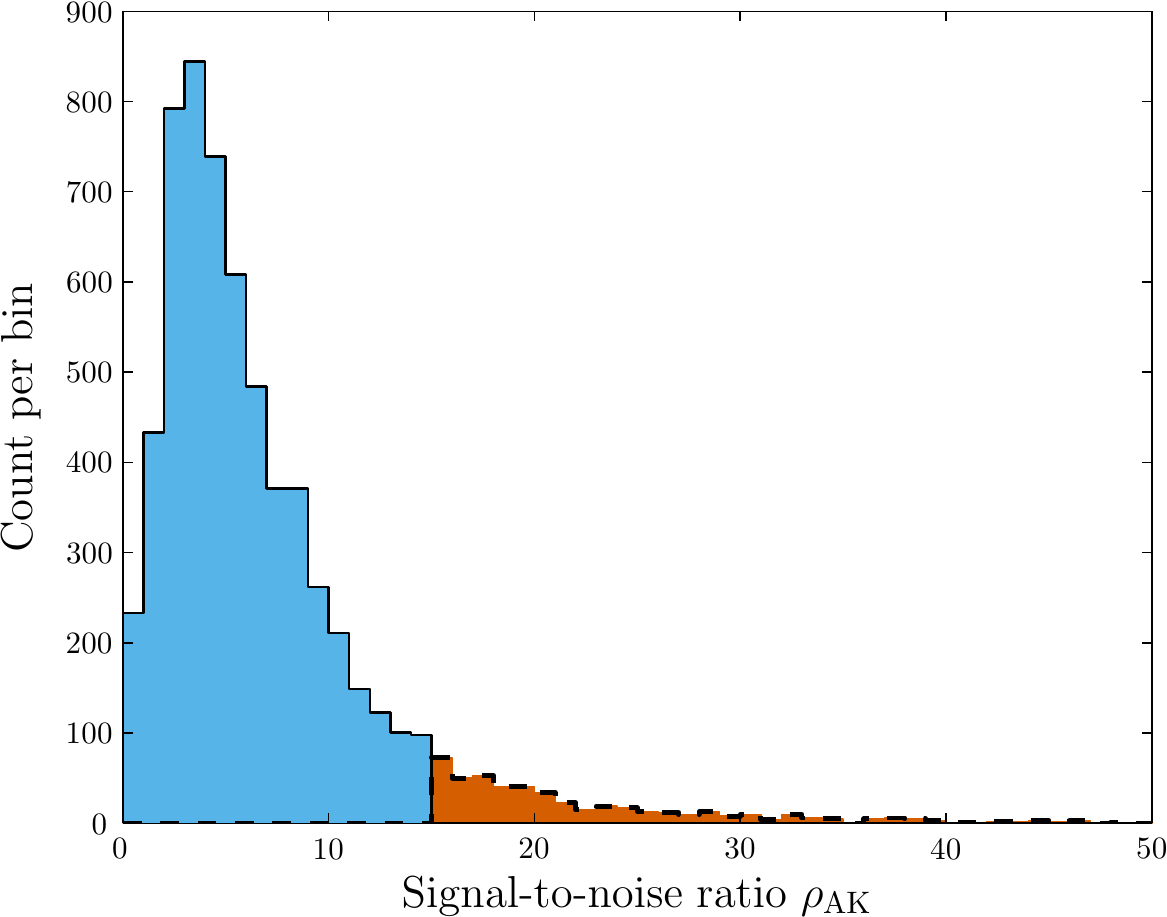}
\caption{\label{fig:EMRIpar-dists}Parameter distributions at plunge for our detectable EMRI systems (dashed outline), alongside those of the undetectable systems (solid outline). For the system parameters, the ordinate-axis values are the probability of a system being found in a particular bin, given that they are either detectable or undetectable. In the final plot for the SNR, we show the number of systems in each bin. The SNRs quoted here are calculated using the AK model and we assume a detection threshold of $\rho\sub{thres} = 15$.}
\end{figure*}

A given EMRI is classified as observable if its SNR exceeds some threshold value $\rho\sub{thres}$. Calculating SNRs from the generated AK waveforms, assuming $6$ laser links and using the eLISA PSD~\cite{Amaro-Seoane2012a}, we find $513$ detectable events across the mission for $\rho\sub{thres} = 15$ (cf.\ \cite{Gair2004,Amaro-Seoane2012a,Mapelli2012}). The parameter distributions for the mass and spin of the SMBH, the orbital shape parameters at plunge, the redshift of the source, and the length of the observation $t\sub{insp}$ are shown in \figref{EMRIpar-dists}, to be contrasted with the distributions of the $5820$ systems with an SNR less than $15$, which are also shown.

The mass distribution for detectable EMRIs is peaked such that the typical GW frequency occurs at the base of the eLISA noise PSD. Systems at higher redshifts start to tail off because the GW amplitude scales inversely to the luminosity distance; by $z = 1.5$, the distribution of detectable EMRIs with eLISA has essentially vanished. Eccentric prograde orbits around SMBHs with larger spins tend to produce larger SNRs because the periapsis in such systems can get much closer to the SMBH, and so the GWs are intrinsically louder. This effect also causes the detectable EMRIs to have smaller values of $p$ at plunge, as observed in its distribution.

EMRI systems within our populations have small eccentricities at plunge than initially due to the circularising effect of GWs~\cite{Peters1964}. For detectable systems, the mean eccentricity is $0.05$ and the maximum is $0.4$, $85\%$ have $e < 0.1$. In \secref{flux-enhance}, we found a strong eccentricity dependence on the magnitude of the resonant flux enhancements. We therefore expect typical resonant jumps in these astrophysical systems to be much less than $1\%$, and so the resultant dephasing to be relatively weak. We now analyse the $513$ systems using our NK models to check for the impact of resonances.

\subsection{Loss of signal-to-noise ratio}
\label{sec:population-SNR}

\begin{figure}
\centering
\includegraphics[width=0.4\textwidth]{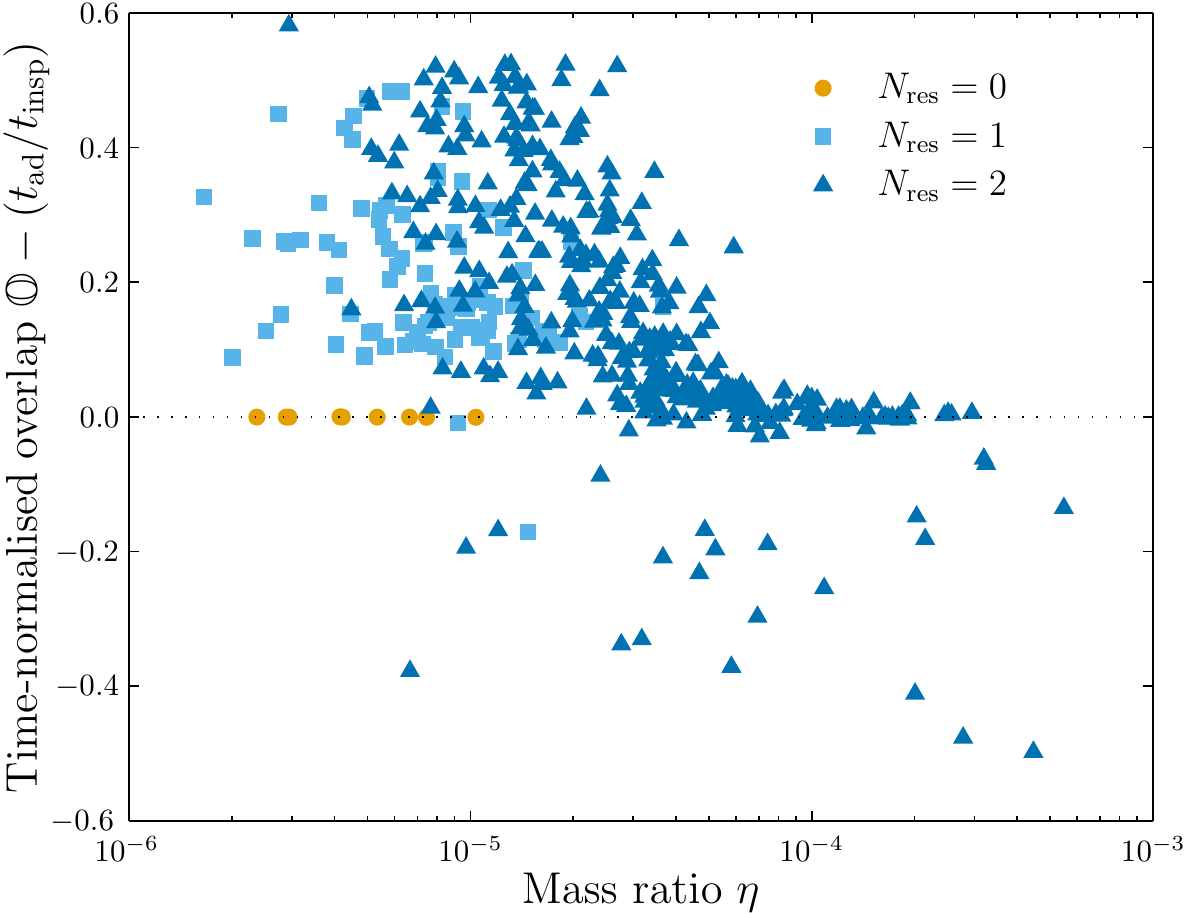}
\caption{\label{fig:pop-adSNR-vs-eta}The difference between the maximum overlap and the expected value $t\sub{ad} / t\sub{insp}$, as a function of the mass ratio for our population of $513$ EMRI signals. Each system encounters either $0$ (circles), $1$ (squares) or $2$ (triangles) resonances during the observation window $t\sub{insp}$, with $t\sub{ad}$ the largest time spent by the inspiral without encountering any resonances.}
\end{figure}

We can study the effect of resonances, by comparing the adiabatic evolution to the full instantaneous evolution; a loss in SNR will reduce the number of detected events. For each inspiral, we denote the longest period of time $t\sub{ad}$ in which neither the $1$:$2$ nor the $2$:$3$ resonance is encountered. From the results of \secref{effres-phase}, we expect the recovered overlap to be approximately given by the proportion of time spent in a resonance-free region, that is $t\sub{ad} / t\sub{insp}$. This assumes that there is perfect overlap in the absence of a resonance and zero overlap across a resonance, with all times during the inspiral contributing equally. 
In \figref{pop-adSNR-vs-eta} we plot the difference between the computed maximum overlap and the value expected from the time between resonances, highlighting the number of resonances $N\sub{res}$ each system encounters.

A small proportion of systems have overlaps below the expected value (approximately $4\%$ have values less than $-0.05$). This might be caused by higher-order resonances disrupting the evolution, in which case $t\sub{ad}$ should be smaller and the systems would lie on the expectation line. However, a more likely explanation is that a suboptimal matching time $t\sub{match}$ was used, and a larger overlap is achievable with a different choice of adiabatic evolution.

Roughly $30\%$ of the systems lie within $0.05$ of the expected value. The vast majority of these are not significantly disrupted by resonances, and produce overlaps approaching unity. For the smallest (most extreme) mass ratios, the inspiral rate is slow, and so the systems do not encounter either the $1$:$2$ or $2$:$3$ resonances during their lifetime. Meanwhile, for the largest (least extreme) mass ratios, the EMRIs encounter both resonances close to plunge. In each case, there is a long resonance-free region, allowing a high overlap to be recovered.

The remaining $66\%$ of systems have overlaps above the level expected if resonances lead to significant dephasing. These occur at low and intermediate values of the mass ratio, where the inspiral rate is slow enough that the low-order resonances are encountered in the middle of the observation window, and the resulting value of $t\sub{ad}$ is small. For these EMRIs, resonances are not as important as expected. Some of these could be because of fortuitous phases on resonance corresponding to small resonance jumps. The most likely explanation for the lack of impact is because of the low eccentricities of our population, this means that the magnitude of the resonant flux enhancements are small (and in most cases negligible).

\begin{figure}
\centering
\includegraphics[width=0.4\textwidth]{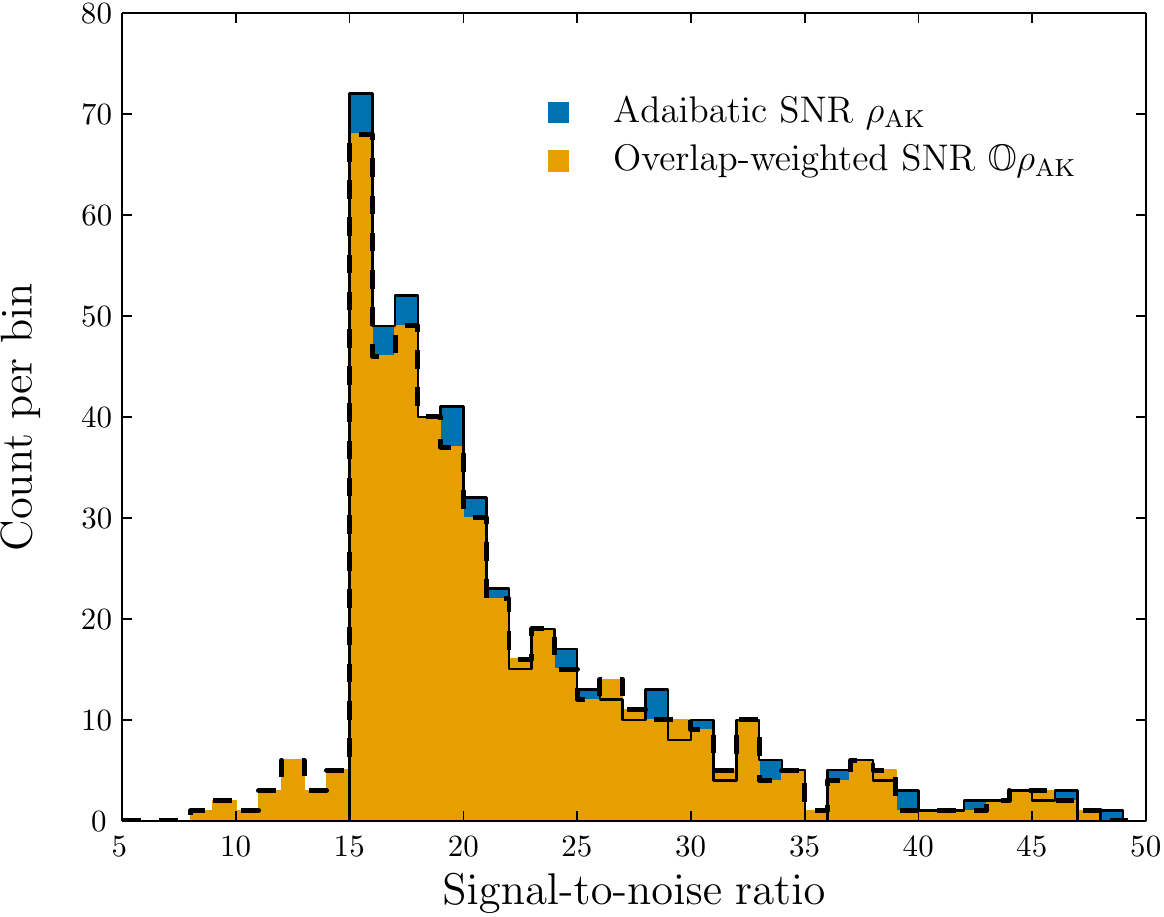}
\caption{\label{fig:pop-SNR-dist}Histogram showing the probability distribution function for the detectable EMRI SNRs, as calculated using the AK formalism (solid outline) and modified by the maximum adiabatic overlap (dashed outline). The tail below $\rho\sub{thres} = 15$ indicates the detections lost because of resonances.}
\end{figure}

Even assuming that all overlap reductions are due to transient resonances (neglecting contributions from imperfectly selected adiabatic waveforms), the overall effect on the population is not significant. To illustrate this, we plot the AK SNRs in \figref{pop-SNR-dist}, multiplied by the maximum recovered overlap to account for the loss in SNR caused by passing through resonance. The total number of detectable systems decreases from $513$ to $492$, a loss of $4\%$. If we increase the threshold $\rho\sub{thres}$, the fractional reduction in the number of detectable systems gets even smaller. We therefore conclude that resonances do not cause sufficient waveform dephasing across a population of EMRIs for the detection rate to be appreciably diminished.

\section{Conclusion}
\label{sec:conclusion}

Transient resonances in EMRIs are an important consideration in waveform modelling due to the high proportion of expected systems encountering a low-order resonance in the later stages of inspiral~\cite{Ruangsri2014}. Passing through resonance can lead to an enhancement or decrement in the radiated fluxes associated with orbital evolution, which in turn leads to a jump in the orbital parameters across the resonance. Including the signature of resonances is necessary for optimal analysis of EMRI signals. 

The duration of resonances and their effect on orbital parameters can be calculated. However, this requires a self-force model. We have made use of a low spin, first-order PN self force. This is of limited validity and so the results should be taken as qualitative estimates; however, the self-force model should suffice for illustrating the potential effects of resonances.

The resonant jump in the orbital parameters depends upon the orbital phase on resonance. This makes it difficult to predict, without detailed calculation, the evolution of an inspiral. The magnitude of the jump depends upon the order of resonance and the orbital eccentricity. The $2$:$3$ and $1$:$2$ resonances have the largest effects, and higher-order resonances are less important. Crucially, jumps are smaller for lower eccentricity orbits. High eccentricity EMRIs may encounter jumps in their orbital parameters of a few percent, leading to rapid dephasing of their waveforms compared to those from an adiabatic evolution.

Amongst a population of sources, unmodelled resonances could diminish detection prospects. However, because of the circularising effects of GW emission, by the time that the most important resonances are encountered, the orbital eccentricity is low. Therefore, the overall effect on SNR recovery is small, and there is not a significant reduction in the number of detectable EMRIs.

While it may not be essential to model resonances to detect (at least a subpopulation of) EMRIs, an unresolved question here is how resonances would affect parameter estimation. Systematic biases may be introduced if inaccurate templates are used; equally, the distinctive features of resonances may allow more precise measurements to be made. Developing a more accurate self-force model is required for a complete quantitative understanding of the effects of transient resonances.

Adiabatic waveform models can still be used for EMRIs away from resonance. Therefore, it may be possible to stitch together waveforms from a sequence of adiabatic evolutions, if the phase on resonance and the magnitude of the self-force can be predicted with sufficient accuracy.\footnote{This may be possible using interpolation schemes~\cite{Warburton2012} if there is sufficient numerical data available for calibration.} Such hybrid models merit further study as relatively simple ways of incorporating resonance effects into adiabatic models.

\begin{acknowledgments}
The authors extend our sincere gratitude to \'{E}anna Flanagan and Tanja Hinderer for providing their PN self-force code, without which this work would not be possible. We are grateful to Stanislav Babak for examining both \cite{BerryThesis2013} and \cite{ColeThesis2015}, and for providing comments on this draft. We thank Tanja Hinderer, Jeandrew Brink, Maarten van de Meent, Leor Barack, Scott Hughes and Nico Yunes for useful conversations, and Christopher Moore with help preparing this manuscript. We are grateful to the anonymous referee for their exceptional careful reading of the paper. RHC was supported by STFC; CPLB thanks STFC and the Cambridge Philosophical Society; PC's work was supported by a Marie Curie Intra-European Fellowship within the 7th European Community Framework Programme (PIEF-GA-2011-299190), and JRG was supported by the Royal Society. This is LIGO Document P1600251. 
\end{acknowledgments}

\appendix

\section{Location of resonances}\label{sec:location}

\begin{figure}
\centering
\includegraphics[width=0.46\textwidth]{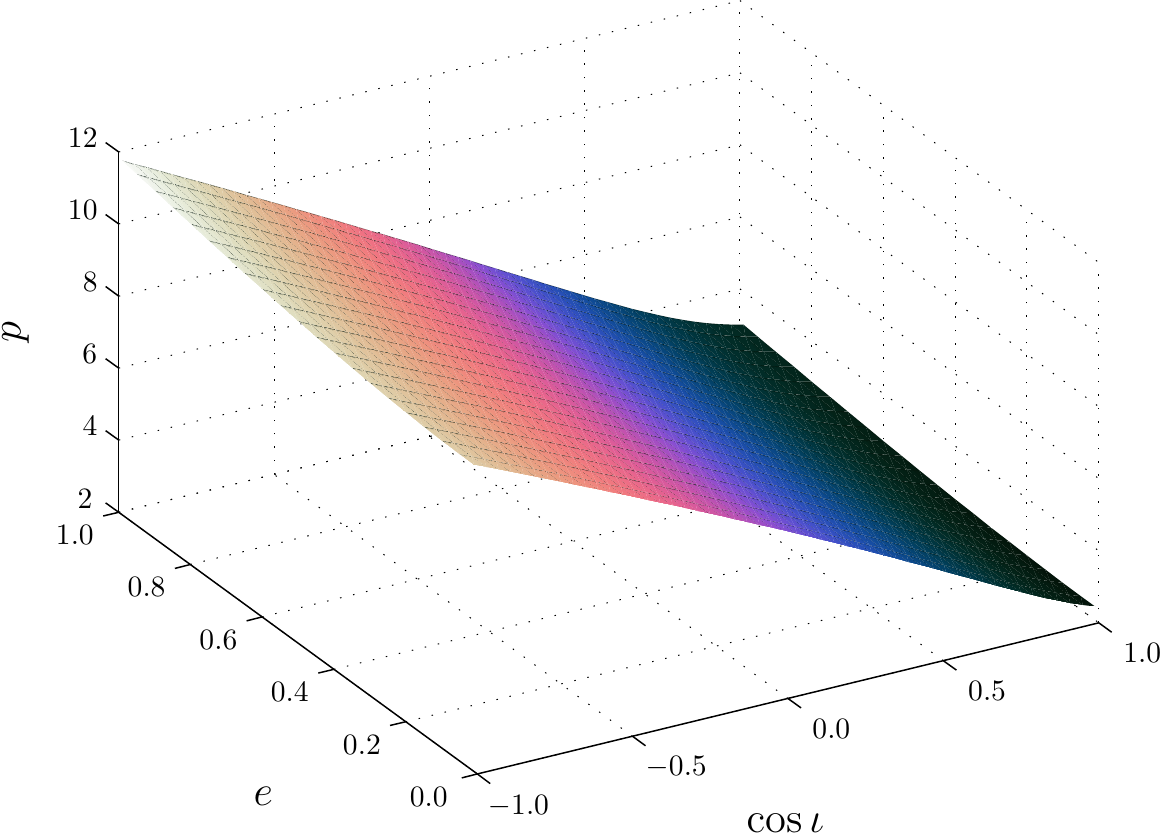}
\caption{\label{fig:res-plane-2-5-95}Location of the $2/5$ resonance surface for an $a_\ast = 0.95$ BH in terms of orbital semilatus rectum $p$, eccentricity $e$ and inclination $\iota$.}
\end{figure}

We can find the location of resonances by numerically solving $\Omega = n_r \Omega_r - n_\theta \Omega_\theta = 0$. \Figref{res-plane-2-5-95} shows the semilatus rectum, eccentricity and (cosine of the) inclination angle of the $\nu = 2/5$ resonance surface for a BH of spin $a_\ast = 0.95$. 
This is almost planar, inspiring us to look for a simple description that can help guide our search for resonance locations. Brink, Geyer and Hinderer~\cite{Brink2013} provide series expansions for the location of resonances in the limit of equatorial orbits for small spin and eccentricity. We do not follow this approach of trying to find analytic expressions for the resonance surface; the expressions become complicated when venturing away from limiting cases. Instead, we build an approximate phenomenological model and fit this to the resonance plane. 
This should be useful for designating the region in which resonance could be expected. To locate them precisely, it is necessary to solve $\Omega = 0$ numerically; the approximate model gives a suitable starting point.

The resonant semilatus rectum for any particular spin and resonance ratio can be well approximated as
\begin{equation}
p(e,\iota;a_\ast,\nu) \simeq A\frac{1 + B e + D \cos\iota}{1 - C\exp(e)}.
\end{equation}
The coefficients $\{A,B,C,D\}$ depend upon the spin and the particular resonance; they can be approximated as
\begin{align} 
A(a_\ast,\nu) \simeq {} & a_0\frac{1 + a_1\nu - a_2 \nu^2 - a_3 \nu a_\ast^2}{1 + a_4\nu - (1 + a_4)\nu^2}, \\
B(a_\ast,\nu) \simeq {} & b_0(1 - b_1\nu)\exp(-b_2\nu)(1 - b_3 a_\ast), \\
C(a_\ast,\nu) \simeq {} & c_0, \\
D(a_\ast,\nu) \simeq {} & d_0\left[1 - \exp(a_\ast)\right]\left[1 - d_1\exp(\nu)\right].
\end{align}
This gives us a total of $12$ parameters for our fit. Whilst this may sound large, if we were fitting an expansion to quadratic order in combinations of $\{e,\iota,a_\ast,\nu\}$ we would have $15$ parameters.\footnote{We find that this does not give as good a fit as our function.} Our optimised parameters are
\begin{equation}
\begin{array}{lll}
a_0 = 5.9854, & a_1 = 3.4116, & a_2 = 0.9253,\\
a_3 = 0.1959, & a_4 = 4.8846, & b_0 = 0.7692,\\
b_1 = 1.4752, & b_2 = 1.4861, & b_3 = 0.5974,\\
c_0 = 0.02332, & d_0 = 0.7968, & d_1 = 0.3115.
\end{array}
\end{equation}
These were fitted for all possible resonances with $n_r = 2$--$7$ as well as the $9$:$10$, $19$:$20$, $49$:$50$ and $99$:$100$ resonances; with SMBH spins of $a_\ast = 0.01$--$0.999$; for orbits with eccentricities $e = 0.01$--$0.99$, and inclinations $\cos\iota = -0.999999$--$0.999999$. 
 
\begin{figure*}[htp]
\centering
\centerline{\includegraphics[width=0.47\textwidth]{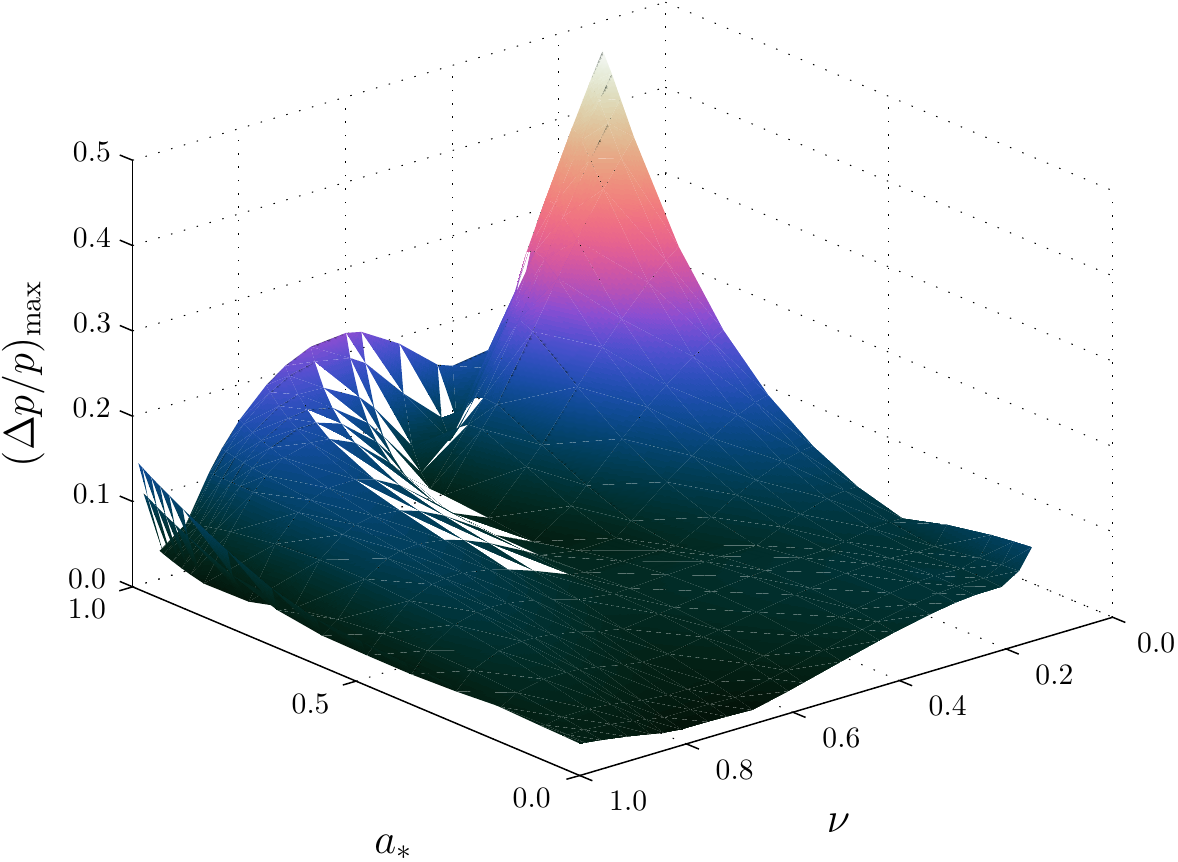}\quad\includegraphics[width=0.47\textwidth]{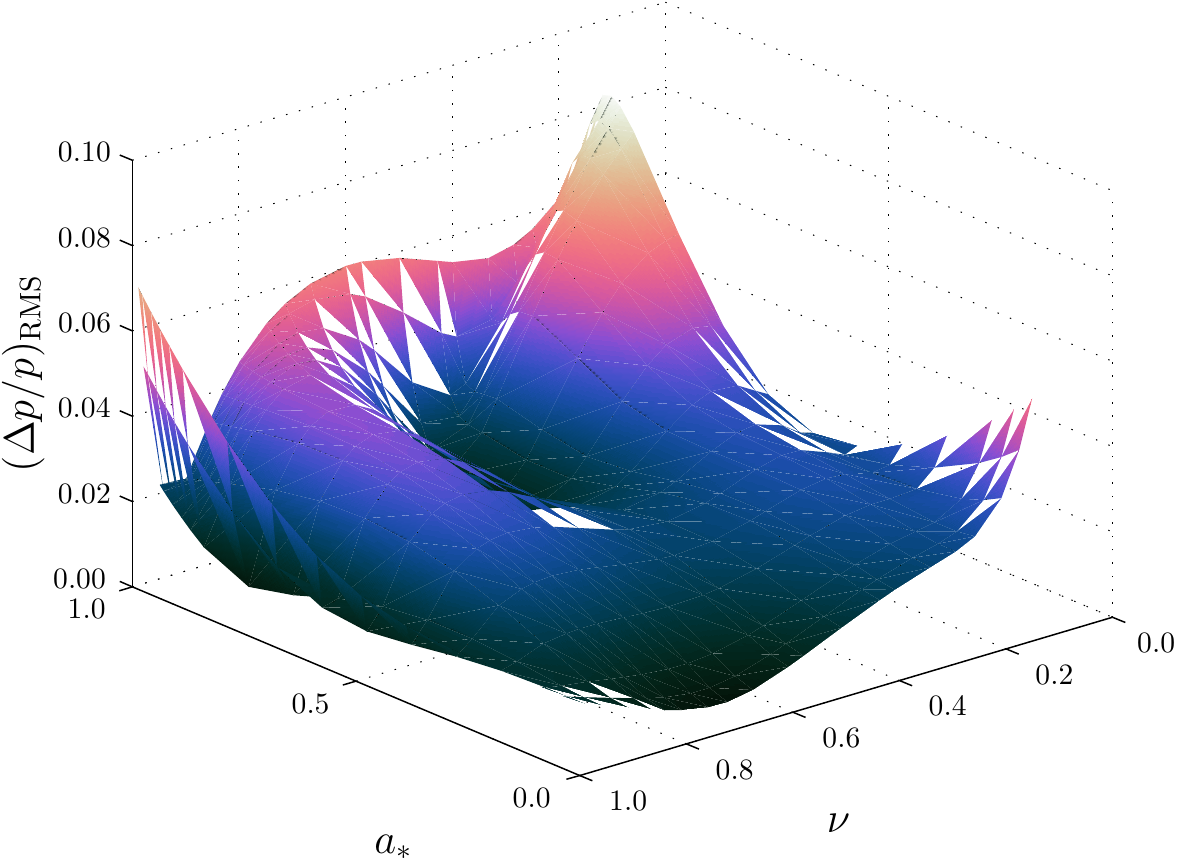}}
\caption{\label{fig:p-error}Relative error in the approximate semilatus rectum compared to the accurate numerical result as a function of BH spin $a_\ast$ and resonance ratio $\nu$. The left panel shows the maximum relative error and the right shows the root-mean-square error; in both cases we are marginalising over eccentricity and inclination.}
\end{figure*}

Using this approximation, the maximum error in $p$ for a given $a_\ast$ and $\nu$ is typically $\sim10\%$ and less than $1$ in absolute terms. The relative error in the semilatus rectum is illustrated in \figref{p-error}. 
The largest fractional error is $\sim50\%$, this is for $a_\ast \rightarrow 1$ and $\nu \rightarrow 0$, and corresponds to small $p$, such that the absolute error is still small. Taking the root-mean-square across $e$ and $\iota$, the fractional error for a given $a_\ast$ and $\nu$ never exceeds $9\%$ and is typically less than $4\%$.

\section{Asymptotic solution for passage through resonance}\label{sec:res-asymptotic}

The impact of passing through resonance on the evolution can be modelled analytically using asymptotic expansions~\cite{Gair2012}. Solutions for the motion are constructed far away from resonance and these are matched to a transition region in the vicinity of resonance~\cite{Kevorkian1971,Bosley1992}. By comparing the matched solution, which incorporates the effects of resonance, with the results of an adiabatic evolution, it is possible to estimate the discrepancy in the orbital parameters. This determines the difference in the orbital phase between the two approaches. If this error is sufficiently small, then it is safe to ignore the effects of the resonance; however, only a small difference is needed to impact the subsequent waveform, since the error accumulates over the subsequent observation of $\sim \order{\eta^{-1}}$ cycles~\cite{Flanagan2012}. We derive formulae which can be used to calculate the discrepancy in the orbital parameters.

The following derivation is reproduced from Berry~\cite{BerryThesis2013}. It is is based upon the analysis of Kevorkian~\cite{Kevorkian1987}; small adjustments have been made to adapt to the specific problem of GW inspiral, but the general argument is unchanged.\footnote{The same two-timescale theory underpins the analysis of Hinderer and Flanagan~\cite{Hinderer2008}, but this explicitly ignores resonances.} A similar derivation can be found in van de Meent~\cite{VanDeMeent2013}.

We model the system using action--angle variables. We are only concerned with the $r$ and $\theta$ motions, so we have a $2$-dimensional system. We perform a canonical transformation  to isolate the resonant combination $q = n_r q_r - n_\theta q_\theta$~\cite{Bosley1992,VanDeMeent2013}. This becomes one of the new angle variables, the other variable $q'$ can be either $q_r$ or $q_\theta$ (as, on resonance, varying one necessarily changes the other). We use $J$ as the conjugate action variable to $q$ and $\omega = n_r \omega_r - n_\theta \omega_\theta$ as its frequency. Similarly, we use $J'$ as the action variable conjugate to $q'$. The system evolves through resonance slowly, on an evolution timescale, so we parametrize it in terms of a slow time parameter
\begin{equation}
\widetilde{\lambda} = \eta\lambda.
\end{equation}
The orbits of $q'$ proceed with the fast time $\lambda$; since this is much more rapid than the evolution we are interested in, it is safe to average over it. We are not interested in the fine-grained fast oscillations caused by changes in $q'$. For this analysis we consider the reduced problem of evolving $q$ and $J$.

At resonance $\widetilde{\lambda} = \widetilde{\lambda}_\star$ and $\omega\left(\widetilde{\lambda}_\star\right) = 0$. We assume that the frequency has a simple zero and can be expanded as
\begin{equation}
\omega\left(\widetilde{\lambda}\right) = \varpi_1\left(\widetilde{\lambda} - \widetilde{\lambda}_\star\right) + \varpi_2\left(\widetilde{\lambda} - \widetilde{\lambda}_\star\right)^2 + \ldots
\label{eq:omega-series}
\end{equation}
The frequency is actually a function of the angle variables, but since these evolve with $\widetilde{\lambda}$ it is safe to write it as a function of the slow time.\footnote{In effect we are defining $\omega\left(\tilde{\lambda}\right) \equiv \omega\left[J\left(\tilde{\lambda}\right)\right]$.}

Using the slow time, the equations of motion (\ref{eq:Mino-E-o-M}) become
\begin{subequations}
\begin{align}
\diff{q}{\widetilde{\lambda}} = {} & \dfrac{\omega(J)}{\eta} + \sum_s g_s^{(1)}(J)\exp(is q)  + \order{\eta}, \\
\diff{J}{\widetilde{\lambda}} = {} & \sum_s G_s^{(1)}(J)\exp(is q) + \order{\eta},
\end{align}
\end{subequations}
where we have rewritten the forcing terms as Fourier series and adapted the forcing functions to those appropriate for $q$ and $J$. We solve these before resonance and then match to solutions in the transition regime about resonance.

\subsection{Solution before resonance}\label{sec:before-res}

To find a solution away from the resonance, we decompose the problem to be a function of two timescales~\cite{Kevorkian1971}. We use the slow time $\widetilde{\lambda}$ and, as a proxy for the fast time,
\begin{equation}
\Psi = \intd{0}{\lambda}{\omega(\eta\tau)}{\tau} = \recip{\eta}\intd{0}{\tilde{\lambda}}{\omega(\widetilde{\tau})}{\widetilde{\tau}}.
\vspace{0.000cm}
\end{equation}
From this
\begin{equation}
\omega = \diff{\Psi}{\lambda}.
\end{equation}
In terms of these two variables, we can build ansatz solutions
\begin{subequations}
\begin{align}
\label{eq:q-series}
q(\lambda;\,\eta) = {} & \Psi + q_0\left(\widetilde{\lambda}\right) + \eta q_1\left(\Psi,\widetilde{\lambda}\right) + \order{\eta^2}, \\
J(\lambda;\,\eta) = {} & J_0\left(\widetilde{\lambda}\right) + \eta J_1\left(\Psi,\widetilde{\lambda}\right) + \order{\eta^2}.
\label{eq:J-series}
\end{align}
\end{subequations}
We can also write a series expansion for the frequency, since it is affected by the self-force too,
\begin{equation}
\omega(\lambda;\,\eta) = \omega_0\left(\widetilde{\lambda}\right) + \eta \omega_1\left(\widetilde{\lambda}\right) + \order{\eta^2}.
\end{equation}
In the limit of $\eta \rightarrow 0$ we are left with a constant frequency $\omega_0(0)$. The higher-order terms are identified below by matching terms in the series expansions of the equations of motion. Taking the two timescales as independent, we may write the time derivative to $\order{\eta}$ as
\begin{equation}
\diffop{\lambda} = \omega_0\partialdiffop{\Psi} + \eta\omega_1\partialdiffop{\Psi} + \eta\partialdiffop{\widetilde{\lambda}}.
\end{equation}

Using the two timescale decomposition to replace the time derivatives in the equations of motion, and substituting in the ansatz expansions gives, to first order,
\begin{widetext}
\begin{subequations}
\begin{align}
\label{eq:q-1}
\omega_0 + \eta\omega_1 + \eta\partialdiff{q_0}{\widetilde{\lambda}} + \eta\omega_0\partialdiff{q_1}{\Psi} = {} & \omega(J_0) + \eta\diff{\omega}{J}J_1 + \eta \sum_s g_s^{(1)}(J_0)\exp\left[is(\Psi + q_0)\right], \\
\eta\partialdiff{J_0}{\widetilde{\lambda}} + \eta\omega_0\partialdiff{J_1}{\Psi} = {} & \eta \sum_s G_s^{(1)}(J_0)\exp\left[is(\Psi + q_0)\right].
\label{eq:J-1}
\end{align}
\end{subequations}
\end{widetext}
Averaging \eqnref{J-1} over $\Psi$ gives\footnote{The ansatz is constructed such that $J_0 \equiv \langle J_0\rangle_\Psi$ and $q_0 \equiv \langle q_0\rangle_\Psi$.}
\begin{equation}
\partialdiff{J_0}{\widetilde{\lambda}} = G_0^{(1)}(J_0).
\label{eq:J-ad}
\end{equation}
This describes the adiabatic evolution, hence we can identify $J_0\left(\widetilde{\lambda}\right)$ with (the lowest-order piece of) the adiabatic solution~\cite{Hinderer2008}. If we similarly average \eqnref{q-1}, we find
\begin{equation}
\omega_0 + \eta\omega_1 +\eta\partialdiff{q_0}{\widetilde{\lambda}} = \omega(J_0) + \eta\partialdiff{\omega}{J}\langle J_1\rangle_\Psi + \eta g_0^{(1)}(J_0).
\end{equation}
From this we can identify the terms that originate from the frequency and, matching by order in $\eta$, obtain
\begin{subequations}
\begin{align}
\omega_0 = {} & \omega(J_0), \\
\omega_1 = {} & \partialdiff{\omega}{J}\langle J_1\rangle_\Psi.
\end{align}
\end{subequations}
This leaves
\begin{align}
\partialdiff{q_0}{\widetilde{\lambda}} = {} & g_0^{(1)}(J_0) \\
q_0 = {} & \kappa_0 + \intd{0}{\tilde{\lambda}}{g_0^{(1)}[J_0(\tau)]}{\tau},
\label{eq:q-0-sol}
\end{align}
where $\kappa_0$ is the constant of integration. We now have expressions for the lowest-order terms in the expansions.

Subtracting the $s = 0$ components from \eqnref{J-1} leaves
\begin{equation}
\omega_0\partialdiff{J_1}{\Psi} = \sum_{s\,\neq\,0} G_s^{(1)}(J_0)\exp\left[is(\Psi + q_0)\right].
\end{equation}
This can be solved to give
\begin{equation}
J_1 = \langle J_1\rangle_\Psi + \recip{\omega_0}\sum_{s\,\neq\,0} \dfrac{G_s^{(1)}(J_0)\exp\left[is(\Psi + q_0)\right]}{is}.
\label{eq:J-1-sol}
\end{equation}
We can do the same for \eqnref{q-1} to obtain
\begin{equation}
q_1 = \langle q_1\rangle_\Psi + \recip{\omega_0}\sum_{s\,\neq\,0} \dfrac{g_s^{(1)}(J_0)\exp\left[is(\Psi + q_0)\right]}{is}.
\label{eq:q-1-sol}
\end{equation}
To find the constants of integration, $\langle q_1\rangle_\Psi$ and $\langle J_1\rangle_\Psi$, it is necessary to extend the analysis to second order in $\eta$. This shows that $\langle J_1\rangle_\Psi$ is the first-order component of the adiabatic solution. We do not need explicit forms for later calculations, so we will not proceed further. We have successfully constructed the pre-resonance solution.

\subsection{Solution near resonance}\label{sec:interior-res}

Near to resonance, we consider an interior layer expansion~\cite{Kevorkian1971}. As explained in \secref{res-time}, evolution near resonance proceeds on a timescale intermediate between the slow and fast times. We therefore introduce a rescaled time
\begin{equation}
\widehat{\lambda} = \dfrac{\widetilde{\lambda} - \widetilde{\lambda}_\star}{\eta^{1/2}} = \eta^{1/2}(\lambda - \lambda_\star).
\end{equation}
As for the before resonance solution, we can create a series expansion; however, now we expand in terms of $\eta^{1/2}$~\cite{Flanagan2012}
\begin{subequations}
\begin{align}
q\left(\widehat{\lambda};\,\eta\right) = {} & \widehat{q}_0\left(\widehat{\lambda}\right) + \eta^{1/2} \widehat{q}_{1/2}\left(\widehat{\lambda}\right) + \order{\eta}, \\
J\left(\widehat{\lambda};\,\eta\right) = {} & \widehat{J}_0 + \eta^{1/2} \widehat{J}_{1/2}\left(\widehat{\lambda}\right) + \order{\eta}.
\end{align}
\end{subequations}
The series expansion for the frequency, \eqnref{omega-series}, can be rewritten as
\begin{equation}
\omega\left(\widehat{\lambda}\right) = \eta^{1/2}\varpi_1\widehat{\lambda} + \eta\varpi_2\widehat{\lambda}^2 + \order{\eta^{3/2}}.
\label{eq:omega-hat}
\end{equation}
Proceeding to write the equations of motion in terms of the rescaled time gives
\begin{subequations}
\begin{align}
\diff{q}{\widehat{\lambda}} = {} & \varpi_1\widehat{\lambda} + \eta^{1/2}\varpi_2\widehat{\lambda}^2 \nonumber \\ 
 {} & + \left. \eta^{1/2}\sum_s g_s^{(1)}\left(\widehat{J}_0,\widetilde{\lambda}_\star\right)\exp(is \widehat{q}_0)  + \order{\eta}, \right. \\
\diff{J}{\widehat{\lambda}} = {} & \eta^{1/2}\sum_s G_s^{(1)}\left(\widehat{J}_0,\widetilde{\lambda}_\star\right)\exp(is \widehat{q}_0) + \order{\eta}.
\end{align}
\end{subequations}

From the equations of motion we can match terms by their order in $\eta^{1/2}$. At zeroth order we find
\begin{equation}
\widehat{J}_0 = \widehat{\varrho}_0
\end{equation}
is constant, and
\begin{equation}
\widehat{q}_0 = \widehat{\kappa}_0 + \dfrac{\varpi_1\widehat{\lambda}^2}{2},
\end{equation}
where $\widehat{\varrho}_0$ and $\widehat{\kappa}_0$ are the constants of integration. Using these, we can build the next-order terms
\begin{align}
\widehat{q}_{1/2} = {} & \widehat{\kappa}_{1/2} + \dfrac{\varpi_2\widehat{\lambda}^3}{3} + g_0^{(1)}(\widehat{\varrho}_0)\widehat{\lambda} \nonumber \\ 
 {} & + \sum_{s\,\neq\,0}g_s^{(1)}(\widehat{\varrho}_0)\exp(is \widehat{\kappa}_0)\intd{0}{\hat{\lambda}}{\exp\left(\dfrac{is \varpi_1\tau^2}{2}\right)}{\tau}, \\
\widehat{J}_{1/2} = {} & \widehat{\varrho}_{1/2} + G_0^{(1)}(\widehat{\varrho}_0)\widehat{\lambda} \nonumber \\
 {} & + \sum_{s\,\neq\,0}G_s^{(1)}(\widehat{\varrho}_0)\exp(is \widehat{\kappa}_0)\intd{0}{\hat{\lambda}}{\exp\left(\dfrac{is \varpi_1\tau^2}{2}\right)}{\tau},
\end{align}
introducing integration constants $\widehat{q}_{1/2}$ and $\widehat{\varrho}_{1/2}$. Both the above expressions involve the complex Fresnel integral~\cite{Olver2010}, the details of which are given in the following section. We have now constructed the interior solution. 

\subsection{The complex Fresnel integral}

The solution for the motion in the interior region near to resonance involves the integral
\begin{equation}
W\left(\widehat{\lambda}\right) = \intd{0}{\hat{\lambda}}{\exp\left(\dfrac{is \varpi_1\tau^2}{2}\right)}{\tau}.
\end{equation}
The complex Fresnel integral is
\begin{equation}
\mathcal{Y}(z) = \intd{0}{z}{\exp\left(\dfrac{i\pi x^2}{2}\right)}{x} = \mathcal{C}(z) + i\mathcal{S}(z),
\end{equation}
where $\mathcal{C}(z)$ and $\mathcal{S}(z)$ are the cosine and sine Fresnel integrals~\cite{Olver2010}, and hence 
\begin{equation}
W\left(\widehat{\lambda}\right) = \sqrt{\dfrac{\pi}{s\varpi_1}}\mathcal{Y}\left(\sqrt{\dfrac{s\varpi_1}{\pi}}\widehat{\lambda}\right).
\end{equation}

We are interested in the asymptotic behaviour for $|\widehat{\lambda}| \rightarrow \infty$. The complex Fresnel integral has the limit~\cite{Olver2010} 
\begin{equation}
\lim_{|z|\,\rightarrow\,\infty} \mathcal{Y}(z) \sim \dfrac{\sgn z}{\sqrt{2}} \exp\left(\dfrac{i\pi}{4}\right) - \dfrac{i}{\pi z}\exp\left(\dfrac{i\pi z^2}{2}\right),
\end{equation}
where 
\begin{equation}
\sgn z = \begin{cases}
1 & z > 0 \\
-1 & z < 0
\end{cases}\,.
\end{equation}
Hence,
\begin{align}
\lim_{|\widehat{\lambda}|\,\rightarrow\,\infty}W\left(\widehat{\lambda}\right) \sim {} & \dfrac{\sgn \widehat{\lambda}}{\sqrt{2}}\sqrt{\dfrac{\pi}{|s\varpi_1|}}\exp\left[\sgn(s\varpi_1)\dfrac{i\pi}{4}\right] \nonumber \\
  {} & + \recip{is\varpi_1 \widehat{\lambda}}\exp\left(\dfrac{is \varpi_1 \widehat{\lambda}^2}{2}\right).
\label{eq:Fres-limit}
\end{align}

\subsection{Matching solutions}

To complete our solution we must match the pre-resonance solution of \secref{before-res} with the near-resonance solution of \secref{interior-res}. This is achieved by rewriting the pre-resonance solution in terms of the rescaled time $\widehat{\lambda}$ and comparing this with the near-resonance solution expanded in the limit of $\widehat{\lambda} \rightarrow -\infty$.

To rewrite the pre-resonance solution, we begin with the fast phase parameter
\begin{equation}
\Psi\left(\widehat{\lambda}\right) = \dfrac{\Psi_\star}{\eta} + \dfrac{\varpi_1\widehat{\lambda}^2}{2} + \eta^{1/2}\dfrac{\varpi_2\widehat{\lambda}^3}{3} + \order{\eta}.
\end{equation}
Using this together with Eqs~(\ref{eq:q-0-sol}) and (\ref{eq:q-1-sol}) in \eqnref{q-series}, the angle variable is
\begin{widetext}
\begin{align}
q\left(\widehat{\lambda};\,\eta\right) = {} & \dfrac{\Psi_\star}{\eta} + \dfrac{\varpi_1\widehat{\lambda}^2}{2} + \kappa_\star + \eta^{1/2}\dfrac{\varpi_2\widehat{\lambda}^3}{3} + \eta^{1/2}g_0^{(1)}(J_\star)\widehat{\lambda} + \dfrac{\eta^{1/2}}{\varpi_1\widehat{\lambda}}\sum_{s\,\neq\,0}\recip{is}g_s^{(1)}(J_\star)\exp\left[is\left(\dfrac{\Psi_\star}{\eta} + \dfrac{\varpi_1\widehat{\lambda}^2}{2} + \kappa_\star\right)\right] + \order{\eta},
\end{align}
where we have defined $J_\star \equiv J_0\left(\widetilde{\lambda}_\star\right)$ and $\kappa_\star = \kappa_0 + \intd{0}{\tilde{\lambda}_\star}{g_0^{(1)}[J_0(\tau)]}{\tau}$, and used \eqnref{omega-hat} to substitute for $\omega$. The action variable is similarly determined by using Eqs~(\ref{eq:J-ad}) and (\ref{eq:J-1-sol}) with \eqnref{J-series} to give
\begin{align}
J\left(\widehat{\lambda};\,\eta\right) = {} & J_\star + \eta^{1/2}G_0^{(1)}(J_\star)\widehat{\lambda} + \dfrac{\eta^{1/2}}{\varpi_1\widehat{\lambda}}\sum_{s\,\neq\,0}\recip{is}G_s^{(1)}(J_\star)\exp\left[is\left(\dfrac{\Psi_\star}{\eta} + \dfrac{\varpi_1\widehat{\lambda}^2}{2} + \kappa_\star\right)\right] + \order{\eta}.
\end{align}
\end{widetext}
We can now compare this to the near-resonance expansion with the integral replaced by the limiting form given in \eqnref{Fres-limit}.

At zeroth order, we immediately obtain
\begin{align}
\widehat{\kappa}_0 = {} & \dfrac{\Psi_\star}{\eta} + \kappa_\star, \\
\widehat{\varrho}_0 = {} & J_\star.
\end{align}
These fix the integration constants. The more interesting result is now found by comparing the $\order{\eta^{1/2}}$ terms. Equating the angle variable expressions and cancelling terms gives
\begin{align}
\widehat{\kappa}_{1/2} = {} & \sum_{s\,\neq\,0}g_s^{(1)}(\widehat{\varrho}_0)\sqrt{\dfrac{\pi}{2|s\varpi_1|}}\exp\left[i\left(s \widehat{\kappa}_0 + \dfrac{\pi}{4}\sgn s\varpi_1\right)\right].
\end{align}
Similarly, for the action variable
\begin{equation}
\widehat{\varrho}_{1/2} = \sum_{s\,\neq\,0}G_s^{(1)}(\widehat{\varrho}_0)\sqrt{\dfrac{\pi}{2|s\varpi_1|}}\exp\left[i\left(s \widehat{\kappa}_0 + \dfrac{\pi}{4}\sgn s\varpi_1\right)\right].
\label{eq:J-1/2}
\end{equation}
We now have a matched solution through resonance.

Having constructed the solution, we see that the lowest-order evolution corresponds to the adiabatic solution; the deviations come in at the following order. When we switch from the pre-resonance solution to the post-resonance solution, there is a change in the sign of $\widehat{\lambda}$. Therefore, when matching the post-resonance solution $\widehat{\varrho}_{1/2}$ and $\widehat{\kappa}_{1/2}$ also change sign: there is a change of
\begin{align}
\Delta q = {} & 2 \eta^{1/2}\widehat{\kappa}_{1/2}, \\
\Delta J = {} & 2 \eta^{1/2}\widehat{\varrho}_{1/2}
\label{eq:jumps}
\end{align}
across the resonance~\cite{Kevorkian1987}. We are not particularly interested in the deviation in $J$, of greater concern is the change in the orbital parameters $\{E,L_z,Q\}$. Assuming that there is a smooth transformation that maps between $J$ and these, then, to lowest order, we can calculate the deviation relative to the adiabatic prescription by substituting the forcing functions $G^{(1)} \rightarrow G_a^{(1)}$, where $G_a^{(1)}$ describes the evolution of $\mathcal{I}^a$ through the effects of the self-force. This result is quoted by Flanagan and Hinderer~\cite{Flanagan2012}. The change in the orbital parameters is determined by the forcing functions, hence it is essential to have an accurate self-force model.

As a final step in understanding our result, we switch from Mino time to coordinate time. An appropriate redefinition of the forcing functions can be done by scaling by $\Gamma$, we define
\begin{equation}
F_a^{(1)} = \dfrac{G_a^{(1)}}{\Gamma},
\end{equation}
such that the equation of motion becomes
\begin{equation}
\left\langle \diff{\mathcal{I}^a}{t}\right\rangle_{q'} =  \eta\sum_s F_{a,\,s}^{(1)}(\boldsymbol{\mathcal{I}})\exp(is q) + \order{\eta^2}.
\end{equation}
Here we have made the averaging over $q'$ explicit to show that the equation is only defined as an orbital average: not only does our asymptotic expansion average out oscillations over an orbit in $q'$, but in converting from $\lambda$ to $t$ we have used $\Gamma$ which is an orbital average.
From \eqnref{omega-series}, we recognise that
\begin{equation}
\varpi_1 = \partialdiff{\omega}{\widetilde{\lambda}} = \dfrac{\Gamma^2}{\eta}\left\langle\dot{\Omega}\right\rangle_{q'}.
\end{equation}
We have used the averaged form of $\dot{\Omega}(t)$ as this is appropriate. Using these to adapt Eqs~(\ref{eq:J-1/2}) and (\ref{eq:jumps}), we obtain
\begin{align}
\Delta \mathcal{I}^a = {} & \eta\sum_{s\,\neq\,0}F_{a,\,s}^{(1)}(\boldsymbol{\mathcal{I}}_\star)\left[\dfrac{2\pi}{\left|s \left\langle\dot{\Omega}\right\rangle_{q'}\right|}\right]^{1/2} \nonumber \\*
 {} & \times \left.\exp\left[i\left(s \widehat{\kappa}_0 + \dfrac{\pi}{4}\sgn s\dot{\Omega}\right)\right]\right. \\
 = {} & \eta\sum_{s\,\neq\,0}F_{a,\,s}^{(1)}(\boldsymbol{\mathcal{I}}_\star)\tau_{\mathrm{res},\,s}\exp\left[i\left(s \widehat{\kappa}_0 + \dfrac{\pi}{4} \sgn s\dot{\Omega} \right)\right],
\end{align}
using \eqnref{T-res-s} and representing the values on resonance of $E$, $L_z$ and $Q$ with $\boldsymbol{\mathcal{I}}_\star$.

\bibliography{Resonances}

\begin{thebibliography}{145}%
\makeatletter
\providecommand \@ifxundefined [1]{%
 \@ifx{#1\undefined}
}%
\providecommand \@ifnum [1]{%
 \ifnum #1\expandafter \@firstoftwo
 \else \expandafter \@secondoftwo
 \fi
}%
\providecommand \@ifx [1]{%
 \ifx #1\expandafter \@firstoftwo
 \else \expandafter \@secondoftwo
 \fi
}%
\providecommand \natexlab [1]{#1}%
\providecommand \enquote  [1]{``#1''}%
\providecommand \bibnamefont  [1]{#1}%
\providecommand \bibfnamefont [1]{#1}%
\providecommand \citenamefont [1]{#1}%
\providecommand \href@noop [0]{\@secondoftwo}%
\providecommand \href [0]{\begingroup \@sanitize@url \@href}%
\providecommand \@href[1]{\@@startlink{#1}\@@href}%
\providecommand \@@href[1]{\endgroup#1\@@endlink}%
\providecommand \@sanitize@url [0]{\catcode `\\12\catcode `\$12\catcode
  `\&12\catcode `\#12\catcode `\^12\catcode `\_12\catcode `\%12\relax}%
\providecommand \@@startlink[1]{}%
\providecommand \@@endlink[0]{}%
\providecommand \url  [0]{\begingroup\@sanitize@url \@url }%
\providecommand \@url [1]{\endgroup\@href {#1}{\urlprefix }}%
\providecommand \urlprefix  [0]{URL }%
\providecommand \Eprint [0]{\href }%
\providecommand \doibase [0]{http://dx.doi.org/}%
\providecommand \selectlanguage [0]{\@gobble}%
\providecommand \bibinfo  [0]{\@secondoftwo}%
\providecommand \bibfield  [0]{\@secondoftwo}%
\providecommand \translation [1]{[#1]}%
\providecommand \BibitemOpen [0]{}%
\providecommand \bibitemStop [0]{}%
\providecommand \bibitemNoStop [0]{.\EOS\space}%
\providecommand \EOS [0]{\spacefactor3000\relax}%
\providecommand \BibitemShut  [1]{\csname bibitem#1\endcsname}%
\let\auto@bib@innerbib\@empty
\bibitem [{\citenamefont {Chandrasekhar}(1992)}]{Chandrasekhar1992}%
  \BibitemOpen
  \bibfield  {author} {\bibinfo {author} {\bibfnamefont {S.}~\bibnamefont
  {Chandrasekhar}},\ }\href@noop {} {\emph {\bibinfo {title} {{The Mathematical
  Theory of Black Holes}}}},\ Oxford Classic Texts in the Physical Sciences\
  (\bibinfo  {publisher} {Oxford University Press},\ \bibinfo {address}
  {Oxford},\ \bibinfo {year} {1992})\BibitemShut {NoStop}%
\bibitem [{\citenamefont {Pretorius}(2005)}]{Pretorius2005}%
  \BibitemOpen
  \bibfield  {author} {\bibinfo {author} {\bibfnamefont {F.}~\bibnamefont
  {Pretorius}},\ }\href {\doibase 10.1103/PhysRevLett.95.121101} {\bibfield
  {journal} {\bibinfo  {journal} {Phys.\ Rev.\ Lett.}\ }\textbf {\bibinfo
  {volume} {95}},\ \bibinfo {pages} {121101} (\bibinfo {year} {2005})},\
  \Eprint {http://arxiv.org/abs/0507014} {arXiv:0507014 [gr-qc]} \BibitemShut
  {NoStop}%
\bibitem [{\citenamefont {Campanelli}\ \emph {et~al.}(2006)\citenamefont
  {Campanelli}, \citenamefont {Lousto}, \citenamefont {Marronetti},\ and\
  \citenamefont {Zlochower}}]{Campanelli2006}%
  \BibitemOpen
  \bibfield  {author} {\bibinfo {author} {\bibfnamefont {M.}~\bibnamefont
  {Campanelli}}, \bibinfo {author} {\bibfnamefont {C.~O.}\ \bibnamefont
  {Lousto}}, \bibinfo {author} {\bibfnamefont {P.}~\bibnamefont {Marronetti}},
  \ and\ \bibinfo {author} {\bibfnamefont {Y.}~\bibnamefont {Zlochower}},\
  }\href {\doibase 10.1103/PhysRevLett.96.111101} {\bibfield  {journal}
  {\bibinfo  {journal} {Phys.\ Rev.\ Lett.}\ }\textbf {\bibinfo {volume}
  {96}},\ \bibinfo {pages} {111101} (\bibinfo {year} {2006})},\ \Eprint
  {http://arxiv.org/abs/0511048} {arXiv:0511048 [gr-qc]} \BibitemShut {NoStop}%
\bibitem [{\citenamefont {Baker}\ \emph {et~al.}(2006)\citenamefont {Baker},
  \citenamefont {Centrella}, \citenamefont {Choi}, \citenamefont {Koppitz},\
  and\ \citenamefont {van Meter}}]{Baker2006}%
  \BibitemOpen
  \bibfield  {author} {\bibinfo {author} {\bibfnamefont {J.~G.}\ \bibnamefont
  {Baker}}, \bibinfo {author} {\bibfnamefont {J.}~\bibnamefont {Centrella}},
  \bibinfo {author} {\bibfnamefont {D.-I.}\ \bibnamefont {Choi}}, \bibinfo
  {author} {\bibfnamefont {M.}~\bibnamefont {Koppitz}}, \ and\ \bibinfo
  {author} {\bibfnamefont {J.}~\bibnamefont {van Meter}},\ }\href {\doibase
  10.1103/PhysRevLett.96.111102} {\bibfield  {journal} {\bibinfo  {journal}
  {Phys.\ Rev.\ Lett.}\ }\textbf {\bibinfo {volume} {96}},\ \bibinfo {pages}
  {111102} (\bibinfo {year} {2006})},\ \Eprint {http://arxiv.org/abs/0511103}
  {arXiv:0511103 [gr-qc]} \BibitemShut {NoStop}%
\bibitem [{\citenamefont {Szil{\'{a}}gyi}\ \emph {et~al.}(2015)\citenamefont
  {Szil{\'{a}}gyi}, \citenamefont {Blackman}, \citenamefont {Buonanno},
  \citenamefont {Taracchini}, \citenamefont {Pfeiffer}, \citenamefont {Scheel},
  \citenamefont {Chu}, \citenamefont {Kidder},\ and\ \citenamefont
  {Pan}}]{Szilagyi2015}%
  \BibitemOpen
  \bibfield  {author} {\bibinfo {author} {\bibfnamefont {B.}~\bibnamefont
  {Szil{\'{a}}gyi}}, \bibinfo {author} {\bibfnamefont {J.}~\bibnamefont
  {Blackman}}, \bibinfo {author} {\bibfnamefont {A.}~\bibnamefont {Buonanno}},
  \bibinfo {author} {\bibfnamefont {A.}~\bibnamefont {Taracchini}}, \bibinfo
  {author} {\bibfnamefont {H.~P.}\ \bibnamefont {Pfeiffer}}, \bibinfo {author}
  {\bibfnamefont {M.~A.}\ \bibnamefont {Scheel}}, \bibinfo {author}
  {\bibfnamefont {T.}~\bibnamefont {Chu}}, \bibinfo {author} {\bibfnamefont
  {L.~E.}\ \bibnamefont {Kidder}}, \ and\ \bibinfo {author} {\bibfnamefont
  {Y.}~\bibnamefont {Pan}},\ }\href {\doibase 10.1103/PhysRevLett.115.031102}
  {\bibfield  {journal} {\bibinfo  {journal} {Phys.\ Rev.\ Lett.}\ }\textbf
  {\bibinfo {volume} {115}},\ \bibinfo {pages} {031102} (\bibinfo {year}
  {2015})},\ \Eprint {http://arxiv.org/abs/1502.04953} {arXiv:1502.04953}
  \BibitemShut {NoStop}%
\bibitem [{\citenamefont {Blanchet}(2014)}]{Blanchet2014}%
  \BibitemOpen
  \bibfield  {author} {\bibinfo {author} {\bibfnamefont {L.}~\bibnamefont
  {Blanchet}},\ }\href {\doibase 10.12942/lrr-2014-2} {\bibfield  {journal}
  {\bibinfo  {journal} {Living Rev.\ Relat.}\ }\textbf {\bibinfo {volume}
  {17}},\ \bibinfo {pages} {2} (\bibinfo {year} {2014})},\ \Eprint
  {http://arxiv.org/abs/1310.1528} {arXiv:1310.1528} \BibitemShut {NoStop}%
\bibitem [{\citenamefont {Buonanno}\ \emph {et~al.}(2009)\citenamefont
  {Buonanno}, \citenamefont {Iyer}, \citenamefont {Ochsner}, \citenamefont
  {Pan},\ and\ \citenamefont {Sathyaprakash}}]{Buonanno2009}%
  \BibitemOpen
  \bibfield  {author} {\bibinfo {author} {\bibfnamefont {A.}~\bibnamefont
  {Buonanno}}, \bibinfo {author} {\bibfnamefont {B.~R.}\ \bibnamefont {Iyer}},
  \bibinfo {author} {\bibfnamefont {E.}~\bibnamefont {Ochsner}}, \bibinfo
  {author} {\bibfnamefont {Y.}~\bibnamefont {Pan}}, \ and\ \bibinfo {author}
  {\bibfnamefont {B.~S.}\ \bibnamefont {Sathyaprakash}},\ }\href {\doibase
  10.1103/PhysRevD.80.084043} {\bibfield  {journal} {\bibinfo  {journal}
  {Phys.\ Rev.\ D}\ }\textbf {\bibinfo {volume} {80}},\ \bibinfo {pages}
  {084043} (\bibinfo {year} {2009})},\ \Eprint {http://arxiv.org/abs/0907.0700}
  {arXiv:0907.0700} \BibitemShut {NoStop}%
\bibitem [{\citenamefont {Buonanno}\ and\ \citenamefont
  {Damour}(1999)}]{Buonanno1999}%
  \BibitemOpen
  \bibfield  {author} {\bibinfo {author} {\bibfnamefont {A.}~\bibnamefont
  {Buonanno}}\ and\ \bibinfo {author} {\bibfnamefont {T.}~\bibnamefont
  {Damour}},\ }\href {\doibase 10.1103/PhysRevD.59.084006} {\bibfield
  {journal} {\bibinfo  {journal} {Phys.\ Rev.\ D}\ }\textbf {\bibinfo {volume}
  {59}},\ \bibinfo {pages} {084006} (\bibinfo {year} {1999})},\ \Eprint
  {http://arxiv.org/abs/9811091} {arXiv:9811091 [gr-qc]} \BibitemShut {NoStop}%
\bibitem [{\citenamefont {Buonanno}\ and\ \citenamefont
  {Damour}(2000)}]{Buonanno2000}%
  \BibitemOpen
  \bibfield  {author} {\bibinfo {author} {\bibfnamefont {A.}~\bibnamefont
  {Buonanno}}\ and\ \bibinfo {author} {\bibfnamefont {T.}~\bibnamefont
  {Damour}},\ }\href {\doibase 10.1103/PhysRevD.62.064015} {\bibfield
  {journal} {\bibinfo  {journal} {Phys.\ Rev.\ D}\ }\textbf {\bibinfo {volume}
  {62}},\ \bibinfo {pages} {064015} (\bibinfo {year} {2000})},\ \Eprint
  {http://arxiv.org/abs/0001013} {arXiv:0001013 [gr-qc]} \BibitemShut {NoStop}%
\bibitem [{\citenamefont {Damour}\ and\ \citenamefont
  {Nagar}(2009)}]{Damour2009}%
  \BibitemOpen
  \bibfield  {author} {\bibinfo {author} {\bibfnamefont {T.}~\bibnamefont
  {Damour}}\ and\ \bibinfo {author} {\bibfnamefont {A.}~\bibnamefont {Nagar}},\
  }\href {\doibase 10.1103/PhysRevD.79.081503} {\bibfield  {journal} {\bibinfo
  {journal} {Phys.\ Rev.\ D}\ }\textbf {\bibinfo {volume} {79}},\ \bibinfo
  {pages} {081503} (\bibinfo {year} {2009})},\ \Eprint
  {http://arxiv.org/abs/0902.0136} {arXiv:0902.0136} \BibitemShut {NoStop}%
\bibitem [{\citenamefont {Barausse}\ and\ \citenamefont
  {Buonanno}(2010)}]{Barausse2010}%
  \BibitemOpen
  \bibfield  {author} {\bibinfo {author} {\bibfnamefont {E.}~\bibnamefont
  {Barausse}}\ and\ \bibinfo {author} {\bibfnamefont {A.}~\bibnamefont
  {Buonanno}},\ }\href {\doibase 10.1103/PhysRevD.81.084024} {\bibfield
  {journal} {\bibinfo  {journal} {Phys.\ Rev.\ D}\ }\textbf {\bibinfo {volume}
  {81}},\ \bibinfo {pages} {084024} (\bibinfo {year} {2010})},\ \Eprint
  {http://arxiv.org/abs/0912.3517} {arXiv:0912.3517} \BibitemShut {NoStop}%
\bibitem [{\citenamefont {Taracchini}\ \emph {et~al.}(2014)\citenamefont
  {Taracchini}, \citenamefont {Buonanno}, \citenamefont {Pan}, \citenamefont
  {Hinderer}, \citenamefont {Boyle}, \citenamefont {Hemberger}, \citenamefont
  {Kidder}, \citenamefont {Lovelace}, \citenamefont {Mrou{\'{e}}},
  \citenamefont {Pfeiffer}, \citenamefont {Scheel}, \citenamefont
  {Szil{\'{a}}gyi}, \citenamefont {Taylor},\ and\ \citenamefont
  {Zenginoglu}}]{Taracchini2014}%
  \BibitemOpen
  \bibfield  {author} {\bibinfo {author} {\bibfnamefont {A.}~\bibnamefont
  {Taracchini}}, \bibinfo {author} {\bibfnamefont {A.}~\bibnamefont
  {Buonanno}}, \bibinfo {author} {\bibfnamefont {Y.}~\bibnamefont {Pan}},
  \bibinfo {author} {\bibfnamefont {T.}~\bibnamefont {Hinderer}}, \bibinfo
  {author} {\bibfnamefont {M.}~\bibnamefont {Boyle}}, \bibinfo {author}
  {\bibfnamefont {D.~A.}\ \bibnamefont {Hemberger}}, \bibinfo {author}
  {\bibfnamefont {L.~E.}\ \bibnamefont {Kidder}}, \bibinfo {author}
  {\bibfnamefont {G.}~\bibnamefont {Lovelace}}, \bibinfo {author}
  {\bibfnamefont {A.~H.}\ \bibnamefont {Mrou{\'{e}}}}, \bibinfo {author}
  {\bibfnamefont {H.~P.}\ \bibnamefont {Pfeiffer}}, \bibinfo {author}
  {\bibfnamefont {M.~A.}\ \bibnamefont {Scheel}}, \bibinfo {author}
  {\bibfnamefont {B.}~\bibnamefont {Szil{\'{a}}gyi}}, \bibinfo {author}
  {\bibfnamefont {N.~W.}\ \bibnamefont {Taylor}}, \ and\ \bibinfo {author}
  {\bibfnamefont {A.}~\bibnamefont {Zenginoglu}},\ }\href {\doibase
  10.1103/PhysRevD.89.061502} {\bibfield  {journal} {\bibinfo  {journal}
  {Phys.\ Rev.\ D}\ }\textbf {\bibinfo {volume} {89}},\ \bibinfo {pages}
  {061502} (\bibinfo {year} {2014})},\ \Eprint {http://arxiv.org/abs/1311.2544}
  {arXiv:1311.2544} \BibitemShut {NoStop}%
\bibitem [{\citenamefont {Pan}\ \emph {et~al.}(2014)\citenamefont {Pan},
  \citenamefont {Buonanno}, \citenamefont {Taracchini}, \citenamefont {Kidder},
  \citenamefont {Mrou{\'{e}}}, \citenamefont {Pfeiffer}, \citenamefont
  {Scheel},\ and\ \citenamefont {Szil{\'{a}}gyi}}]{Pan2014}%
  \BibitemOpen
  \bibfield  {author} {\bibinfo {author} {\bibfnamefont {Y.}~\bibnamefont
  {Pan}}, \bibinfo {author} {\bibfnamefont {A.}~\bibnamefont {Buonanno}},
  \bibinfo {author} {\bibfnamefont {A.}~\bibnamefont {Taracchini}}, \bibinfo
  {author} {\bibfnamefont {L.~E.}\ \bibnamefont {Kidder}}, \bibinfo {author}
  {\bibfnamefont {A.~H.}\ \bibnamefont {Mrou{\'{e}}}}, \bibinfo {author}
  {\bibfnamefont {H.~P.}\ \bibnamefont {Pfeiffer}}, \bibinfo {author}
  {\bibfnamefont {M.~A.}\ \bibnamefont {Scheel}}, \ and\ \bibinfo {author}
  {\bibfnamefont {B.}~\bibnamefont {Szil{\'{a}}gyi}},\ }\href {\doibase
  10.1103/PhysRevD.89.084006} {\bibfield  {journal} {\bibinfo  {journal}
  {Phys.\ Rev.\ D}\ }\textbf {\bibinfo {volume} {89}},\ \bibinfo {pages}
  {084006} (\bibinfo {year} {2014})},\ \Eprint {http://arxiv.org/abs/1307.6232}
  {arXiv:1307.6232} \BibitemShut {NoStop}%
\bibitem [{\citenamefont {Husa}\ \emph {et~al.}(2016)\citenamefont {Husa},
  \citenamefont {Khan}, \citenamefont {Hannam}, \citenamefont {P{\"{u}}rrer},
  \citenamefont {Ohme}, \citenamefont {Forteza},\ and\ \citenamefont
  {Boh{\'{e}}}}]{Husa2015}%
  \BibitemOpen
  \bibfield  {author} {\bibinfo {author} {\bibfnamefont {S.}~\bibnamefont
  {Husa}}, \bibinfo {author} {\bibfnamefont {S.}~\bibnamefont {Khan}}, \bibinfo
  {author} {\bibfnamefont {M.}~\bibnamefont {Hannam}}, \bibinfo {author}
  {\bibfnamefont {M.}~\bibnamefont {P{\"{u}}rrer}}, \bibinfo {author}
  {\bibfnamefont {F.}~\bibnamefont {Ohme}}, \bibinfo {author} {\bibfnamefont
  {X.~J.}\ \bibnamefont {Forteza}}, \ and\ \bibinfo {author} {\bibfnamefont
  {A.}~\bibnamefont {Boh{\'{e}}}},\ }\href {\doibase
  10.1103/PhysRevD.93.044006} {\bibfield  {journal} {\bibinfo  {journal}
  {Phys.\ Rev.\ D}\ }\textbf {\bibinfo {volume} {93}},\ \bibinfo {pages}
  {044006} (\bibinfo {year} {2016})},\ \Eprint
  {http://arxiv.org/abs/1508.07250} {arXiv:1508.07250} \BibitemShut {NoStop}%
\bibitem [{\citenamefont {Khan}\ \emph {et~al.}(2016)\citenamefont {Khan},
  \citenamefont {Husa}, \citenamefont {Hannam}, \citenamefont {Ohme},
  \citenamefont {P{\"{u}}rrer}, \citenamefont {Forteza},\ and\ \citenamefont
  {Boh{\'{e}}}}]{Khan2015}%
  \BibitemOpen
  \bibfield  {author} {\bibinfo {author} {\bibfnamefont {S.}~\bibnamefont
  {Khan}}, \bibinfo {author} {\bibfnamefont {S.}~\bibnamefont {Husa}}, \bibinfo
  {author} {\bibfnamefont {M.}~\bibnamefont {Hannam}}, \bibinfo {author}
  {\bibfnamefont {F.}~\bibnamefont {Ohme}}, \bibinfo {author} {\bibfnamefont
  {M.}~\bibnamefont {P{\"{u}}rrer}}, \bibinfo {author} {\bibfnamefont {X.~J.}\
  \bibnamefont {Forteza}}, \ and\ \bibinfo {author} {\bibfnamefont
  {A.}~\bibnamefont {Boh{\'{e}}}},\ }\href {\doibase
  10.1103/PhysRevD.93.044007} {\bibfield  {journal} {\bibinfo  {journal}
  {Phys.\ Rev.\ D}\ }\textbf {\bibinfo {volume} {93}},\ \bibinfo {pages}
  {044007} (\bibinfo {year} {2016})},\ \Eprint
  {http://arxiv.org/abs/1508.07253} {arXiv:1508.07253} \BibitemShut {NoStop}%
\bibitem [{\citenamefont {Schmidt}\ \emph {et~al.}(2015)\citenamefont
  {Schmidt}, \citenamefont {Ohme},\ and\ \citenamefont {Hannam}}]{Schmidt2015}%
  \BibitemOpen
  \bibfield  {author} {\bibinfo {author} {\bibfnamefont {P.}~\bibnamefont
  {Schmidt}}, \bibinfo {author} {\bibfnamefont {F.}~\bibnamefont {Ohme}}, \
  and\ \bibinfo {author} {\bibfnamefont {M.}~\bibnamefont {Hannam}},\ }\href
  {\doibase 10.1103/PhysRevD.91.024043} {\bibfield  {journal} {\bibinfo
  {journal} {Phys.\ Rev.\ D}\ }\textbf {\bibinfo {volume} {91}},\ \bibinfo
  {pages} {024043} (\bibinfo {year} {2015})},\ \Eprint
  {http://arxiv.org/abs/1408.1810} {arXiv:1408.1810} \BibitemShut {NoStop}%
\bibitem [{\citenamefont {Aasi}\ \emph {et~al.}(2015)\citenamefont {Aasi} \emph
  {et~al.}}]{Aasi2015}%
  \BibitemOpen
  \bibfield  {author} {\bibinfo {author} {\bibfnamefont {J.}~\bibnamefont
  {Aasi}} \emph {et~al.},\ }\href {\doibase 10.1088/0264-9381/32/7/074001}
  {\bibfield  {journal} {\bibinfo  {journal} {Class.\ Quantum Grav.}\ }\textbf
  {\bibinfo {volume} {32}},\ \bibinfo {pages} {074001} (\bibinfo {year}
  {2015})},\ \Eprint {http://arxiv.org/abs/1411.4547} {arXiv:1411.4547}
  \BibitemShut {NoStop}%
\bibitem [{\citenamefont {Acernese}\ \emph {et~al.}(2015)\citenamefont
  {Acernese} \emph {et~al.}}]{Acernese2015}%
  \BibitemOpen
  \bibfield  {author} {\bibinfo {author} {\bibfnamefont {F.}~\bibnamefont
  {Acernese}} \emph {et~al.},\ }\href {\doibase 10.1088/0264-9381/32/2/024001}
  {\bibfield  {journal} {\bibinfo  {journal} {Class.\ Quantum Grav.}\ }\textbf
  {\bibinfo {volume} {32}},\ \bibinfo {pages} {024001} (\bibinfo {year}
  {2015})},\ \Eprint {http://arxiv.org/abs/1408.3978} {arXiv:1408.3978}
  \BibitemShut {NoStop}%
\bibitem [{\citenamefont {Aso}\ \emph {et~al.}(2013)\citenamefont {Aso},
  \citenamefont {Michimura}, \citenamefont {Somiya}, \citenamefont {Ando},
  \citenamefont {Miyakawa}, \citenamefont {Sekiguchi}, \citenamefont
  {Tatsumi},\ and\ \citenamefont {Yamamoto}}]{Aso2013}%
  \BibitemOpen
  \bibfield  {author} {\bibinfo {author} {\bibfnamefont {Y.}~\bibnamefont
  {Aso}}, \bibinfo {author} {\bibfnamefont {Y.}~\bibnamefont {Michimura}},
  \bibinfo {author} {\bibfnamefont {K.}~\bibnamefont {Somiya}}, \bibinfo
  {author} {\bibfnamefont {M.}~\bibnamefont {Ando}}, \bibinfo {author}
  {\bibfnamefont {O.}~\bibnamefont {Miyakawa}}, \bibinfo {author}
  {\bibfnamefont {T.}~\bibnamefont {Sekiguchi}}, \bibinfo {author}
  {\bibfnamefont {D.}~\bibnamefont {Tatsumi}}, \ and\ \bibinfo {author}
  {\bibfnamefont {H.}~\bibnamefont {Yamamoto}},\ }\href {\doibase
  10.1103/PhysRevD.88.043007} {\bibfield  {journal} {\bibinfo  {journal}
  {Phys.\ Rev.\ D}\ }\textbf {\bibinfo {volume} {88}},\ \bibinfo {pages}
  {043007} (\bibinfo {year} {2013})},\ \Eprint {http://arxiv.org/abs/1306.6747}
  {arXiv:1306.6747} \BibitemShut {NoStop}%
\bibitem [{\citenamefont {Punturo}\ \emph {et~al.}(2010)\citenamefont {Punturo}
  \emph {et~al.}}]{Punturo2010a}%
  \BibitemOpen
  \bibfield  {author} {\bibinfo {author} {\bibfnamefont {M.}~\bibnamefont
  {Punturo}} \emph {et~al.},\ }\href {\doibase 10.1088/0264-9381/27/19/194002}
  {\bibfield  {journal} {\bibinfo  {journal} {Class.\ Quantum Grav.}\ }\textbf
  {\bibinfo {volume} {27}},\ \bibinfo {pages} {194002} (\bibinfo {year}
  {2010})}\BibitemShut {NoStop}%
\bibitem [{\citenamefont {Abbott}\ \emph
  {et~al.}(2016{\natexlab{a}})\citenamefont {Abbott} \emph
  {et~al.}}]{Abbott2016}%
  \BibitemOpen
  \bibfield  {author} {\bibinfo {author} {\bibfnamefont {B.}~\bibnamefont
  {Abbott}} \emph {et~al.},\ }\href {\doibase 10.1103/PhysRevLett.116.061102}
  {\bibfield  {journal} {\bibinfo  {journal} {Phys.\ Rev.\ Lett.}\ }\textbf
  {\bibinfo {volume} {116}},\ \bibinfo {pages} {061102} (\bibinfo {year}
  {2016}{\natexlab{a}})},\ \Eprint {http://arxiv.org/abs/1602.03837}
  {arXiv:1602.03837} \BibitemShut {NoStop}%
\bibitem [{\citenamefont {Abbott}\ \emph
  {et~al.}(2016{\natexlab{b}})\citenamefont {Abbott} \emph
  {et~al.}}]{Abbott2016e}%
  \BibitemOpen
  \bibfield  {author} {\bibinfo {author} {\bibfnamefont {B.~P.}\ \bibnamefont
  {Abbott}} \emph {et~al.},\ }\href {\doibase 10.1103/PhysRevLett.116.241103}
  {\bibfield  {journal} {\bibinfo  {journal} {Phys.\ Rev.\ Lett.}\ }\textbf
  {\bibinfo {volume} {116}},\ \bibinfo {pages} {241103} (\bibinfo {year}
  {2016}{\natexlab{b}})},\ \Eprint {http://arxiv.org/abs/1606.04855}
  {arXiv:1606.04855} \BibitemShut {NoStop}%
\bibitem [{\citenamefont {Abbott}\ \emph
  {et~al.}(2016{\natexlab{c}})\citenamefont {Abbott} \emph
  {et~al.}}]{Abbott2016d}%
  \BibitemOpen
  \bibfield  {author} {\bibinfo {author} {\bibfnamefont {B.~P.}\ \bibnamefont
  {Abbott}} \emph {et~al.},\ }\href {\doibase 10.1103/PhysRevX.6.041015}
  {\bibfield  {journal} {\bibinfo  {journal} {Phys.\ Rev.\ X}\ }\textbf
  {\bibinfo {volume} {6}},\ \bibinfo {pages} {041015} (\bibinfo {year}
  {2016}{\natexlab{c}})},\ \Eprint {http://arxiv.org/abs/1606.04856}
  {arXiv:1606.04856} \BibitemShut {NoStop}%
\bibitem [{\citenamefont {Abbott}\ \emph
  {et~al.}(2016{\natexlab{d}})\citenamefont {Abbott} \emph
  {et~al.}}]{Abbott2016f}%
  \BibitemOpen
  \bibfield  {author} {\bibinfo {author} {\bibfnamefont {B.~P.}\ \bibnamefont
  {Abbott}} \emph {et~al.},\ }\href {\doibase 10.1103/PhysRevLett.116.241102}
  {\bibfield  {journal} {\bibinfo  {journal} {Phys.\ Rev.\ Lett.}\ }\textbf
  {\bibinfo {volume} {116}},\ \bibinfo {pages} {241102} (\bibinfo {year}
  {2016}{\natexlab{d}})},\ \Eprint {http://arxiv.org/abs/1602.03840}
  {arXiv:1602.03840} \BibitemShut {NoStop}%
\bibitem [{\citenamefont {Abbott}\ \emph
  {et~al.}(2016{\natexlab{e}})\citenamefont {Abbott} \emph
  {et~al.}}]{Abbott2016h}%
  \BibitemOpen
  \bibfield  {author} {\bibinfo {author} {\bibfnamefont {B.~P.}\ \bibnamefont
  {Abbott}} \emph {et~al.},\ }\href {\doibase 10.1103/PhysRevX.6.041014}
  {\bibfield  {journal} {\bibinfo  {journal} {Phys.\ Rev.\ X}\ }\textbf
  {\bibinfo {volume} {6}},\ \bibinfo {pages} {041014} (\bibinfo {year}
  {2016}{\natexlab{e}})},\ \Eprint {http://arxiv.org/abs/1606.01210}
  {arXiv:1606.01210} \BibitemShut {NoStop}%
\bibitem [{\citenamefont {Abbott}\ \emph
  {et~al.}(2016{\natexlab{f}})\citenamefont {Abbott} \emph
  {et~al.}}]{Abbott2016i}%
  \BibitemOpen
  \bibfield  {author} {\bibinfo {author} {\bibfnamefont {B.~P.}\ \bibnamefont
  {Abbott}} \emph {et~al.},\ }\href {\doibase 10.3847/2041-8205/818/2/L22}
  {\bibfield  {journal} {\bibinfo  {journal} {Astrophys.\ J.}\ }\textbf
  {\bibinfo {volume} {818}},\ \bibinfo {pages} {L22} (\bibinfo {year}
  {2016}{\natexlab{f}})},\ \Eprint {http://arxiv.org/abs/1602.03846}
  {arXiv:1602.03846} \BibitemShut {NoStop}%
\bibitem [{\citenamefont {Abbott}\ \emph
  {et~al.}(2016{\natexlab{g}})\citenamefont {Abbott} \emph
  {et~al.}}]{Abbott2016j}%
  \BibitemOpen
  \bibfield  {author} {\bibinfo {author} {\bibfnamefont {B.~P.}\ \bibnamefont
  {Abbott}} \emph {et~al.},\ }\href {\doibase 10.1103/PhysRevLett.116.221101}
  {\bibfield  {journal} {\bibinfo  {journal} {Phys.\ Rev.\ Lett.}\ }\textbf
  {\bibinfo {volume} {116}},\ \bibinfo {pages} {221101} (\bibinfo {year}
  {2016}{\natexlab{g}})},\ \Eprint {http://arxiv.org/abs/1602.03841}
  {arXiv:1602.03841} \BibitemShut {NoStop}%
\bibitem [{\citenamefont {Yunes}\ \emph {et~al.}(2016)\citenamefont {Yunes},
  \citenamefont {Yagi},\ and\ \citenamefont {Pretorius}}]{Yunes2016}%
  \BibitemOpen
  \bibfield  {author} {\bibinfo {author} {\bibfnamefont {N.}~\bibnamefont
  {Yunes}}, \bibinfo {author} {\bibfnamefont {K.}~\bibnamefont {Yagi}}, \ and\
  \bibinfo {author} {\bibfnamefont {F.}~\bibnamefont {Pretorius}},\ }\href
  {\doibase 10.1103/PhysRevD.94.084002} {\bibfield  {journal} {\bibinfo
  {journal} {Phys.\ Rev.\ D}\ }\textbf {\bibinfo {volume} {94}},\ \bibinfo
  {pages} {084002} (\bibinfo {year} {2016})},\ \Eprint
  {http://arxiv.org/abs/1603.08955} {arXiv:1603.08955} \BibitemShut {NoStop}%
\bibitem [{\citenamefont {Kormendy}\ and\ \citenamefont
  {Richstone}(1995)}]{Kormendy1995}%
  \BibitemOpen
  \bibfield  {author} {\bibinfo {author} {\bibfnamefont {J.}~\bibnamefont
  {Kormendy}}\ and\ \bibinfo {author} {\bibfnamefont {D.}~\bibnamefont
  {Richstone}},\ }\href {\doibase 10.1146/annurev.aa.33.090195.003053}
  {\bibfield  {journal} {\bibinfo  {journal} {Annu.\ Rev.\ Astron.\
  Astrophys.}\ }\textbf {\bibinfo {volume} {33}},\ \bibinfo {pages} {581}
  (\bibinfo {year} {1995})}\BibitemShut {NoStop}%
\bibitem [{\citenamefont {Ferrarese}\ and\ \citenamefont
  {Ford}(2005)}]{Ferrarese2005}%
  \BibitemOpen
  \bibfield  {author} {\bibinfo {author} {\bibfnamefont {L.}~\bibnamefont
  {Ferrarese}}\ and\ \bibinfo {author} {\bibfnamefont {H.}~\bibnamefont
  {Ford}},\ }\href {\doibase 10.1007/s11214-005-3947-6} {\bibfield  {journal}
  {\bibinfo  {journal} {Space Sci.\ Rev.}\ }\textbf {\bibinfo {volume} {116}},\
  \bibinfo {pages} {523} (\bibinfo {year} {2005})},\ \Eprint
  {http://arxiv.org/abs/0411247} {arXiv:0411247 [astro-ph]} \BibitemShut
  {NoStop}%
\bibitem [{\citenamefont {Boehle}\ \emph {et~al.}(2016)\citenamefont {Boehle},
  \citenamefont {Ghez}, \citenamefont {Sch{\"{o}}del}, \citenamefont {Meyer},
  \citenamefont {Yelda}, \citenamefont {Albers}, \citenamefont {Martinez},
  \citenamefont {Becklin}, \citenamefont {Do}, \citenamefont {Lu},
  \citenamefont {Matthews}, \citenamefont {Morris}, \citenamefont {Sitarski},\
  and\ \citenamefont {Witzel}}]{Boehle2016}%
  \BibitemOpen
  \bibfield  {author} {\bibinfo {author} {\bibfnamefont {A.}~\bibnamefont
  {Boehle}}, \bibinfo {author} {\bibfnamefont {A.~M.}\ \bibnamefont {Ghez}},
  \bibinfo {author} {\bibfnamefont {R.}~\bibnamefont {Sch{\"{o}}del}}, \bibinfo
  {author} {\bibfnamefont {L.}~\bibnamefont {Meyer}}, \bibinfo {author}
  {\bibfnamefont {S.}~\bibnamefont {Yelda}}, \bibinfo {author} {\bibfnamefont
  {S.}~\bibnamefont {Albers}}, \bibinfo {author} {\bibfnamefont {G.~D.}\
  \bibnamefont {Martinez}}, \bibinfo {author} {\bibfnamefont {E.~E.}\
  \bibnamefont {Becklin}}, \bibinfo {author} {\bibfnamefont {T.}~\bibnamefont
  {Do}}, \bibinfo {author} {\bibfnamefont {J.~R.}\ \bibnamefont {Lu}}, \bibinfo
  {author} {\bibfnamefont {K.}~\bibnamefont {Matthews}}, \bibinfo {author}
  {\bibfnamefont {M.~R.}\ \bibnamefont {Morris}}, \bibinfo {author}
  {\bibfnamefont {B.}~\bibnamefont {Sitarski}}, \ and\ \bibinfo {author}
  {\bibfnamefont {G.}~\bibnamefont {Witzel}},\ }\href {\doibase
  10.3847/0004-637X/830/1/17} {\bibfield  {journal} {\bibinfo  {journal}
  {Astrophys.\ J.}\ }\textbf {\bibinfo {volume} {830}},\ \bibinfo {pages} {17}
  (\bibinfo {year} {2016})},\ \Eprint {http://arxiv.org/abs/1607.05726}
  {arXiv:1607.05726} \BibitemShut {NoStop}%
\bibitem [{\citenamefont {Amaro-Seoane}\ \emph {et~al.}(2007)\citenamefont
  {Amaro-Seoane}, \citenamefont {Gair}, \citenamefont {Freitag}, \citenamefont
  {Miller}, \citenamefont {Mandel}, \citenamefont {Cutler},\ and\ \citenamefont
  {Babak}}]{Amaro-Seoane2007}%
  \BibitemOpen
  \bibfield  {author} {\bibinfo {author} {\bibfnamefont {P.}~\bibnamefont
  {Amaro-Seoane}}, \bibinfo {author} {\bibfnamefont {J.~R.}\ \bibnamefont
  {Gair}}, \bibinfo {author} {\bibfnamefont {M.}~\bibnamefont {Freitag}},
  \bibinfo {author} {\bibfnamefont {M.~C.}\ \bibnamefont {Miller}}, \bibinfo
  {author} {\bibfnamefont {I.}~\bibnamefont {Mandel}}, \bibinfo {author}
  {\bibfnamefont {C.~J.}\ \bibnamefont {Cutler}}, \ and\ \bibinfo {author}
  {\bibfnamefont {S.}~\bibnamefont {Babak}},\ }\href {\doibase
  10.1088/0264-9381/24/17/R01} {\bibfield  {journal} {\bibinfo  {journal}
  {Class.\ Quantum Grav.}\ }\textbf {\bibinfo {volume} {24}},\ \bibinfo {pages}
  {R113} (\bibinfo {year} {2007})},\ \Eprint {http://arxiv.org/abs/0703495}
  {arXiv:0703495 [astro-ph]} \BibitemShut {NoStop}%
\bibitem [{\citenamefont {Amaro-Seoane}\ \emph {et~al.}(2012)\citenamefont
  {Amaro-Seoane} \emph {et~al.}}]{Amaro-Seoane2012a}%
  \BibitemOpen
  \bibfield  {author} {\bibinfo {author} {\bibfnamefont {P.}~\bibnamefont
  {Amaro-Seoane}} \emph {et~al.},\ }\href {\doibase
  10.1088/0264-9381/29/12/124016} {\bibfield  {journal} {\bibinfo  {journal}
  {Class.\ Quantum Grav.}\ }\textbf {\bibinfo {volume} {29}},\ \bibinfo {pages}
  {124016} (\bibinfo {year} {2012})},\ \Eprint {http://arxiv.org/abs/1202.0839}
  {arXiv:1202.0839} \BibitemShut {NoStop}%
\bibitem [{\citenamefont {Barack}\ and\ \citenamefont
  {Cutler}(2004)}]{Barack2004}%
  \BibitemOpen
  \bibfield  {author} {\bibinfo {author} {\bibfnamefont {L.}~\bibnamefont
  {Barack}}\ and\ \bibinfo {author} {\bibfnamefont {C.}~\bibnamefont
  {Cutler}},\ }\href {\doibase 10.1103/PhysRevD.69.082005} {\bibfield
  {journal} {\bibinfo  {journal} {Phys.\ Rev.\ D}\ }\textbf {\bibinfo {volume}
  {69}},\ \bibinfo {pages} {082005} (\bibinfo {year} {2004})},\ \Eprint
  {http://arxiv.org/abs/0310125} {arXiv:0310125 [gr-qc]} \BibitemShut {NoStop}%
\bibitem [{\citenamefont {Arun}\ \emph {et~al.}(2009)\citenamefont {Arun},
  \citenamefont {Babak}, \citenamefont {Berti}, \citenamefont {Cornish},
  \citenamefont {Cutler}, \citenamefont {Gair}, \citenamefont {Hughes},
  \citenamefont {Iyer}, \citenamefont {Lang}, \citenamefont {Mandel},
  \citenamefont {Porter}, \citenamefont {Sathyaprakash}, \citenamefont {Sinha},
  \citenamefont {Sintes}, \citenamefont {Trias}, \citenamefont {{Van Den
  Broeck}},\ and\ \citenamefont {Volonteri}}]{Arun2009}%
  \BibitemOpen
  \bibfield  {author} {\bibinfo {author} {\bibfnamefont {K.~G.}\ \bibnamefont
  {Arun}}, \bibinfo {author} {\bibfnamefont {S.}~\bibnamefont {Babak}},
  \bibinfo {author} {\bibfnamefont {E.}~\bibnamefont {Berti}}, \bibinfo
  {author} {\bibfnamefont {N.}~\bibnamefont {Cornish}}, \bibinfo {author}
  {\bibfnamefont {C.}~\bibnamefont {Cutler}}, \bibinfo {author} {\bibfnamefont
  {J.}~\bibnamefont {Gair}}, \bibinfo {author} {\bibfnamefont {S.~A.}\
  \bibnamefont {Hughes}}, \bibinfo {author} {\bibfnamefont {B.~R.}\
  \bibnamefont {Iyer}}, \bibinfo {author} {\bibfnamefont {R.~N.}\ \bibnamefont
  {Lang}}, \bibinfo {author} {\bibfnamefont {I.}~\bibnamefont {Mandel}},
  \bibinfo {author} {\bibfnamefont {E.~K.}\ \bibnamefont {Porter}}, \bibinfo
  {author} {\bibfnamefont {B.~S.}\ \bibnamefont {Sathyaprakash}}, \bibinfo
  {author} {\bibfnamefont {S.}~\bibnamefont {Sinha}}, \bibinfo {author}
  {\bibfnamefont {A.~M.}\ \bibnamefont {Sintes}}, \bibinfo {author}
  {\bibfnamefont {M.}~\bibnamefont {Trias}}, \bibinfo {author} {\bibfnamefont
  {C.}~\bibnamefont {{Van Den Broeck}}}, \ and\ \bibinfo {author}
  {\bibfnamefont {M.}~\bibnamefont {Volonteri}},\ }\href {\doibase
  10.1088/0264-9381/26/9/094027} {\bibfield  {journal} {\bibinfo  {journal}
  {Class.\ Quantum Grav.}\ }\textbf {\bibinfo {volume} {26}},\ \bibinfo {pages}
  {094027} (\bibinfo {year} {2009})},\ \Eprint {http://arxiv.org/abs/0811.1011}
  {arXiv:0811.1011} \BibitemShut {NoStop}%
\bibitem [{\citenamefont {Gair}\ \emph {et~al.}(2010)\citenamefont {Gair},
  \citenamefont {Tang},\ and\ \citenamefont {Volonteri}}]{Gair2010b}%
  \BibitemOpen
  \bibfield  {author} {\bibinfo {author} {\bibfnamefont {J.~R.}\ \bibnamefont
  {Gair}}, \bibinfo {author} {\bibfnamefont {C.}~\bibnamefont {Tang}}, \ and\
  \bibinfo {author} {\bibfnamefont {M.}~\bibnamefont {Volonteri}},\ }\href
  {\doibase 10.1103/PhysRevD.81.104014} {\bibfield  {journal} {\bibinfo
  {journal} {Phys.\ Rev.\ D}\ }\textbf {\bibinfo {volume} {81}},\ \bibinfo
  {pages} {104014} (\bibinfo {year} {2010})},\ \Eprint
  {http://arxiv.org/abs/1004.1921} {arXiv:1004.1921} \BibitemShut {NoStop}%
\bibitem [{\citenamefont {Gair}\ \emph
  {et~al.}(2011{\natexlab{a}})\citenamefont {Gair}, \citenamefont {Sesana},
  \citenamefont {Berti},\ and\ \citenamefont {Volonteri}}]{Gair2010a}%
  \BibitemOpen
  \bibfield  {author} {\bibinfo {author} {\bibfnamefont {J.~R.}\ \bibnamefont
  {Gair}}, \bibinfo {author} {\bibfnamefont {A.}~\bibnamefont {Sesana}},
  \bibinfo {author} {\bibfnamefont {E.}~\bibnamefont {Berti}}, \ and\ \bibinfo
  {author} {\bibfnamefont {M.}~\bibnamefont {Volonteri}},\ }\href {\doibase
  10.1088/0264-9381/28/9/094018} {\bibfield  {journal} {\bibinfo  {journal}
  {Class.\ Quantum Grav.}\ }\textbf {\bibinfo {volume} {28}},\ \bibinfo {pages}
  {094018} (\bibinfo {year} {2011}{\natexlab{a}})},\ \Eprint
  {http://arxiv.org/abs/1009.6172} {arXiv:1009.6172} \BibitemShut {NoStop}%
\bibitem [{\citenamefont {Yunes}\ \emph {et~al.}(2011)\citenamefont {Yunes},
  \citenamefont {Kocsis}, \citenamefont {Loeb},\ and\ \citenamefont
  {Haiman}}]{Yunes2011a}%
  \BibitemOpen
  \bibfield  {author} {\bibinfo {author} {\bibfnamefont {N.}~\bibnamefont
  {Yunes}}, \bibinfo {author} {\bibfnamefont {B.}~\bibnamefont {Kocsis}},
  \bibinfo {author} {\bibfnamefont {A.}~\bibnamefont {Loeb}}, \ and\ \bibinfo
  {author} {\bibfnamefont {Z.}~\bibnamefont {Haiman}},\ }\href {\doibase
  10.1103/PhysRevLett.107.171103} {\bibfield  {journal} {\bibinfo  {journal}
  {Phys.\ Rev.\ Lett.}\ }\textbf {\bibinfo {volume} {107}},\ \bibinfo {pages}
  {171103} (\bibinfo {year} {2011})},\ \Eprint {http://arxiv.org/abs/1103.4609}
  {arXiv:1103.4609} \BibitemShut {NoStop}%
\bibitem [{\citenamefont {Barausse}\ \emph {et~al.}(2014)\citenamefont
  {Barausse}, \citenamefont {Cardoso},\ and\ \citenamefont
  {Pani}}]{Barausse2014b}%
  \BibitemOpen
  \bibfield  {author} {\bibinfo {author} {\bibfnamefont {E.}~\bibnamefont
  {Barausse}}, \bibinfo {author} {\bibfnamefont {V.}~\bibnamefont {Cardoso}}, \
  and\ \bibinfo {author} {\bibfnamefont {P.}~\bibnamefont {Pani}},\ }\href
  {\doibase 10.1103/PhysRevD.89.104059} {\bibfield  {journal} {\bibinfo
  {journal} {Phys.\ Rev.\ D}\ }\textbf {\bibinfo {volume} {89}},\ \bibinfo
  {pages} {104059} (\bibinfo {year} {2014})},\ \Eprint
  {http://arxiv.org/abs/1404.7149} {arXiv:1404.7149} \BibitemShut {NoStop}%
\bibitem [{\citenamefont {Barack}\ and\ \citenamefont
  {Cutler}(2007)}]{Barack2007}%
  \BibitemOpen
  \bibfield  {author} {\bibinfo {author} {\bibfnamefont {L.}~\bibnamefont
  {Barack}}\ and\ \bibinfo {author} {\bibfnamefont {C.}~\bibnamefont
  {Cutler}},\ }\href {\doibase 10.1103/PhysRevD.75.042003} {\bibfield
  {journal} {\bibinfo  {journal} {Phys.\ Rev.\ D}\ }\textbf {\bibinfo {volume}
  {75}},\ \bibinfo {pages} {042003} (\bibinfo {year} {2007})},\ \Eprint
  {http://arxiv.org/abs/0612029} {arXiv:0612029 [gr-qc]} \BibitemShut {NoStop}%
\bibitem [{\citenamefont {Gair}\ \emph {et~al.}(2013)\citenamefont {Gair},
  \citenamefont {Vallisneri}, \citenamefont {Larson},\ and\ \citenamefont
  {Baker}}]{Gair2012a}%
  \BibitemOpen
  \bibfield  {author} {\bibinfo {author} {\bibfnamefont {J.}~\bibnamefont
  {Gair}}, \bibinfo {author} {\bibfnamefont {M.}~\bibnamefont {Vallisneri}},
  \bibinfo {author} {\bibfnamefont {S.~L.}\ \bibnamefont {Larson}}, \ and\
  \bibinfo {author} {\bibfnamefont {J.~G.}\ \bibnamefont {Baker}},\ }\href
  {\doibase 10.12942/lrr-2013-7} {\bibfield  {journal} {\bibinfo  {journal}
  {Living Rev.\ Relat.}\ }\textbf {\bibinfo {volume} {16}},\ \bibinfo {pages}
  {7} (\bibinfo {year} {2013})},\ \Eprint {http://arxiv.org/abs/1212.5575}
  {arXiv:1212.5575} \BibitemShut {NoStop}%
\bibitem [{\citenamefont {Barack}(2009)}]{Barack2009}%
  \BibitemOpen
  \bibfield  {author} {\bibinfo {author} {\bibfnamefont {L.}~\bibnamefont
  {Barack}},\ }\href {\doibase 10.1088/0264-9381/26/21/213001} {\bibfield
  {journal} {\bibinfo  {journal} {Class.\ Quantum Grav.}\ }\textbf {\bibinfo
  {volume} {26}},\ \bibinfo {pages} {213001} (\bibinfo {year} {2009})},\
  \Eprint {http://arxiv.org/abs/0908.1664} {arXiv:0908.1664} \BibitemShut
  {NoStop}%
\bibitem [{\citenamefont {Poisson}(2004)}]{Poisson2004}%
  \BibitemOpen
  \bibfield  {author} {\bibinfo {author} {\bibfnamefont {E.}~\bibnamefont
  {Poisson}},\ }\href {\doibase 10.12942/lrr-2004-6} {\bibfield  {journal}
  {\bibinfo  {journal} {Living Rev.\ Relat.}\ }\textbf {\bibinfo {volume} {7}}
  (\bibinfo {year} {2004}),\ 10.12942/lrr-2004-6},\ \Eprint
  {http://arxiv.org/abs/0306052} {arXiv:0306052 [gr-qc]} \BibitemShut {NoStop}%
\bibitem [{\citenamefont {Pound}(2015)}]{Pound2015}%
  \BibitemOpen
  \bibfield  {author} {\bibinfo {author} {\bibfnamefont {A.}~\bibnamefont
  {Pound}},\ }in\ \href {\doibase 10.1007/978-3-319-18335-0_13} {\emph
  {\bibinfo {booktitle} {Equations of Motion in Relativistic Gravity}}}\
  (\bibinfo  {publisher} {Springer International Publishing},\ \bibinfo
  {address} {Cham},\ \bibinfo {year} {2015})\ pp.\ \bibinfo {pages}
  {399--486}\BibitemShut {NoStop}%
\bibitem [{\citenamefont {Sago}\ \emph {et~al.}(2008)\citenamefont {Sago},
  \citenamefont {Barack},\ and\ \citenamefont {Detweiler}}]{Sago2008}%
  \BibitemOpen
  \bibfield  {author} {\bibinfo {author} {\bibfnamefont {N.}~\bibnamefont
  {Sago}}, \bibinfo {author} {\bibfnamefont {L.}~\bibnamefont {Barack}}, \ and\
  \bibinfo {author} {\bibfnamefont {S.}~\bibnamefont {Detweiler}},\ }\href
  {\doibase 10.1103/PhysRevD.78.124024} {\bibfield  {journal} {\bibinfo
  {journal} {Phys.\ Rev.\ D}\ }\textbf {\bibinfo {volume} {78}},\ \bibinfo
  {pages} {124024} (\bibinfo {year} {2008})},\ \Eprint
  {http://arxiv.org/abs/0810.2530} {arXiv:0810.2530} \BibitemShut {NoStop}%
\bibitem [{\citenamefont {Flanagan}\ and\ \citenamefont
  {Hinderer}(2012)}]{Flanagan2012}%
  \BibitemOpen
  \bibfield  {author} {\bibinfo {author} {\bibfnamefont {{\'E}.~{\'E}.}\
  \bibnamefont {Flanagan}}\ and\ \bibinfo {author} {\bibfnamefont
  {T.}~\bibnamefont {Hinderer}},\ }\href {\doibase
  10.1103/PhysRevLett.109.071102} {\bibfield  {journal} {\bibinfo  {journal}
  {Phys.\ Rev.\ Lett.}\ }\textbf {\bibinfo {volume} {109}},\ \bibinfo {pages}
  {071102} (\bibinfo {year} {2012})},\ \Eprint {http://arxiv.org/abs/1009.4923}
  {arXiv:1009.4923} \BibitemShut {NoStop}%
\bibitem [{\citenamefont {Warburton}\ \emph {et~al.}(2013)\citenamefont
  {Warburton}, \citenamefont {Barack},\ and\ \citenamefont
  {Sago}}]{Warburton2013}%
  \BibitemOpen
  \bibfield  {author} {\bibinfo {author} {\bibfnamefont {N.}~\bibnamefont
  {Warburton}}, \bibinfo {author} {\bibfnamefont {L.}~\bibnamefont {Barack}}, \
  and\ \bibinfo {author} {\bibfnamefont {N.}~\bibnamefont {Sago}},\ }\href
  {\doibase 10.1103/PhysRevD.87.084012} {\bibfield  {journal} {\bibinfo
  {journal} {Phys.\ Rev.\ D}\ }\textbf {\bibinfo {volume} {87}},\ \bibinfo
  {pages} {084012} (\bibinfo {year} {2013})},\ \Eprint
  {http://arxiv.org/abs/1301.3918} {arXiv:1301.3918} \BibitemShut {NoStop}%
\bibitem [{\citenamefont {Goldstein}\ \emph {et~al.}(2002)\citenamefont
  {Goldstein}, \citenamefont {Poole},\ and\ \citenamefont
  {Safko}}]{Goldstein2002}%
  \BibitemOpen
  \bibfield  {author} {\bibinfo {author} {\bibfnamefont {H.}~\bibnamefont
  {Goldstein}}, \bibinfo {author} {\bibfnamefont {C.}~\bibnamefont {Poole}}, \
  and\ \bibinfo {author} {\bibfnamefont {J.}~\bibnamefont {Safko}},\
  }\href@noop {} {\emph {\bibinfo {title} {{Classical Mechanics}}}},\ \bibinfo
  {edition} {3rd}\ ed.\ (\bibinfo  {publisher} {Pearson Education
  International},\ \bibinfo {address} {Upper Saddle River, New Jersey},\
  \bibinfo {year} {2002})\BibitemShut {NoStop}%
\bibitem [{\citenamefont {Pound}\ and\ \citenamefont
  {Poisson}(2008)}]{Pound2008}%
  \BibitemOpen
  \bibfield  {author} {\bibinfo {author} {\bibfnamefont {A.}~\bibnamefont
  {Pound}}\ and\ \bibinfo {author} {\bibfnamefont {E.}~\bibnamefont
  {Poisson}},\ }\href {\doibase 10.1103/PhysRevD.77.044013} {\bibfield
  {journal} {\bibinfo  {journal} {Phys.\ Rev.\ D}\ }\textbf {\bibinfo {volume}
  {77}},\ \bibinfo {pages} {044013(18)} (\bibinfo {year} {2008})},\ \Eprint
  {http://arxiv.org/abs/0708.3033} {arXiv:0708.3033} \BibitemShut {NoStop}%
\bibitem [{\citenamefont {Gair}\ \emph
  {et~al.}(2011{\natexlab{b}})\citenamefont {Gair}, \citenamefont {Flanagan},
  \citenamefont {Drasco}, \citenamefont {Hinderer},\ and\ \citenamefont
  {Babak}}]{Gair2011a}%
  \BibitemOpen
  \bibfield  {author} {\bibinfo {author} {\bibfnamefont {J.~R.}\ \bibnamefont
  {Gair}}, \bibinfo {author} {\bibfnamefont {{\'E}.~{\'E}.}\ \bibnamefont
  {Flanagan}}, \bibinfo {author} {\bibfnamefont {S.}~\bibnamefont {Drasco}},
  \bibinfo {author} {\bibfnamefont {T.}~\bibnamefont {Hinderer}}, \ and\
  \bibinfo {author} {\bibfnamefont {S.}~\bibnamefont {Babak}},\ }\href
  {\doibase 10.1103/PhysRevD.83.044037} {\bibfield  {journal} {\bibinfo
  {journal} {Phys.\ Rev.\ D}\ }\textbf {\bibinfo {volume} {83}},\ \bibinfo
  {pages} {044037} (\bibinfo {year} {2011}{\natexlab{b}})},\ \Eprint
  {http://arxiv.org/abs/1012.5111} {arXiv:1012.5111} \BibitemShut {NoStop}%
\bibitem [{\citenamefont {Flanagan}\ \emph {et~al.}(2014)\citenamefont
  {Flanagan}, \citenamefont {Hughes},\ and\ \citenamefont
  {Ruangsri}}]{Flanagan2012a}%
  \BibitemOpen
  \bibfield  {author} {\bibinfo {author} {\bibfnamefont {E.~E.}\ \bibnamefont
  {Flanagan}}, \bibinfo {author} {\bibfnamefont {S.~A.}\ \bibnamefont
  {Hughes}}, \ and\ \bibinfo {author} {\bibfnamefont {U.}~\bibnamefont
  {Ruangsri}},\ }\href {\doibase 10.1103/PhysRevD.89.084028} {\bibfield
  {journal} {\bibinfo  {journal} {Phys.\ Rev.\ D}\ }\textbf {\bibinfo {volume}
  {89}},\ \bibinfo {pages} {084028} (\bibinfo {year} {2014})},\ \Eprint
  {http://arxiv.org/abs/1208.3906} {arXiv:1208.3906} \BibitemShut {NoStop}%
\bibitem [{\citenamefont {Hirata}(2011)}]{Hirata2011}%
  \BibitemOpen
  \bibfield  {author} {\bibinfo {author} {\bibfnamefont {C.~M.}\ \bibnamefont
  {Hirata}},\ }\href {\doibase 10.1103/PhysRevD.83.104024} {\bibfield
  {journal} {\bibinfo  {journal} {Phys.\ Rev.\ D}\ }\textbf {\bibinfo {volume}
  {83}},\ \bibinfo {pages} {104024} (\bibinfo {year} {2011})},\ \Eprint
  {http://arxiv.org/abs/1011.4987} {arXiv:1011.4987} \BibitemShut {NoStop}%
\bibitem [{\citenamefont {van~de Meent}(2014{\natexlab{a}})}]{VanDeMeent2013}%
  \BibitemOpen
  \bibfield  {author} {\bibinfo {author} {\bibfnamefont {M.}~\bibnamefont
  {van~de Meent}},\ }\href {\doibase 10.1103/PhysRevD.89.084033} {\bibfield
  {journal} {\bibinfo  {journal} {Phys.\ Rev.\ D}\ }\textbf {\bibinfo {volume}
  {89}},\ \bibinfo {pages} {084033} (\bibinfo {year} {2014}{\natexlab{a}})},\
  \Eprint {http://arxiv.org/abs/1311.4457} {arXiv:1311.4457} \BibitemShut
  {NoStop}%
\bibitem [{\citenamefont {van~de Meent}(2014{\natexlab{b}})}]{VanDeMeent2014}%
  \BibitemOpen
  \bibfield  {author} {\bibinfo {author} {\bibfnamefont {M.}~\bibnamefont
  {van~de Meent}},\ }\href {\doibase 10.1103/PhysRevD.90.044027} {\bibfield
  {journal} {\bibinfo  {journal} {Phys.\ Rev.\ D}\ }\textbf {\bibinfo {volume}
  {90}},\ \bibinfo {pages} {044027} (\bibinfo {year} {2014}{\natexlab{b}})},\
  \Eprint {http://arxiv.org/abs/1406.2594} {arXiv:1406.2594} \BibitemShut
  {NoStop}%
\bibitem [{\citenamefont {Boyer}\ and\ \citenamefont
  {Lindquist}(1967)}]{Boyer1967}%
  \BibitemOpen
  \bibfield  {author} {\bibinfo {author} {\bibfnamefont {R.~H.}\ \bibnamefont
  {Boyer}}\ and\ \bibinfo {author} {\bibfnamefont {R.~W.}\ \bibnamefont
  {Lindquist}},\ }\href {\doibase 10.1063/1.1705193} {\bibfield  {journal}
  {\bibinfo  {journal} {J.\ Math.\ Phys.}\ }\textbf {\bibinfo {volume} {8}},\
  \bibinfo {pages} {265} (\bibinfo {year} {1967})}\BibitemShut {NoStop}%
\bibitem [{\citenamefont {Hinderer}\ and\ \citenamefont
  {Flanagan}(2008)}]{Hinderer2008}%
  \BibitemOpen
  \bibfield  {author} {\bibinfo {author} {\bibfnamefont {T.}~\bibnamefont
  {Hinderer}}\ and\ \bibinfo {author} {\bibfnamefont {{\'E}.~{\'E}.}\
  \bibnamefont {Flanagan}},\ }\href {\doibase 10.1103/PhysRevD.78.064028}
  {\bibfield  {journal} {\bibinfo  {journal} {Phys.\ Rev.\ D}\ }\textbf
  {\bibinfo {volume} {78}},\ \bibinfo {pages} {064028} (\bibinfo {year}
  {2008})},\ \Eprint {http://arxiv.org/abs/0805.3337} {arXiv:0805.3337}
  \BibitemShut {NoStop}%
\bibitem [{\citenamefont {Carter}(1968)}]{Carter1968}%
  \BibitemOpen
  \bibfield  {author} {\bibinfo {author} {\bibfnamefont {B.}~\bibnamefont
  {Carter}},\ }\href {\doibase 10.1103/PhysRev.174.1559} {\bibfield  {journal}
  {\bibinfo  {journal} {Phys.\ Rev.}\ }\textbf {\bibinfo {volume} {174}},\
  \bibinfo {pages} {1559} (\bibinfo {year} {1968})}\BibitemShut {NoStop}%
\bibitem [{\citenamefont {Mino}(2003)}]{Mino2003}%
  \BibitemOpen
  \bibfield  {author} {\bibinfo {author} {\bibfnamefont {Y.}~\bibnamefont
  {Mino}},\ }\href {\doibase 10.1103/PhysRevD.67.084027} {\bibfield  {journal}
  {\bibinfo  {journal} {Phys.\ Rev.\ D}\ }\textbf {\bibinfo {volume} {67}},\
  \bibinfo {pages} {084027} (\bibinfo {year} {2003})},\ \Eprint
  {http://arxiv.org/abs/0302075} {arXiv:0302075 [gr-qc]} \BibitemShut {NoStop}%
\bibitem [{\citenamefont {Rosenthal}(2006)}]{Rosenthal2006}%
  \BibitemOpen
  \bibfield  {author} {\bibinfo {author} {\bibfnamefont {E.}~\bibnamefont
  {Rosenthal}},\ }\href {\doibase 10.1103/PhysRevD.74.084018} {\bibfield
  {journal} {\bibinfo  {journal} {Phys.\ Rev.\ D}\ }\textbf {\bibinfo {volume}
  {74}},\ \bibinfo {pages} {084018} (\bibinfo {year} {2006})},\ \Eprint
  {http://arxiv.org/abs/0609069} {arXiv:0609069 [gr-qc]} \BibitemShut {NoStop}%
\bibitem [{\citenamefont {Pound}(2012)}]{Pound2012}%
  \BibitemOpen
  \bibfield  {author} {\bibinfo {author} {\bibfnamefont {A.}~\bibnamefont
  {Pound}},\ }\href {\doibase 10.1103/PhysRevLett.109.051101} {\bibfield
  {journal} {\bibinfo  {journal} {Phys.\ Rev.\ Lett.}\ }\textbf {\bibinfo
  {volume} {109}},\ \bibinfo {pages} {051101} (\bibinfo {year} {2012})},\
  \Eprint {http://arxiv.org/abs/1201.5089} {arXiv:1201.5089} \BibitemShut
  {NoStop}%
\bibitem [{\citenamefont {Gralla}(2012)}]{Gralla2012}%
  \BibitemOpen
  \bibfield  {author} {\bibinfo {author} {\bibfnamefont {S.~E.}\ \bibnamefont
  {Gralla}},\ }\href {\doibase 10.1103/PhysRevD.85.124011} {\bibfield
  {journal} {\bibinfo  {journal} {Phys.\ Rev.\ D}\ }\textbf {\bibinfo {volume}
  {85}},\ \bibinfo {pages} {124011} (\bibinfo {year} {2012})},\ \Eprint
  {http://arxiv.org/abs/1203.3189} {arXiv:1203.3189} \BibitemShut {NoStop}%
\bibitem [{\citenamefont {Drasco}\ \emph {et~al.}(2005)\citenamefont {Drasco},
  \citenamefont {Flanagan},\ and\ \citenamefont {Hughes}}]{Drasco2005}%
  \BibitemOpen
  \bibfield  {author} {\bibinfo {author} {\bibfnamefont {S.}~\bibnamefont
  {Drasco}}, \bibinfo {author} {\bibfnamefont {{\'E}.~{\'E}.}\ \bibnamefont
  {Flanagan}}, \ and\ \bibinfo {author} {\bibfnamefont {S.~A.}\ \bibnamefont
  {Hughes}},\ }\href {\doibase 10.1088/0264-9381/22/15/011} {\bibfield
  {journal} {\bibinfo  {journal} {Class.\ Quantum Grav.}\ }\textbf {\bibinfo
  {volume} {22}},\ \bibinfo {pages} {S801} (\bibinfo {year} {2005})},\ \Eprint
  {http://arxiv.org/abs/0505075} {arXiv:0505075 [gr-qc]} \BibitemShut {NoStop}%
\bibitem [{\citenamefont {Arnold}\ \emph {et~al.}(1988)\citenamefont {Arnold},
  \citenamefont {Kozlov},\ and\ \citenamefont {Neishtadt}}]{Arnold1988}%
  \BibitemOpen
  \bibfield  {author} {\bibinfo {author} {\bibfnamefont {V.~I.}\ \bibnamefont
  {Arnold}}, \bibinfo {author} {\bibfnamefont {V.}~\bibnamefont {Kozlov}}, \
  and\ \bibinfo {author} {\bibfnamefont {A.~I.}\ \bibnamefont {Neishtadt}},\
  }in\ \href@noop {} {\emph {\bibinfo {booktitle} {Dynamical Systems III}}},\
  \bibinfo {series and number} {Encyclopaedia of Mathematical Sciences},\
  \bibinfo {editor} {edited by\ \bibinfo {editor} {\bibfnamefont {V.~I.}\
  \bibnamefont {Arnold}}}\ (\bibinfo  {publisher} {Springer--Verlag},\ \bibinfo
  {address} {New York},\ \bibinfo {year} {1988})\BibitemShut {NoStop}%
\bibitem [{\citenamefont {Grossman}\ \emph {et~al.}(2012)\citenamefont
  {Grossman}, \citenamefont {Levin},\ and\ \citenamefont
  {Perez-Giz}}]{Grossman2012}%
  \BibitemOpen
  \bibfield  {author} {\bibinfo {author} {\bibfnamefont {R.}~\bibnamefont
  {Grossman}}, \bibinfo {author} {\bibfnamefont {J.}~\bibnamefont {Levin}}, \
  and\ \bibinfo {author} {\bibfnamefont {G.}~\bibnamefont {Perez-Giz}},\ }\href
  {\doibase 10.1103/PhysRevD.85.023012} {\bibfield  {journal} {\bibinfo
  {journal} {Phys.\ Rev.\ D}\ }\textbf {\bibinfo {volume} {85}},\ \bibinfo
  {pages} {023012} (\bibinfo {year} {2012})},\ \Eprint
  {http://arxiv.org/abs/1105.5811} {arXiv:1105.5811} \BibitemShut {NoStop}%
\bibitem [{\citenamefont {Mapelli}\ \emph {et~al.}(2012)\citenamefont
  {Mapelli}, \citenamefont {Ripamonti}, \citenamefont {Vecchio}, \citenamefont
  {Graham},\ and\ \citenamefont {Gualandris}}]{Mapelli2012}%
  \BibitemOpen
  \bibfield  {author} {\bibinfo {author} {\bibfnamefont {M.}~\bibnamefont
  {Mapelli}}, \bibinfo {author} {\bibfnamefont {E.}~\bibnamefont {Ripamonti}},
  \bibinfo {author} {\bibfnamefont {A.}~\bibnamefont {Vecchio}}, \bibinfo
  {author} {\bibfnamefont {A.~W.}\ \bibnamefont {Graham}}, \ and\ \bibinfo
  {author} {\bibfnamefont {A.}~\bibnamefont {Gualandris}},\ }\href {\doibase
  10.1051/0004-6361/201118444} {\bibfield  {journal} {\bibinfo  {journal}
  {Astron.\ Astrophys.}\ }\textbf {\bibinfo {volume} {542}},\ \bibinfo {pages}
  {A102} (\bibinfo {year} {2012})},\ \Eprint {http://arxiv.org/abs/1205.2702}
  {arXiv:1205.2702} \BibitemShut {NoStop}%
\bibitem [{\citenamefont {Wilkins}(1972)}]{Wilkins1972}%
  \BibitemOpen
  \bibfield  {author} {\bibinfo {author} {\bibfnamefont {D.}~\bibnamefont
  {Wilkins}},\ }\href {\doibase 10.1103/PhysRevD.5.814} {\bibfield  {journal}
  {\bibinfo  {journal} {Phys.\ Rev.\ D}\ }\textbf {\bibinfo {volume} {5}},\
  \bibinfo {pages} {814} (\bibinfo {year} {1972})}\BibitemShut {NoStop}%
\bibitem [{\citenamefont {Darwin}(1961)}]{Darwin1961}%
  \BibitemOpen
  \bibfield  {author} {\bibinfo {author} {\bibfnamefont {C.}~\bibnamefont
  {Darwin}},\ }\href {\doibase 10.1098/rspa.1961.0142} {\bibfield  {journal}
  {\bibinfo  {journal} {Proc.\ R.\ Soc.\ A}\ }\textbf {\bibinfo {volume}
  {263}},\ \bibinfo {pages} {39} (\bibinfo {year} {1961})}\BibitemShut
  {NoStop}%
\bibitem [{\citenamefont {Drasco}\ and\ \citenamefont
  {Hughes}(2004)}]{Drasco2004}%
  \BibitemOpen
  \bibfield  {author} {\bibinfo {author} {\bibfnamefont {S.}~\bibnamefont
  {Drasco}}\ and\ \bibinfo {author} {\bibfnamefont {S.~A.}\ \bibnamefont
  {Hughes}},\ }\href {\doibase 10.1103/PhysRevD.69.044015} {\bibfield
  {journal} {\bibinfo  {journal} {Phys.\ Rev.\ D}\ }\textbf {\bibinfo {volume}
  {69}},\ \bibinfo {pages} {044015} (\bibinfo {year} {2004})},\ \Eprint
  {http://arxiv.org/abs/0308479} {arXiv:0308479 [astro-ph]} \BibitemShut
  {NoStop}%
\bibitem [{\citenamefont {Hughes}(2000)}]{Hughes2000}%
  \BibitemOpen
  \bibfield  {author} {\bibinfo {author} {\bibfnamefont {S.~A.}\ \bibnamefont
  {Hughes}},\ }\href {\doibase 10.1103/PhysRevD.61.084004} {\bibfield
  {journal} {\bibinfo  {journal} {Phys.\ Rev.\ D}\ }\textbf {\bibinfo {volume}
  {61}},\ \bibinfo {pages} {084004} (\bibinfo {year} {2000})},\ \Eprint
  {http://arxiv.org/abs/9910091} {arXiv:9910091 [gr-qc]} \BibitemShut {NoStop}%
\bibitem [{\citenamefont {Schmidt}(2002)}]{Schmidt2002}%
  \BibitemOpen
  \bibfield  {author} {\bibinfo {author} {\bibfnamefont {W.}~\bibnamefont
  {Schmidt}},\ }\href {\doibase 10.1088/0264-9381/19/10/314} {\bibfield
  {journal} {\bibinfo  {journal} {Class.\ Quantum Grav.}\ }\textbf {\bibinfo
  {volume} {19}},\ \bibinfo {pages} {2743} (\bibinfo {year} {2002})},\ \Eprint
  {http://arxiv.org/abs/0202090} {arXiv:0202090 [gr-qc]} \BibitemShut {NoStop}%
\bibitem [{\citenamefont {Ryan}(1996)}]{Ryan1996}%
  \BibitemOpen
  \bibfield  {author} {\bibinfo {author} {\bibfnamefont {F.~D.}\ \bibnamefont
  {Ryan}},\ }\href {\doibase 10.1103/PhysRevD.53.3064} {\bibfield  {journal}
  {\bibinfo  {journal} {Phys.\ Rev.\ D}\ }\textbf {\bibinfo {volume} {53}},\
  \bibinfo {pages} {3064} (\bibinfo {year} {1996})}\BibitemShut {NoStop}%
\bibitem [{\citenamefont {Glampedakis}\ \emph {et~al.}(2002)\citenamefont
  {Glampedakis}, \citenamefont {Hughes},\ and\ \citenamefont
  {Kennefick}}]{Glampedakis2002}%
  \BibitemOpen
  \bibfield  {author} {\bibinfo {author} {\bibfnamefont {K.}~\bibnamefont
  {Glampedakis}}, \bibinfo {author} {\bibfnamefont {S.~A.}\ \bibnamefont
  {Hughes}}, \ and\ \bibinfo {author} {\bibfnamefont {D.}~\bibnamefont
  {Kennefick}},\ }\href {\doibase 10.1103/PhysRevD.66.064005} {\bibfield
  {journal} {\bibinfo  {journal} {Phys.\ Rev.\ D}\ }\textbf {\bibinfo {volume}
  {66}},\ \bibinfo {pages} {064005} (\bibinfo {year} {2002})},\ \Eprint
  {http://arxiv.org/abs/0205033} {arXiv:0205033 [gr-qc]} \BibitemShut {NoStop}%
\bibitem [{\citenamefont {Fujita}\ and\ \citenamefont
  {Hikida}(2009)}]{Fujita2009}%
  \BibitemOpen
  \bibfield  {author} {\bibinfo {author} {\bibfnamefont {R.}~\bibnamefont
  {Fujita}}\ and\ \bibinfo {author} {\bibfnamefont {W.}~\bibnamefont
  {Hikida}},\ }\href {\doibase 10.1088/0264-9381/26/13/135002} {\bibfield
  {journal} {\bibinfo  {journal} {Class.\ Quantum Grav.}\ }\textbf {\bibinfo
  {volume} {26}},\ \bibinfo {pages} {135002} (\bibinfo {year} {2009})},\
  \Eprint {http://arxiv.org/abs/0906.1420} {arXiv:0906.1420} \BibitemShut
  {NoStop}%
\bibitem [{\citenamefont {Bosley}\ and\ \citenamefont
  {Kevorkian}(1992)}]{Bosley1992}%
  \BibitemOpen
  \bibfield  {author} {\bibinfo {author} {\bibfnamefont {D.~L.}\ \bibnamefont
  {Bosley}}\ and\ \bibinfo {author} {\bibfnamefont {J.}~\bibnamefont
  {Kevorkian}},\ }\href {\doibase 10.1137/0152028} {\bibfield  {journal}
  {\bibinfo  {journal} {SIAM J.\ Appl.\ Math.}\ }\textbf {\bibinfo {volume}
  {52}},\ \bibinfo {pages} {494} (\bibinfo {year} {1992})}\BibitemShut
  {NoStop}%
\bibitem [{\citenamefont {Warburton}\ \emph {et~al.}(2012)\citenamefont
  {Warburton}, \citenamefont {Akcay}, \citenamefont {Barack}, \citenamefont
  {Gair},\ and\ \citenamefont {Sago}}]{Warburton2012}%
  \BibitemOpen
  \bibfield  {author} {\bibinfo {author} {\bibfnamefont {N.}~\bibnamefont
  {Warburton}}, \bibinfo {author} {\bibfnamefont {S.}~\bibnamefont {Akcay}},
  \bibinfo {author} {\bibfnamefont {L.}~\bibnamefont {Barack}}, \bibinfo
  {author} {\bibfnamefont {J.~R.}\ \bibnamefont {Gair}}, \ and\ \bibinfo
  {author} {\bibfnamefont {N.}~\bibnamefont {Sago}},\ }\href {\doibase
  10.1103/PhysRevD.85.061501} {\bibfield  {journal} {\bibinfo  {journal}
  {Phys.\ Rev.\ D}\ }\textbf {\bibinfo {volume} {85}},\ \bibinfo {pages}
  {061501(R)} (\bibinfo {year} {2012})},\ \Eprint
  {http://arxiv.org/abs/1111.6908} {arXiv:1111.6908} \BibitemShut {NoStop}%
\bibitem [{\citenamefont {Arnol'd}(1963)}]{Arnold1963}%
  \BibitemOpen
  \bibfield  {author} {\bibinfo {author} {\bibfnamefont {V.~I.}\ \bibnamefont
  {Arnol'd}},\ }\href {\doibase 10.1070/RM1963v018n05ABEH004130} {\bibfield
  {journal} {\bibinfo  {journal} {Russ.\ Math.\ Surv.}\ }\textbf {\bibinfo
  {volume} {18}},\ \bibinfo {pages} {9} (\bibinfo {year} {1963})}\BibitemShut
  {NoStop}%
\bibitem [{\citenamefont {Moser}(1973)}]{Moser1973}%
  \BibitemOpen
  \bibfield  {author} {\bibinfo {author} {\bibfnamefont {J.}~\bibnamefont
  {Moser}},\ }\href@noop {} {\emph {\bibinfo {title} {{Stable and Random
  Motions in Dynamical Systems: With Special Emphasis on Celestial
  Mechanics}}}},\ Annals of mathematical studies\ (\bibinfo  {publisher}
  {Princeton University Press},\ \bibinfo {address} {Princeton, New Jersey},\
  \bibinfo {year} {1973})\BibitemShut {NoStop}%
\bibitem [{\citenamefont {Flanagan}\ and\ \citenamefont
  {Hinderer}(2007)}]{Flanagan2007}%
  \BibitemOpen
  \bibfield  {author} {\bibinfo {author} {\bibfnamefont {{\'E}.~{\'E}.}\
  \bibnamefont {Flanagan}}\ and\ \bibinfo {author} {\bibfnamefont
  {T.}~\bibnamefont {Hinderer}},\ }\href {\doibase 10.1103/PhysRevD.75.124007}
  {\bibfield  {journal} {\bibinfo  {journal} {Phys.\ Rev.\ D}\ }\textbf
  {\bibinfo {volume} {75}},\ \bibinfo {pages} {124007} (\bibinfo {year}
  {2007})},\ \Eprint {http://arxiv.org/abs/0704.0389} {arXiv:0704.0389}
  \BibitemShut {NoStop}%
\bibitem [{\citenamefont {Iyer}\ and\ \citenamefont {Will}(1993)}]{Iyer1993}%
  \BibitemOpen
  \bibfield  {author} {\bibinfo {author} {\bibfnamefont {B.~R.}\ \bibnamefont
  {Iyer}}\ and\ \bibinfo {author} {\bibfnamefont {C.~M.}\ \bibnamefont
  {Will}},\ }\href {\doibase 10.1103/PhysRevLett.70.113} {\bibfield  {journal}
  {\bibinfo  {journal} {Phys.\ Rev.\ Lett.}\ }\textbf {\bibinfo {volume}
  {70}},\ \bibinfo {pages} {113} (\bibinfo {year} {1993})}\BibitemShut
  {NoStop}%
\bibitem [{\citenamefont {Kidder}(1995)}]{Kidder1995}%
  \BibitemOpen
  \bibfield  {author} {\bibinfo {author} {\bibfnamefont {L.~E.}\ \bibnamefont
  {Kidder}},\ }\href {\doibase 10.1103/PhysRevD.52.821} {\bibfield  {journal}
  {\bibinfo  {journal} {Phys.\ Rev.\ D}\ }\textbf {\bibinfo {volume} {52}},\
  \bibinfo {pages} {821} (\bibinfo {year} {1995})},\ \Eprint
  {http://arxiv.org/abs/9506022} {arXiv:9506022 [gr-qc]} \BibitemShut {NoStop}%
\bibitem [{\citenamefont {Grossman}\ \emph {et~al.}(2013)\citenamefont
  {Grossman}, \citenamefont {Levin},\ and\ \citenamefont
  {Perez-Giz}}]{Grossman2011}%
  \BibitemOpen
  \bibfield  {author} {\bibinfo {author} {\bibfnamefont {R.}~\bibnamefont
  {Grossman}}, \bibinfo {author} {\bibfnamefont {J.}~\bibnamefont {Levin}}, \
  and\ \bibinfo {author} {\bibfnamefont {G.}~\bibnamefont {Perez-Giz}},\ }\href
  {\doibase 10.1103/PhysRevD.88.023002} {\bibfield  {journal} {\bibinfo
  {journal} {Phys.\ Rev.\ D}\ }\textbf {\bibinfo {volume} {88}},\ \bibinfo
  {pages} {023002} (\bibinfo {year} {2013})},\ \Eprint
  {http://arxiv.org/abs/1108.1819} {arXiv:1108.1819} \BibitemShut {NoStop}%
\bibitem [{\citenamefont {Gal'tsov}(1982)}]{Galtsov1982}%
  \BibitemOpen
  \bibfield  {author} {\bibinfo {author} {\bibfnamefont {D.~V.}\ \bibnamefont
  {Gal'tsov}},\ }\href {\doibase 10.1088/0305-4470/15/12/025} {\bibfield
  {journal} {\bibinfo  {journal} {J.\ Phys.\ A}\ }\textbf {\bibinfo {volume}
  {15}},\ \bibinfo {pages} {3737} (\bibinfo {year} {1982})}\BibitemShut
  {NoStop}%
\bibitem [{\citenamefont {Sago}\ \emph {et~al.}(2006)\citenamefont {Sago},
  \citenamefont {Tanaka}, \citenamefont {Hikida}, \citenamefont {Ganz},\ and\
  \citenamefont {Nakano}}]{Sago2006}%
  \BibitemOpen
  \bibfield  {author} {\bibinfo {author} {\bibfnamefont {N.}~\bibnamefont
  {Sago}}, \bibinfo {author} {\bibfnamefont {T.}~\bibnamefont {Tanaka}},
  \bibinfo {author} {\bibfnamefont {W.}~\bibnamefont {Hikida}}, \bibinfo
  {author} {\bibfnamefont {K.}~\bibnamefont {Ganz}}, \ and\ \bibinfo {author}
  {\bibfnamefont {H.}~\bibnamefont {Nakano}},\ }\href {\doibase
  10.1143/PTP.115.873} {\bibfield  {journal} {\bibinfo  {journal} {Prog.\
  Theor.\ Phys.}\ }\textbf {\bibinfo {volume} {115}},\ \bibinfo {pages} {873}
  (\bibinfo {year} {2006})},\ \Eprint {http://arxiv.org/abs/0511151}
  {arXiv:0511151 [gr-qc]} \BibitemShut {NoStop}%
\bibitem [{\citenamefont {Ganz}\ \emph {et~al.}(2007)\citenamefont {Ganz},
  \citenamefont {Hikida}, \citenamefont {Nakano}, \citenamefont {Sago},\ and\
  \citenamefont {Tanaka}}]{Ganz2007}%
  \BibitemOpen
  \bibfield  {author} {\bibinfo {author} {\bibfnamefont {K.}~\bibnamefont
  {Ganz}}, \bibinfo {author} {\bibfnamefont {W.}~\bibnamefont {Hikida}},
  \bibinfo {author} {\bibfnamefont {H.}~\bibnamefont {Nakano}}, \bibinfo
  {author} {\bibfnamefont {N.}~\bibnamefont {Sago}}, \ and\ \bibinfo {author}
  {\bibfnamefont {T.}~\bibnamefont {Tanaka}},\ }\href {\doibase
  10.1143/PTP.117.1041} {\bibfield  {journal} {\bibinfo  {journal} {Prog.\
  Theor.\ Phys.}\ }\textbf {\bibinfo {volume} {117}},\ \bibinfo {pages} {1041}
  (\bibinfo {year} {2007})},\ \Eprint {http://arxiv.org/abs/0702054}
  {arXiv:0702054 [gr-qc]} \BibitemShut {NoStop}%
\bibitem [{\citenamefont {Pound}\ \emph {et~al.}(2005)\citenamefont {Pound},
  \citenamefont {Poisson},\ and\ \citenamefont {Nickel}}]{Pound2005}%
  \BibitemOpen
  \bibfield  {author} {\bibinfo {author} {\bibfnamefont {A.}~\bibnamefont
  {Pound}}, \bibinfo {author} {\bibfnamefont {E.}~\bibnamefont {Poisson}}, \
  and\ \bibinfo {author} {\bibfnamefont {B.~G.}\ \bibnamefont {Nickel}},\
  }\href {\doibase 10.1103/PhysRevD.72.124001} {\bibfield  {journal} {\bibinfo
  {journal} {Phys.\ Rev.\ D}\ }\textbf {\bibinfo {volume} {72}},\ \bibinfo
  {pages} {124001} (\bibinfo {year} {2005})},\ \Eprint
  {http://arxiv.org/abs/0509122} {arXiv:0509122 [gr-qc]} \BibitemShut {NoStop}%
\bibitem [{\citenamefont {Ruangsri}\ and\ \citenamefont
  {Hughes}(2014)}]{Ruangsri2014}%
  \BibitemOpen
  \bibfield  {author} {\bibinfo {author} {\bibfnamefont {U.}~\bibnamefont
  {Ruangsri}}\ and\ \bibinfo {author} {\bibfnamefont {S.~A.}\ \bibnamefont
  {Hughes}},\ }\href {\doibase 10.1103/PhysRevD.89.084036} {\bibfield
  {journal} {\bibinfo  {journal} {Phys.\ Rev.\ D}\ }\textbf {\bibinfo {volume}
  {89}},\ \bibinfo {pages} {084036} (\bibinfo {year} {2014})},\ \Eprint
  {http://arxiv.org/abs/1307.6483} {arXiv:1307.6483} \BibitemShut {NoStop}%
\bibitem [{\citenamefont {Kevorkian}(1987)}]{Kevorkian1987}%
  \BibitemOpen
  \bibfield  {author} {\bibinfo {author} {\bibfnamefont {J.}~\bibnamefont
  {Kevorkian}},\ }\href {\doibase 10.1137/1029076} {\bibfield  {journal}
  {\bibinfo  {journal} {SIAM Rev.}\ }\textbf {\bibinfo {volume} {29}},\
  \bibinfo {pages} {391} (\bibinfo {year} {1987})}\BibitemShut {NoStop}%
\bibitem [{\citenamefont {Babak}\ \emph {et~al.}(2007)\citenamefont {Babak},
  \citenamefont {Fang}, \citenamefont {Gair}, \citenamefont {Glampedakis},\
  and\ \citenamefont {Hughes}}]{Babak2007}%
  \BibitemOpen
  \bibfield  {author} {\bibinfo {author} {\bibfnamefont {S.}~\bibnamefont
  {Babak}}, \bibinfo {author} {\bibfnamefont {H.}~\bibnamefont {Fang}},
  \bibinfo {author} {\bibfnamefont {J.~R.}\ \bibnamefont {Gair}}, \bibinfo
  {author} {\bibfnamefont {K.}~\bibnamefont {Glampedakis}}, \ and\ \bibinfo
  {author} {\bibfnamefont {S.~A.}\ \bibnamefont {Hughes}},\ }\href {\doibase
  10.1103/PhysRevD.75.024005} {\bibfield  {journal} {\bibinfo  {journal}
  {Phys.\ Rev.\ D}\ }\textbf {\bibinfo {volume} {75}},\ \bibinfo {pages}
  {024005} (\bibinfo {year} {2007})},\ \Eprint {http://arxiv.org/abs/0607007}
  {arXiv:0607007 [gr-qc]} \BibitemShut {NoStop}%
\bibitem [{\citenamefont {Berry}\ and\ \citenamefont {Gair}(2013)}]{Berry2013}%
  \BibitemOpen
  \bibfield  {author} {\bibinfo {author} {\bibfnamefont {C.~P.~L.}\
  \bibnamefont {Berry}}\ and\ \bibinfo {author} {\bibfnamefont {J.~R.}\
  \bibnamefont {Gair}},\ }\href {\doibase 10.1093/mnras/sts360} {\bibfield
  {journal} {\bibinfo  {journal} {Mon.\ Not.\ R.\ Astron.\ Soc.}\ }\textbf
  {\bibinfo {volume} {429}},\ \bibinfo {pages} {589} (\bibinfo {year}
  {2013})},\ \Eprint {http://arxiv.org/abs/1210.2778} {arXiv:1210.2778}
  \BibitemShut {NoStop}%
\bibitem [{\citenamefont {Misner}\ \emph {et~al.}(1973)\citenamefont {Misner},
  \citenamefont {Thorne},\ and\ \citenamefont {Wheeler}}]{Misner1973}%
  \BibitemOpen
  \bibfield  {author} {\bibinfo {author} {\bibfnamefont {C.~W.}\ \bibnamefont
  {Misner}}, \bibinfo {author} {\bibfnamefont {K.~S.}\ \bibnamefont {Thorne}},
  \ and\ \bibinfo {author} {\bibfnamefont {J.~A.}\ \bibnamefont {Wheeler}},\
  }\href@noop {} {\emph {\bibinfo {title} {{Gravitation}}}}\ (\bibinfo
  {publisher} {W.H. Freeman},\ \bibinfo {address} {New York},\ \bibinfo {year}
  {1973})\BibitemShut {NoStop}%
\bibitem [{\citenamefont {Finn}(1992)}]{Finn1992}%
  \BibitemOpen
  \bibfield  {author} {\bibinfo {author} {\bibfnamefont {L.~S.}\ \bibnamefont
  {Finn}},\ }\href {\doibase 10.1103/PhysRevD.46.5236} {\bibfield  {journal}
  {\bibinfo  {journal} {Phys.\ Rev.\ D}\ }\textbf {\bibinfo {volume} {46}},\
  \bibinfo {pages} {5236} (\bibinfo {year} {1992})},\ \Eprint
  {http://arxiv.org/abs/9209010} {arXiv:9209010 [gr-qc]} \BibitemShut {NoStop}%
\bibitem [{\citenamefont {Moore}\ \emph {et~al.}(2015)\citenamefont {Moore},
  \citenamefont {Cole},\ and\ \citenamefont {Berry}}]{Moore2014a}%
  \BibitemOpen
  \bibfield  {author} {\bibinfo {author} {\bibfnamefont {C.~J.}\ \bibnamefont
  {Moore}}, \bibinfo {author} {\bibfnamefont {R.~H.}\ \bibnamefont {Cole}}, \
  and\ \bibinfo {author} {\bibfnamefont {C.~P.~L.}\ \bibnamefont {Berry}},\
  }\href {\doibase 10.1088/0264-9381/32/1/015014} {\bibfield  {journal}
  {\bibinfo  {journal} {Class.\ Quantum Grav.}\ }\textbf {\bibinfo {volume}
  {32}},\ \bibinfo {pages} {015014} (\bibinfo {year} {2015})},\ \Eprint
  {http://arxiv.org/abs/1408.0740} {arXiv:1408.0740} \BibitemShut {NoStop}%
\bibitem [{\citenamefont {McKechan}\ \emph {et~al.}(2010)\citenamefont
  {McKechan}, \citenamefont {Robinson},\ and\ \citenamefont
  {Sathyaprakash}}]{McKechan2010}%
  \BibitemOpen
  \bibfield  {author} {\bibinfo {author} {\bibfnamefont {D.~J.~A.}\
  \bibnamefont {McKechan}}, \bibinfo {author} {\bibfnamefont {C.}~\bibnamefont
  {Robinson}}, \ and\ \bibinfo {author} {\bibfnamefont {B.~S.}\ \bibnamefont
  {Sathyaprakash}},\ }\href {\doibase 10.1088/0264-9381/27/8/084020} {\bibfield
   {journal} {\bibinfo  {journal} {Class.\ Quantum Grav.}\ }\textbf {\bibinfo
  {volume} {27}},\ \bibinfo {pages} {084020} (\bibinfo {year} {2010})},\
  \Eprint {http://arxiv.org/abs/1003.2939} {arXiv:1003.2939} \BibitemShut
  {NoStop}%
\bibitem [{\citenamefont {Armano}\ \emph {et~al.}(2016)\citenamefont {Armano}
  \emph {et~al.}}]{Armano2016}%
  \BibitemOpen
  \bibfield  {author} {\bibinfo {author} {\bibfnamefont {M.}~\bibnamefont
  {Armano}} \emph {et~al.},\ }\href {\doibase 10.1103/PhysRevLett.116.231101}
  {\bibfield  {journal} {\bibinfo  {journal} {Phys.\ Rev.\ Lett.}\ }\textbf
  {\bibinfo {volume} {116}},\ \bibinfo {pages} {231101} (\bibinfo {year}
  {2016})}\BibitemShut {NoStop}%
\bibitem [{\citenamefont {Cutler}\ and\ \citenamefont
  {Vallisneri}(2007)}]{Cutler2007}%
  \BibitemOpen
  \bibfield  {author} {\bibinfo {author} {\bibfnamefont {C.}~\bibnamefont
  {Cutler}}\ and\ \bibinfo {author} {\bibfnamefont {M.}~\bibnamefont
  {Vallisneri}},\ }\href {\doibase 10.1103/PhysRevD.76.104018} {\bibfield
  {journal} {\bibinfo  {journal} {Phys.\ Rev.\ D}\ }\textbf {\bibinfo {volume}
  {76}},\ \bibinfo {pages} {104018} (\bibinfo {year} {2007})},\ \Eprint
  {http://arxiv.org/abs/0707.2982} {arXiv:0707.2982} \BibitemShut {NoStop}%
\bibitem [{\citenamefont {Mandel}\ \emph {et~al.}(2014)\citenamefont {Mandel},
  \citenamefont {Berry}, \citenamefont {Ohme}, \citenamefont {Fairhurst},\ and\
  \citenamefont {Farr}}]{Mandel2014}%
  \BibitemOpen
  \bibfield  {author} {\bibinfo {author} {\bibfnamefont {I.}~\bibnamefont
  {Mandel}}, \bibinfo {author} {\bibfnamefont {C.~P.~L.}\ \bibnamefont
  {Berry}}, \bibinfo {author} {\bibfnamefont {F.}~\bibnamefont {Ohme}},
  \bibinfo {author} {\bibfnamefont {S.}~\bibnamefont {Fairhurst}}, \ and\
  \bibinfo {author} {\bibfnamefont {W.~M.}\ \bibnamefont {Farr}},\ }\href
  {\doibase 10.1088/0264-9381/31/15/155005} {\bibfield  {journal} {\bibinfo
  {journal} {Class.\ Quantum Grav.}\ }\textbf {\bibinfo {volume} {31}},\
  \bibinfo {pages} {155005} (\bibinfo {year} {2014})},\ \Eprint
  {http://arxiv.org/abs/1404.2382} {arXiv:1404.2382} \BibitemShut {NoStop}%
\bibitem [{\citenamefont {Cole}(2015)}]{ColeThesis2015}%
  \BibitemOpen
  \bibfield  {author} {\bibinfo {author} {\bibfnamefont {R.~H.}\ \bibnamefont
  {Cole}},\ }\emph {\bibinfo {title} {Gravitational waves from
  extreme-mass-ratio inspirals}},\ \href@noop {} {\bibinfo {type} {{Ph.{D}.}
  dissertation}},\ \bibinfo  {school} {University of Cambridge} (\bibinfo
  {year} {2015})\BibitemShut {NoStop}%
\bibitem [{\citenamefont {Alexander}(2005)}]{Alexander2005}%
  \BibitemOpen
  \bibfield  {author} {\bibinfo {author} {\bibfnamefont {T.}~\bibnamefont
  {Alexander}},\ }\href {\doibase 10.1016/j.physrep.2005.08.002} {\bibfield
  {journal} {\bibinfo  {journal} {Phys.\ Rep.}\ }\textbf {\bibinfo {volume}
  {419}},\ \bibinfo {pages} {65} (\bibinfo {year} {2005})},\ \Eprint
  {http://arxiv.org/abs/0508106} {arXiv:0508106 [astro-ph]} \BibitemShut
  {NoStop}%
\bibitem [{\citenamefont {Rees}(1988)}]{Rees1988}%
  \BibitemOpen
  \bibfield  {author} {\bibinfo {author} {\bibfnamefont {M.~J.}\ \bibnamefont
  {Rees}},\ }\href {\doibase 10.1038/333523a0} {\bibfield  {journal} {\bibinfo
  {journal} {Nature}\ }\textbf {\bibinfo {volume} {333}},\ \bibinfo {pages}
  {523} (\bibinfo {year} {1988})}\BibitemShut {NoStop}%
\bibitem [{\citenamefont {Sigurdsson}\ and\ \citenamefont
  {Rees}(1997)}]{Sigurdsson1997}%
  \BibitemOpen
  \bibfield  {author} {\bibinfo {author} {\bibfnamefont {S.}~\bibnamefont
  {Sigurdsson}}\ and\ \bibinfo {author} {\bibfnamefont {M.~J.}\ \bibnamefont
  {Rees}},\ }\href {\doibase 10.1093/mnras/284.2.318} {\bibfield  {journal}
  {\bibinfo  {journal} {Mon.\ Not.\ R.\ Astron.\ Soc.}\ }\textbf {\bibinfo
  {volume} {284}},\ \bibinfo {pages} {318} (\bibinfo {year} {1997})},\ \Eprint
  {http://arxiv.org/abs/9608093} {arXiv:9608093 [astro-ph]} \BibitemShut
  {NoStop}%
\bibitem [{\citenamefont {Gair}\ \emph {et~al.}(2004)\citenamefont {Gair},
  \citenamefont {Barack}, \citenamefont {Creighton}, \citenamefont {Cutler},
  \citenamefont {Larson}, \citenamefont {Phinney},\ and\ \citenamefont
  {Vallisneri}}]{Gair2004}%
  \BibitemOpen
  \bibfield  {author} {\bibinfo {author} {\bibfnamefont {J.~R.}\ \bibnamefont
  {Gair}}, \bibinfo {author} {\bibfnamefont {L.}~\bibnamefont {Barack}},
  \bibinfo {author} {\bibfnamefont {T.}~\bibnamefont {Creighton}}, \bibinfo
  {author} {\bibfnamefont {C.}~\bibnamefont {Cutler}}, \bibinfo {author}
  {\bibfnamefont {S.~L.}\ \bibnamefont {Larson}}, \bibinfo {author}
  {\bibfnamefont {E.~S.}\ \bibnamefont {Phinney}}, \ and\ \bibinfo {author}
  {\bibfnamefont {M.}~\bibnamefont {Vallisneri}},\ }\href {\doibase
  10.1088/0264-9381/21/20/003} {\bibfield  {journal} {\bibinfo  {journal}
  {Class.\ Quantum Grav.}\ }\textbf {\bibinfo {volume} {21}},\ \bibinfo {pages}
  {S1595} (\bibinfo {year} {2004})},\ \Eprint {http://arxiv.org/abs/0405137}
  {arXiv:0405137 [gr-qc]} \BibitemShut {NoStop}%
\bibitem [{\citenamefont {Peters}\ and\ \citenamefont
  {Mathews}(1963)}]{Peters1963}%
  \BibitemOpen
  \bibfield  {author} {\bibinfo {author} {\bibfnamefont {P.~C.}\ \bibnamefont
  {Peters}}\ and\ \bibinfo {author} {\bibfnamefont {J.}~\bibnamefont
  {Mathews}},\ }\href {\doibase 10.1103/PhysRev.131.435} {\bibfield  {journal}
  {\bibinfo  {journal} {Phys.\ Rev.}\ }\textbf {\bibinfo {volume} {131}},\
  \bibinfo {pages} {435} (\bibinfo {year} {1963})}\BibitemShut {NoStop}%
\bibitem [{\citenamefont {Chandrasekhar}(1960)}]{Chandrasekhar1960}%
  \BibitemOpen
  \bibfield  {author} {\bibinfo {author} {\bibfnamefont {S.}~\bibnamefont
  {Chandrasekhar}},\ }\href@noop {} {\emph {\bibinfo {title} {{Principles of
  Stellar Dynamics}}}},\ \bibinfo {edition} {enlarged}\ ed.\ (\bibinfo
  {publisher} {Dover Publications},\ \bibinfo {address} {New York},\ \bibinfo
  {year} {1960})\BibitemShut {NoStop}%
\bibitem [{\citenamefont {Antonini}\ and\ \citenamefont
  {Merritt}(2012)}]{Antonini2011}%
  \BibitemOpen
  \bibfield  {author} {\bibinfo {author} {\bibfnamefont {F.}~\bibnamefont
  {Antonini}}\ and\ \bibinfo {author} {\bibfnamefont {D.}~\bibnamefont
  {Merritt}},\ }\href {\doibase 10.1088/0004-637X/745/1/83} {\bibfield
  {journal} {\bibinfo  {journal} {Astrophys.\ J.}\ }\textbf {\bibinfo {volume}
  {745}},\ \bibinfo {pages} {83} (\bibinfo {year} {2012})},\ \Eprint
  {http://arxiv.org/abs/1108.1163} {arXiv:1108.1163} \BibitemShut {NoStop}%
\bibitem [{\citenamefont {Bahcall}\ and\ \citenamefont
  {Wolf}(1977)}]{Bahcall1977}%
  \BibitemOpen
  \bibfield  {author} {\bibinfo {author} {\bibfnamefont {J.~N.}\ \bibnamefont
  {Bahcall}}\ and\ \bibinfo {author} {\bibfnamefont {R.~A.}\ \bibnamefont
  {Wolf}},\ }\href {\doibase 10.1086/155534} {\bibfield  {journal} {\bibinfo
  {journal} {Astrophys.\ J.}\ }\textbf {\bibinfo {volume} {216}},\ \bibinfo
  {pages} {883} (\bibinfo {year} {1977})}\BibitemShut {NoStop}%
\bibitem [{\citenamefont {Freitag}\ \emph {et~al.}(2006)\citenamefont
  {Freitag}, \citenamefont {Amaro-Seoane},\ and\ \citenamefont
  {Kalogera}}]{Freitag2006}%
  \BibitemOpen
  \bibfield  {author} {\bibinfo {author} {\bibfnamefont {M.}~\bibnamefont
  {Freitag}}, \bibinfo {author} {\bibfnamefont {P.}~\bibnamefont
  {Amaro-Seoane}}, \ and\ \bibinfo {author} {\bibfnamefont {V.}~\bibnamefont
  {Kalogera}},\ }\href {\doibase 10.1086/506193} {\bibfield  {journal}
  {\bibinfo  {journal} {Astrophys.\ J.}\ }\textbf {\bibinfo {volume} {649}},\
  \bibinfo {pages} {91} (\bibinfo {year} {2006})},\ \Eprint
  {http://arxiv.org/abs/0603280} {arXiv:0603280 [astro-ph]} \BibitemShut
  {NoStop}%
\bibitem [{\citenamefont {Alexander}\ and\ \citenamefont
  {Hopman}(2009)}]{Alexander2009}%
  \BibitemOpen
  \bibfield  {author} {\bibinfo {author} {\bibfnamefont {T.}~\bibnamefont
  {Alexander}}\ and\ \bibinfo {author} {\bibfnamefont {C.}~\bibnamefont
  {Hopman}},\ }\href {\doibase 10.1088/0004-637X/697/2/1861} {\bibfield
  {journal} {\bibinfo  {journal} {Astrophys.\ J.}\ }\textbf {\bibinfo {volume}
  {697}},\ \bibinfo {pages} {1861} (\bibinfo {year} {2009})},\ \Eprint
  {http://arxiv.org/abs/0808.3150} {arXiv:0808.3150} \BibitemShut {NoStop}%
\bibitem [{\citenamefont {Casares}\ and\ \citenamefont
  {Jonker}(2014)}]{Casares2014}%
  \BibitemOpen
  \bibfield  {author} {\bibinfo {author} {\bibfnamefont {J.}~\bibnamefont
  {Casares}}\ and\ \bibinfo {author} {\bibfnamefont {P.~G.}\ \bibnamefont
  {Jonker}},\ }\href {\doibase 10.1007/s11214-013-0030-6} {\bibfield  {journal}
  {\bibinfo  {journal} {Space Sci.\ Rev.}\ }\textbf {\bibinfo {volume} {183}},\
  \bibinfo {pages} {223} (\bibinfo {year} {2014})},\ \Eprint
  {http://arxiv.org/abs/1311.5118} {arXiv:1311.5118} \BibitemShut {NoStop}%
\bibitem [{\citenamefont {Corral-Santana}\ \emph {et~al.}(2016)\citenamefont
  {Corral-Santana}, \citenamefont {Casares}, \citenamefont
  {Mu{\~{n}}oz-Darias}, \citenamefont {Bauer}, \citenamefont
  {Mart{\'{i}}nez-Pais},\ and\ \citenamefont {Russell}}]{Corral-Santana2016}%
  \BibitemOpen
  \bibfield  {author} {\bibinfo {author} {\bibfnamefont {J.~M.}\ \bibnamefont
  {Corral-Santana}}, \bibinfo {author} {\bibfnamefont {J.}~\bibnamefont
  {Casares}}, \bibinfo {author} {\bibfnamefont {T.}~\bibnamefont
  {Mu{\~{n}}oz-Darias}}, \bibinfo {author} {\bibfnamefont {F.~E.}\ \bibnamefont
  {Bauer}}, \bibinfo {author} {\bibfnamefont {I.~G.}\ \bibnamefont
  {Mart{\'{i}}nez-Pais}}, \ and\ \bibinfo {author} {\bibfnamefont {D.~M.}\
  \bibnamefont {Russell}},\ }\href {\doibase 10.1051/0004-6361/201527130}
  {\bibfield  {journal} {\bibinfo  {journal} {Astron.\ Astrophys.}\ }\textbf
  {\bibinfo {volume} {587}},\ \bibinfo {pages} {A61} (\bibinfo {year}
  {2016})},\ \Eprint {http://arxiv.org/abs/1510.08869} {arXiv:1510.08869}
  \BibitemShut {NoStop}%
\bibitem [{\citenamefont {Tetarenko}\ \emph {et~al.}(2016)\citenamefont
  {Tetarenko}, \citenamefont {Sivakoff}, \citenamefont {Heinke},\ and\
  \citenamefont {Gladstone}}]{Tetarenko2016}%
  \BibitemOpen
  \bibfield  {author} {\bibinfo {author} {\bibfnamefont {B.~E.}\ \bibnamefont
  {Tetarenko}}, \bibinfo {author} {\bibfnamefont {G.~R.}\ \bibnamefont
  {Sivakoff}}, \bibinfo {author} {\bibfnamefont {C.~O.}\ \bibnamefont
  {Heinke}}, \ and\ \bibinfo {author} {\bibfnamefont {J.~C.}\ \bibnamefont
  {Gladstone}},\ }\href {\doibase 10.3847/0067-0049/222/2/15} {\bibfield
  {journal} {\bibinfo  {journal} {Astrophys.\ J. Supp.\ S.}\ }\textbf {\bibinfo
  {volume} {222}},\ \bibinfo {pages} {15} (\bibinfo {year} {2016})},\ \Eprint
  {http://arxiv.org/abs/1512.00778} {arXiv:1512.00778} \BibitemShut {NoStop}%
\bibitem [{\citenamefont {Lynden-Bell}\ and\ \citenamefont
  {Rees}(1971)}]{Lynden-Bell1971}%
  \BibitemOpen
  \bibfield  {author} {\bibinfo {author} {\bibfnamefont {D.}~\bibnamefont
  {Lynden-Bell}}\ and\ \bibinfo {author} {\bibfnamefont {M.~J.}\ \bibnamefont
  {Rees}},\ }\href {\doibase 10.1093/mnras/152.4.461} {\bibfield  {journal}
  {\bibinfo  {journal} {Mon.\ Not.\ R.\ Astron.\ Soc.}\ }\textbf {\bibinfo
  {volume} {152}},\ \bibinfo {pages} {461} (\bibinfo {year}
  {1971})}\BibitemShut {NoStop}%
\bibitem [{\citenamefont {So{\l}tan}(1982)}]{Soltan1982}%
  \BibitemOpen
  \bibfield  {author} {\bibinfo {author} {\bibfnamefont {A.}~\bibnamefont
  {So{\l}tan}},\ }\href {\doibase 10.1093/mnras/200.1.115} {\bibfield
  {journal} {\bibinfo  {journal} {Mon.\ Not.\ R.\ Astron.\ Soc.}\ }\textbf
  {\bibinfo {volume} {200}},\ \bibinfo {pages} {115} (\bibinfo {year}
  {1982})}\BibitemShut {NoStop}%
\bibitem [{\citenamefont {Volonteri}(2010)}]{Volonteri2010}%
  \BibitemOpen
  \bibfield  {author} {\bibinfo {author} {\bibfnamefont {M.}~\bibnamefont
  {Volonteri}},\ }\href {\doibase 10.1007/s00159-010-0029-x} {\bibfield
  {journal} {\bibinfo  {journal} {Astron.\ Astrophys.\ Rev.}\ }\textbf
  {\bibinfo {volume} {18}},\ \bibinfo {pages} {279} (\bibinfo {year} {2010})},\
  \Eprint {http://arxiv.org/abs/1003.4404} {arXiv:1003.4404} \BibitemShut
  {NoStop}%
\bibitem [{\citenamefont {Dotti}\ \emph {et~al.}(2013)\citenamefont {Dotti},
  \citenamefont {Colpi}, \citenamefont {Pallini}, \citenamefont {Perego},\ and\
  \citenamefont {Volonteri}}]{Dotti2013}%
  \BibitemOpen
  \bibfield  {author} {\bibinfo {author} {\bibfnamefont {M.}~\bibnamefont
  {Dotti}}, \bibinfo {author} {\bibfnamefont {M.}~\bibnamefont {Colpi}},
  \bibinfo {author} {\bibfnamefont {S.}~\bibnamefont {Pallini}}, \bibinfo
  {author} {\bibfnamefont {A.}~\bibnamefont {Perego}}, \ and\ \bibinfo {author}
  {\bibfnamefont {M.}~\bibnamefont {Volonteri}},\ }\href {\doibase
  10.1088/0004-637X/762/2/68} {\bibfield  {journal} {\bibinfo  {journal}
  {Astrophys.\ J.}\ }\textbf {\bibinfo {volume} {762}},\ \bibinfo {pages} {68}
  (\bibinfo {year} {2013})},\ \Eprint {http://arxiv.org/abs/1211.4871}
  {arXiv:1211.4871} \BibitemShut {NoStop}%
\bibitem [{\citenamefont {Sesana}\ \emph {et~al.}(2014)\citenamefont {Sesana},
  \citenamefont {Barausse}, \citenamefont {Dotti},\ and\ \citenamefont
  {Rossi}}]{Sesana2014}%
  \BibitemOpen
  \bibfield  {author} {\bibinfo {author} {\bibfnamefont {A.}~\bibnamefont
  {Sesana}}, \bibinfo {author} {\bibfnamefont {E.}~\bibnamefont {Barausse}},
  \bibinfo {author} {\bibfnamefont {M.}~\bibnamefont {Dotti}}, \ and\ \bibinfo
  {author} {\bibfnamefont {E.~M.}\ \bibnamefont {Rossi}},\ }\href {\doibase
  10.1088/0004-637X/794/2/104} {\bibfield  {journal} {\bibinfo  {journal}
  {Astrophys.\ J.}\ }\textbf {\bibinfo {volume} {794}},\ \bibinfo {pages} {104}
  (\bibinfo {year} {2014})},\ \Eprint {http://arxiv.org/abs/1402.7088}
  {arXiv:1402.7088} \BibitemShut {NoStop}%
\bibitem [{\citenamefont {Greene}\ and\ \citenamefont {Ho}(2007)}]{Greene2007}%
  \BibitemOpen
  \bibfield  {author} {\bibinfo {author} {\bibfnamefont {J.~E.}\ \bibnamefont
  {Greene}}\ and\ \bibinfo {author} {\bibfnamefont {L.~C.}\ \bibnamefont
  {Ho}},\ }\href {\doibase 10.1086/520497 10.1088/0004-637X/704/2/1743}
  {\bibfield  {journal} {\bibinfo  {journal} {Astrophys.\ J.}\ }\textbf
  {\bibinfo {volume} {667}},\ \bibinfo {pages} {131} (\bibinfo {year}
  {2007})},\ \Eprint {http://arxiv.org/abs/0705.0020} {arXiv:0705.0020}
  \BibitemShut {NoStop}%
\bibitem [{\citenamefont {Merritt}(2013)}]{Merritt2013}%
  \BibitemOpen
  \bibfield  {author} {\bibinfo {author} {\bibfnamefont {D.}~\bibnamefont
  {Merritt}},\ }\href {\doibase 10.1088/0264-9381/30/24/244005} {\bibfield
  {journal} {\bibinfo  {journal} {Class.\ Quantum Grav.}\ }\textbf {\bibinfo
  {volume} {30}},\ \bibinfo {pages} {244005} (\bibinfo {year} {2013})},\
  \Eprint {http://arxiv.org/abs/1307.3268} {arXiv:1307.3268} \BibitemShut
  {NoStop}%
\bibitem [{\citenamefont {Hopman}(2009)}]{Hopman2009a}%
  \BibitemOpen
  \bibfield  {author} {\bibinfo {author} {\bibfnamefont {C.}~\bibnamefont
  {Hopman}},\ }\href {\doibase 10.1088/0264-9381/26/9/094028} {\bibfield
  {journal} {\bibinfo  {journal} {Class.\ Quantum Grav.}\ }\textbf {\bibinfo
  {volume} {26}},\ \bibinfo {pages} {094028} (\bibinfo {year} {2009})},\
  \Eprint {http://arxiv.org/abs/0901.1667} {arXiv:0901.1667} \BibitemShut
  {NoStop}%
\bibitem [{\citenamefont {Amaro-Seoane}\ and\ \citenamefont
  {Preto}(2011)}]{Amaro-Seoane2011d}%
  \BibitemOpen
  \bibfield  {author} {\bibinfo {author} {\bibfnamefont {P.}~\bibnamefont
  {Amaro-Seoane}}\ and\ \bibinfo {author} {\bibfnamefont {M.}~\bibnamefont
  {Preto}},\ }\href {\doibase 10.1088/0264-9381/28/9/094017} {\bibfield
  {journal} {\bibinfo  {journal} {Class.\ Quantum Grav.}\ }\textbf {\bibinfo
  {volume} {28}},\ \bibinfo {pages} {094017} (\bibinfo {year} {2011})},\
  \Eprint {http://arxiv.org/abs/1010.5781} {arXiv:1010.5781} \BibitemShut
  {NoStop}%
\bibitem [{\citenamefont {Rauch}\ and\ \citenamefont
  {Tremaine}(1996)}]{Rauch1996}%
  \BibitemOpen
  \bibfield  {author} {\bibinfo {author} {\bibfnamefont {K.~P.}\ \bibnamefont
  {Rauch}}\ and\ \bibinfo {author} {\bibfnamefont {S.}~\bibnamefont
  {Tremaine}},\ }\href {\doibase 10.1016/S1384-1076(96)00012-7} {\bibfield
  {journal} {\bibinfo  {journal} {New Astron.}\ }\textbf {\bibinfo {volume}
  {1}},\ \bibinfo {pages} {149} (\bibinfo {year} {1996})},\ \Eprint
  {http://arxiv.org/abs/9603018} {arXiv:9603018 [astro-ph]} \BibitemShut
  {NoStop}%
\bibitem [{\citenamefont {Rauch}\ and\ \citenamefont
  {Ingalls}(1998)}]{Rauch1998}%
  \BibitemOpen
  \bibfield  {author} {\bibinfo {author} {\bibfnamefont {K.~P.}\ \bibnamefont
  {Rauch}}\ and\ \bibinfo {author} {\bibfnamefont {B.}~\bibnamefont
  {Ingalls}},\ }\href {\doibase 10.1046/j.1365-8711.1998.01889.x} {\bibfield
  {journal} {\bibinfo  {journal} {Mon.\ Not.\ R.\ Astron.\ Soc.}\ }\textbf
  {\bibinfo {volume} {299}},\ \bibinfo {pages} {1231} (\bibinfo {year}
  {1998})},\ \Eprint {http://arxiv.org/abs/9710288} {arXiv:9710288 [astro-ph]}
  \BibitemShut {NoStop}%
\bibitem [{\citenamefont {Merritt}\ \emph {et~al.}(2011)\citenamefont
  {Merritt}, \citenamefont {Alexander}, \citenamefont {Mikkola},\ and\
  \citenamefont {Will}}]{Merritt2011}%
  \BibitemOpen
  \bibfield  {author} {\bibinfo {author} {\bibfnamefont {D.}~\bibnamefont
  {Merritt}}, \bibinfo {author} {\bibfnamefont {T.}~\bibnamefont {Alexander}},
  \bibinfo {author} {\bibfnamefont {S.}~\bibnamefont {Mikkola}}, \ and\
  \bibinfo {author} {\bibfnamefont {C.~M.}\ \bibnamefont {Will}},\ }\href
  {\doibase 10.1103/PhysRevD.84.044024} {\bibfield  {journal} {\bibinfo
  {journal} {Phys.\ Rev.\ D}\ }\textbf {\bibinfo {volume} {84}},\ \bibinfo
  {pages} {044024} (\bibinfo {year} {2011})},\ \Eprint
  {http://arxiv.org/abs/1102.3180} {arXiv:1102.3180} \BibitemShut {NoStop}%
\bibitem [{\citenamefont {Hamers}\ \emph {et~al.}(2014)\citenamefont {Hamers},
  \citenamefont {{Portegies Zwart}},\ and\ \citenamefont
  {Merritt}}]{Hamers2014}%
  \BibitemOpen
  \bibfield  {author} {\bibinfo {author} {\bibfnamefont {A.~S.}\ \bibnamefont
  {Hamers}}, \bibinfo {author} {\bibfnamefont {S.~F.}\ \bibnamefont {{Portegies
  Zwart}}}, \ and\ \bibinfo {author} {\bibfnamefont {D.}~\bibnamefont
  {Merritt}},\ }\href {\doibase 10.1093/mnras/stu1126} {\bibfield  {journal}
  {\bibinfo  {journal} {Mon.\ Not.\ R.\ Astron.\ Soc.}\ }\textbf {\bibinfo
  {volume} {443}},\ \bibinfo {pages} {355} (\bibinfo {year} {2014})},\ \Eprint
  {http://arxiv.org/abs/1406.2846} {arXiv:1406.2846} \BibitemShut {NoStop}%
\bibitem [{\citenamefont {Merritt}(2015)}]{Merritt2015c}%
  \BibitemOpen
  \bibfield  {author} {\bibinfo {author} {\bibfnamefont {D.}~\bibnamefont
  {Merritt}},\ }\href {\doibase 10.1088/0004-637X/814/1/57} {\bibfield
  {journal} {\bibinfo  {journal} {Astrophys.\ J.}\ }\textbf {\bibinfo {volume}
  {814}},\ \bibinfo {pages} {57} (\bibinfo {year} {2015})},\ \Eprint
  {http://arxiv.org/abs/1511.08169} {arXiv:1511.08169} \BibitemShut {NoStop}%
\bibitem [{\citenamefont {Amaro-Seoane}\ \emph {et~al.}(2013)\citenamefont
  {Amaro-Seoane}, \citenamefont {Sopuerta},\ and\ \citenamefont
  {Freitag}}]{Amaro-Seoane2012b}%
  \BibitemOpen
  \bibfield  {author} {\bibinfo {author} {\bibfnamefont {P.}~\bibnamefont
  {Amaro-Seoane}}, \bibinfo {author} {\bibfnamefont {C.~F.}\ \bibnamefont
  {Sopuerta}}, \ and\ \bibinfo {author} {\bibfnamefont {M.~D.}\ \bibnamefont
  {Freitag}},\ }\href {\doibase 10.1093/mnras/sts572} {\bibfield  {journal}
  {\bibinfo  {journal} {Mon.\ Not.\ R.\ Astron.\ Soc.}\ }\textbf {\bibinfo
  {volume} {429}},\ \bibinfo {pages} {3155} (\bibinfo {year} {2013})},\ \Eprint
  {http://arxiv.org/abs/1205.4713} {arXiv:1205.4713} \BibitemShut {NoStop}%
\bibitem [{\citenamefont {Ferrarese}\ and\ \citenamefont
  {Merritt}(2000)}]{Ferrarese2000}%
  \BibitemOpen
  \bibfield  {author} {\bibinfo {author} {\bibfnamefont {L.}~\bibnamefont
  {Ferrarese}}\ and\ \bibinfo {author} {\bibfnamefont {D.}~\bibnamefont
  {Merritt}},\ }\href {\doibase 10.1086/312838} {\bibfield  {journal} {\bibinfo
   {journal} {Astrophys.\ J.}\ }\textbf {\bibinfo {volume} {539}},\ \bibinfo
  {pages} {L9} (\bibinfo {year} {2000})},\ \Eprint
  {http://arxiv.org/abs/0006053} {arXiv:0006053 [astro-ph]} \BibitemShut
  {NoStop}%
\bibitem [{\citenamefont {Graham}(2016)}]{Graham2016}%
  \BibitemOpen
  \bibfield  {author} {\bibinfo {author} {\bibfnamefont {A.~W.}\ \bibnamefont
  {Graham}},\ }in\ \href {\doibase 10.1007/978-3-319-19378-6_11} {\emph
  {\bibinfo {booktitle} {Galactic Bulges}}},\ \bibinfo {editor} {edited by\
  \bibinfo {editor} {\bibfnamefont {E.}~\bibnamefont {Laurikainen}}, \bibinfo
  {editor} {\bibfnamefont {R.}~\bibnamefont {Peletier}}, \ and\ \bibinfo
  {editor} {\bibfnamefont {D.}~\bibnamefont {Gadotti}}}\ (\bibinfo  {publisher}
  {Springer},\ \bibinfo {year} {2016})\ Chap.~\bibinfo {chapter} {11}, pp.\
  \bibinfo {pages} {263--313},\ \Eprint {http://arxiv.org/abs/1501.02937}
  {arXiv:1501.02937} \BibitemShut {NoStop}%
\bibitem [{\citenamefont {Hlavacek-Larrondo}\ \emph {et~al.}(2012)\citenamefont
  {Hlavacek-Larrondo}, \citenamefont {Fabian}, \citenamefont {Edge},\ and\
  \citenamefont {Hogan}}]{Hlavacek-Larrondo2012}%
  \BibitemOpen
  \bibfield  {author} {\bibinfo {author} {\bibfnamefont {J.}~\bibnamefont
  {Hlavacek-Larrondo}}, \bibinfo {author} {\bibfnamefont {A.~C.}\ \bibnamefont
  {Fabian}}, \bibinfo {author} {\bibfnamefont {A.~C.}\ \bibnamefont {Edge}}, \
  and\ \bibinfo {author} {\bibfnamefont {M.~T.}\ \bibnamefont {Hogan}},\ }\href
  {\doibase 10.1111/j.1365-2966.2012.21187.x} {\bibfield  {journal} {\bibinfo
  {journal} {Mon.\ Not.\ R.\ Astron.\ Soc.}\ }\textbf {\bibinfo {volume}
  {424}},\ \bibinfo {pages} {224} (\bibinfo {year} {2012})},\ \Eprint
  {http://arxiv.org/abs/1204.5759} {arXiv:1204.5759} \BibitemShut {NoStop}%
\bibitem [{\citenamefont {Ferr{\'{e}}-Mateu}\ \emph {et~al.}(2015)\citenamefont
  {Ferr{\'{e}}-Mateu}, \citenamefont {Mezcua}, \citenamefont {Trujillo},
  \citenamefont {Balcells},\ and\ \citenamefont {van~den
  Bosch}}]{Ferre-Mateu2015}%
  \BibitemOpen
  \bibfield  {author} {\bibinfo {author} {\bibfnamefont {A.}~\bibnamefont
  {Ferr{\'{e}}-Mateu}}, \bibinfo {author} {\bibfnamefont {M.}~\bibnamefont
  {Mezcua}}, \bibinfo {author} {\bibfnamefont {I.}~\bibnamefont {Trujillo}},
  \bibinfo {author} {\bibfnamefont {M.}~\bibnamefont {Balcells}}, \ and\
  \bibinfo {author} {\bibfnamefont {R.~C.~E.}\ \bibnamefont {van~den Bosch}},\
  }\href {\doibase 10.1088/0004-637X/808/1/79} {\bibfield  {journal} {\bibinfo
  {journal} {Astrophys.\ J.}\ }\textbf {\bibinfo {volume} {808}},\ \bibinfo
  {pages} {79} (\bibinfo {year} {2015})},\ \Eprint
  {http://arxiv.org/abs/1506.02663} {arXiv:1506.02663} \BibitemShut {NoStop}%
\bibitem [{\citenamefont {van Loon}\ and\ \citenamefont
  {Sansom}(2015)}]{VanLoon2015}%
  \BibitemOpen
  \bibfield  {author} {\bibinfo {author} {\bibfnamefont {J.~T.}\ \bibnamefont
  {van Loon}}\ and\ \bibinfo {author} {\bibfnamefont {A.~E.}\ \bibnamefont
  {Sansom}},\ }\href {\doibase 10.1093/mnras/stv1787} {\bibfield  {journal}
  {\bibinfo  {journal} {Mon.\ Not.\ R.\ Astron.\ Soc.}\ }\textbf {\bibinfo
  {volume} {453}},\ \bibinfo {pages} {2342} (\bibinfo {year} {2015})},\ \Eprint
  {http://arxiv.org/abs/1508.00698} {arXiv:1508.00698} \BibitemShut {NoStop}%
\bibitem [{\citenamefont {Trakhtenbrot}\ \emph {et~al.}(2015)\citenamefont
  {Trakhtenbrot}, \citenamefont {Urry}, \citenamefont {Civano}, \citenamefont
  {Rosario}, \citenamefont {Elvis}, \citenamefont {Schawinski}, \citenamefont
  {Suh}, \citenamefont {Bongiorno},\ and\ \citenamefont
  {Simmons}}]{Trakhtenbrot2015}%
  \BibitemOpen
  \bibfield  {author} {\bibinfo {author} {\bibfnamefont {B.}~\bibnamefont
  {Trakhtenbrot}}, \bibinfo {author} {\bibfnamefont {C.~M.}\ \bibnamefont
  {Urry}}, \bibinfo {author} {\bibfnamefont {F.}~\bibnamefont {Civano}},
  \bibinfo {author} {\bibfnamefont {D.~J.}\ \bibnamefont {Rosario}}, \bibinfo
  {author} {\bibfnamefont {M.}~\bibnamefont {Elvis}}, \bibinfo {author}
  {\bibfnamefont {K.}~\bibnamefont {Schawinski}}, \bibinfo {author}
  {\bibfnamefont {H.}~\bibnamefont {Suh}}, \bibinfo {author} {\bibfnamefont
  {A.}~\bibnamefont {Bongiorno}}, \ and\ \bibinfo {author} {\bibfnamefont
  {B.~D.}\ \bibnamefont {Simmons}},\ }\href {\doibase 10.1126/science.aaa4506}
  {\bibfield  {journal} {\bibinfo  {journal} {Science}\ }\textbf {\bibinfo
  {volume} {349}},\ \bibinfo {pages} {168} (\bibinfo {year} {2015})},\ \Eprint
  {http://arxiv.org/abs/1507.02290} {arXiv:1507.02290} \BibitemShut {NoStop}%
\bibitem [{\citenamefont {King}(2016)}]{King2016}%
  \BibitemOpen
  \bibfield  {author} {\bibinfo {author} {\bibfnamefont {A.}~\bibnamefont
  {King}},\ }\href {\doibase 10.1093/mnrasl/slv186} {\bibfield  {journal}
  {\bibinfo  {journal} {Mon.\ Not.\ R.\ Astron.\ Soc.\ Lett.}\ }\textbf
  {\bibinfo {volume} {456}},\ \bibinfo {pages} {L109} (\bibinfo {year}
  {2016})},\ \Eprint {http://arxiv.org/abs/1511.08502} {arXiv:1511.08502}
  \BibitemShut {NoStop}%
\bibitem [{\citenamefont {Reynolds}(2013)}]{Reynolds2013a}%
  \BibitemOpen
  \bibfield  {author} {\bibinfo {author} {\bibfnamefont {C.~S.}\ \bibnamefont
  {Reynolds}},\ }\href {\doibase 10.1088/0264-9381/30/24/244004} {\bibfield
  {journal} {\bibinfo  {journal} {Class.\ Quantum Grav.}\ }\textbf {\bibinfo
  {volume} {30}},\ \bibinfo {pages} {244004} (\bibinfo {year} {2013})},\
  \Eprint {http://arxiv.org/abs/1307.3246} {arXiv:1307.3246} \BibitemShut
  {NoStop}%
\bibitem [{\citenamefont {Patrick}\ \emph {et~al.}(2012)\citenamefont
  {Patrick}, \citenamefont {Reeves}, \citenamefont {Porquet}, \citenamefont
  {Markowitz}, \citenamefont {Braito},\ and\ \citenamefont
  {Lobban}}]{Patrick2012}%
  \BibitemOpen
  \bibfield  {author} {\bibinfo {author} {\bibfnamefont {A.~R.}\ \bibnamefont
  {Patrick}}, \bibinfo {author} {\bibfnamefont {J.~N.}\ \bibnamefont {Reeves}},
  \bibinfo {author} {\bibfnamefont {D.}~\bibnamefont {Porquet}}, \bibinfo
  {author} {\bibfnamefont {A.~G.}\ \bibnamefont {Markowitz}}, \bibinfo {author}
  {\bibfnamefont {V.}~\bibnamefont {Braito}}, \ and\ \bibinfo {author}
  {\bibfnamefont {A.~P.}\ \bibnamefont {Lobban}},\ }\href {\doibase
  10.1111/j.1365-2966.2012.21868.x} {\bibfield  {journal} {\bibinfo  {journal}
  {Mon.\ Not.\ R.\ Astron.\ Soc.}\ }\textbf {\bibinfo {volume} {426}},\
  \bibinfo {pages} {2522} (\bibinfo {year} {2012})},\ \Eprint
  {http://arxiv.org/abs/1208.1150} {arXiv:1208.1150} \BibitemShut {NoStop}%
\bibitem [{\citenamefont {Walton}\ \emph {et~al.}(2013)\citenamefont {Walton},
  \citenamefont {Nardini}, \citenamefont {Fabian}, \citenamefont {Gallo},\ and\
  \citenamefont {Reis}}]{Walton2013}%
  \BibitemOpen
  \bibfield  {author} {\bibinfo {author} {\bibfnamefont {D.~J.}\ \bibnamefont
  {Walton}}, \bibinfo {author} {\bibfnamefont {E.}~\bibnamefont {Nardini}},
  \bibinfo {author} {\bibfnamefont {A.~C.}\ \bibnamefont {Fabian}}, \bibinfo
  {author} {\bibfnamefont {L.~C.}\ \bibnamefont {Gallo}}, \ and\ \bibinfo
  {author} {\bibfnamefont {R.~C.}\ \bibnamefont {Reis}},\ }\href {\doibase
  10.1093/mnras/sts227} {\bibfield  {journal} {\bibinfo  {journal} {Mon.\ Not.\
  R.\ Astron.\ Soc.}\ }\textbf {\bibinfo {volume} {428}},\ \bibinfo {pages}
  {2901} (\bibinfo {year} {2013})},\ \Eprint {http://arxiv.org/abs/1210.4593}
  {arXiv:1210.4593} \BibitemShut {NoStop}%
\bibitem [{\citenamefont {Berry}(2013)}]{BerryThesis2013}%
  \BibitemOpen
  \bibfield  {author} {\bibinfo {author} {\bibfnamefont {C.~P.~L.}\
  \bibnamefont {Berry}},\ }\emph {\bibinfo {title} {Exploring gravity}},\ \href
  {\doibase 10.17863/CAM.1199} {\bibinfo {type} {{Ph.{D}.} dissertation}},\
  \bibinfo  {school} {University of Cambridge} (\bibinfo {year}
  {2013})\BibitemShut {NoStop}%
\bibitem [{\citenamefont {Brenneman}\ \emph {et~al.}(2011)\citenamefont
  {Brenneman}, \citenamefont {Reynolds}, \citenamefont {Nowak}, \citenamefont
  {Reis}, \citenamefont {Trippe}, \citenamefont {Fabian}, \citenamefont
  {Iwasawa}, \citenamefont {Lee}, \citenamefont {Miller}, \citenamefont
  {Mushotzky}, \citenamefont {Nandra},\ and\ \citenamefont
  {Volonteri}}]{Brenneman2011}%
  \BibitemOpen
  \bibfield  {author} {\bibinfo {author} {\bibfnamefont {L.~W.}\ \bibnamefont
  {Brenneman}}, \bibinfo {author} {\bibfnamefont {C.~S.}\ \bibnamefont
  {Reynolds}}, \bibinfo {author} {\bibfnamefont {M.~A.}\ \bibnamefont {Nowak}},
  \bibinfo {author} {\bibfnamefont {R.~C.}\ \bibnamefont {Reis}}, \bibinfo
  {author} {\bibfnamefont {M.}~\bibnamefont {Trippe}}, \bibinfo {author}
  {\bibfnamefont {A.~C.}\ \bibnamefont {Fabian}}, \bibinfo {author}
  {\bibfnamefont {K.}~\bibnamefont {Iwasawa}}, \bibinfo {author} {\bibfnamefont
  {J.~C.}\ \bibnamefont {Lee}}, \bibinfo {author} {\bibfnamefont {J.~M.}\
  \bibnamefont {Miller}}, \bibinfo {author} {\bibfnamefont {R.~F.}\
  \bibnamefont {Mushotzky}}, \bibinfo {author} {\bibfnamefont {K.}~\bibnamefont
  {Nandra}}, \ and\ \bibinfo {author} {\bibfnamefont {M.}~\bibnamefont
  {Volonteri}},\ }\href {\doibase 10.1088/0004-637X/736/2/103} {\bibfield
  {journal} {\bibinfo  {journal} {Astrophys.\ J.}\ }\textbf {\bibinfo {volume}
  {736}},\ \bibinfo {pages} {103} (\bibinfo {year} {2011})},\ \Eprint
  {http://arxiv.org/abs/1104.1172} {arXiv:1104.1172} \BibitemShut {NoStop}%
\bibitem [{\citenamefont {Hopman}\ and\ \citenamefont
  {Alexander}(2005)}]{Hopman2005}%
  \BibitemOpen
  \bibfield  {author} {\bibinfo {author} {\bibfnamefont {C.}~\bibnamefont
  {Hopman}}\ and\ \bibinfo {author} {\bibfnamefont {T.}~\bibnamefont
  {Alexander}},\ }\href {\doibase 10.1086/431475} {\bibfield  {journal}
  {\bibinfo  {journal} {Astrophys.\ J.}\ }\textbf {\bibinfo {volume} {629}},\
  \bibinfo {pages} {362} (\bibinfo {year} {2005})},\ \Eprint
  {http://arxiv.org/abs/0503672} {arXiv:0503672 [astro-ph]} \BibitemShut
  {NoStop}%
\bibitem [{\citenamefont {Isoyama}\ \emph {et~al.}(2014)\citenamefont
  {Isoyama}, \citenamefont {Barack}, \citenamefont {Dolan}, \citenamefont {{Le
  Tiec}}, \citenamefont {Nakano}, \citenamefont {Shah}, \citenamefont
  {Tanaka},\ and\ \citenamefont {Warburton}}]{Isoyama2014}%
  \BibitemOpen
  \bibfield  {author} {\bibinfo {author} {\bibfnamefont {S.}~\bibnamefont
  {Isoyama}}, \bibinfo {author} {\bibfnamefont {L.}~\bibnamefont {Barack}},
  \bibinfo {author} {\bibfnamefont {S.~R.}\ \bibnamefont {Dolan}}, \bibinfo
  {author} {\bibfnamefont {A.}~\bibnamefont {{Le Tiec}}}, \bibinfo {author}
  {\bibfnamefont {H.}~\bibnamefont {Nakano}}, \bibinfo {author} {\bibfnamefont
  {A.~G.}\ \bibnamefont {Shah}}, \bibinfo {author} {\bibfnamefont
  {T.}~\bibnamefont {Tanaka}}, \ and\ \bibinfo {author} {\bibfnamefont
  {N.}~\bibnamefont {Warburton}},\ }\href {\doibase
  10.1103/PhysRevLett.113.161101} {\bibfield  {journal} {\bibinfo  {journal}
  {Phys.\ Rev.\ Lett.}\ }\textbf {\bibinfo {volume} {113}},\ \bibinfo {pages}
  {161101} (\bibinfo {year} {2014})},\ \Eprint {http://arxiv.org/abs/1404.6133}
  {arXiv:1404.6133} \BibitemShut {NoStop}%
\bibitem [{\citenamefont {Gair}(2009)}]{Gair2009}%
  \BibitemOpen
  \bibfield  {author} {\bibinfo {author} {\bibfnamefont {J.~R.}\ \bibnamefont
  {Gair}},\ }\href {\doibase 10.1088/0264-9381/26/9/094034} {\bibfield
  {journal} {\bibinfo  {journal} {Class.\ Quantum Grav.}\ }\textbf {\bibinfo
  {volume} {26}},\ \bibinfo {pages} {094034} (\bibinfo {year} {2009})},\
  \Eprint {http://arxiv.org/abs/0811.0188} {arXiv:0811.0188} \BibitemShut
  {NoStop}%
\bibitem [{\citenamefont {Peters}(1964)}]{Peters1964}%
  \BibitemOpen
  \bibfield  {author} {\bibinfo {author} {\bibfnamefont {P.~C.}\ \bibnamefont
  {Peters}},\ }\href {\doibase 10.1103/PhysRev.136.B1224} {\bibfield  {journal}
  {\bibinfo  {journal} {Phys.\ Rev.}\ }\textbf {\bibinfo {volume} {136}},\
  \bibinfo {pages} {B1224} (\bibinfo {year} {1964})}\BibitemShut {NoStop}%
\bibitem [{\citenamefont {Brink}\ \emph {et~al.}(2015)\citenamefont {Brink},
  \citenamefont {Geyer},\ and\ \citenamefont {Hinderer}}]{Brink2013}%
  \BibitemOpen
  \bibfield  {author} {\bibinfo {author} {\bibfnamefont {J.}~\bibnamefont
  {Brink}}, \bibinfo {author} {\bibfnamefont {M.}~\bibnamefont {Geyer}}, \ and\
  \bibinfo {author} {\bibfnamefont {T.}~\bibnamefont {Hinderer}},\ }\href
  {\doibase 10.1103/PhysRevLett.114.081102} {\bibfield  {journal} {\bibinfo
  {journal} {Phys.\ Rev.\ Lett.}\ }\textbf {\bibinfo {volume} {114}},\ \bibinfo
  {pages} {081102} (\bibinfo {year} {2015})},\ \Eprint
  {http://arxiv.org/abs/1304.0330} {arXiv:1304.0330} \BibitemShut {NoStop}%
\bibitem [{\citenamefont {Gair}\ \emph {et~al.}(2012)\citenamefont {Gair},
  \citenamefont {Yunes},\ and\ \citenamefont {Bender}}]{Gair2012}%
  \BibitemOpen
  \bibfield  {author} {\bibinfo {author} {\bibfnamefont {J.}~\bibnamefont
  {Gair}}, \bibinfo {author} {\bibfnamefont {N.}~\bibnamefont {Yunes}}, \ and\
  \bibinfo {author} {\bibfnamefont {C.~M.}\ \bibnamefont {Bender}},\ }\href
  {\doibase 10.1063/1.3691226} {\bibfield  {journal} {\bibinfo  {journal} {J.\
  Math.\ Phys.}\ }\textbf {\bibinfo {volume} {53}},\ \bibinfo {pages} {032503}
  (\bibinfo {year} {2012})},\ \Eprint {http://arxiv.org/abs/1111.3605}
  {arXiv:1111.3605} \BibitemShut {NoStop}%
\bibitem [{\citenamefont {Kevorkian}(1971)}]{Kevorkian1971}%
  \BibitemOpen
  \bibfield  {author} {\bibinfo {author} {\bibfnamefont {J.}~\bibnamefont
  {Kevorkian}},\ }\href {\doibase 10.1137/0120039} {\bibfield  {journal}
  {\bibinfo  {journal} {SIAM J.\ Appl.\ Math.}\ }\textbf {\bibinfo {volume}
  {20}},\ \bibinfo {pages} {364} (\bibinfo {year} {1971})}\BibitemShut
  {NoStop}%
\bibitem [{\citenamefont {Olver}\ \emph {et~al.}(2010)\citenamefont {Olver},
  \citenamefont {Lozier}, \citenamefont {Boisvert},\ and\ \citenamefont
  {Clark}}]{Olver2010}%
  \BibitemOpen
  \bibinfo {editor} {\bibfnamefont {F.~W.~J.}\ \bibnamefont {Olver}}, \bibinfo
  {editor} {\bibfnamefont {D.~W.}\ \bibnamefont {Lozier}}, \bibinfo {editor}
  {\bibfnamefont {R.~F.}\ \bibnamefont {Boisvert}}, \ and\ \bibinfo {editor}
  {\bibfnamefont {C.~W.}\ \bibnamefont {Clark}},\ eds.,\ \href@noop {} {\emph
  {\bibinfo {title} {{NIST Handbook of Mathematical Functions}}}}\ (\bibinfo
  {publisher} {Cambridge University Press},\ \bibinfo {address} {Cambridge},\
  \bibinfo {year} {2010})\BibitemShut {NoStop}%
\end{thebibliography}%

\end{document}